\documentclass[openacc]{rstransa}

\titlehead{Research}

\usepackage[dvipsnames]{xcolor}
\usepackage{tikz}
\usetikzlibrary{calc}
\usetikzlibrary{trees}
\usepackage{color,hyperref,listings}
\usepackage[shortlabels]{enumitem}
\usepackage{algorithm}
\usepackage{algorithmic}
\usepackage{amssymb}
\usepackage{tkz-euclide}
\usepackage{siunitx}
\usepackage{physics}
\usepackage{multicol}
\usepackage{pgfplots}
\pgfplotsset{compat=1.18}
\usepackage{adjustbox}
\usepackage{cancel}
\usepackage{commath}
\usepackage{graphicx}
\usepackage{mathdots}
\usepackage{tabularx}
\usepackage{mathtools}
\usepackage{booktabs}
\usepackage{environ}
\usepackage{cleveref}
\usepackage{fancyhdr}
\usepackage{centernot}
\usepackage[american,siunitx,arrowmos]{circuitikz}
\usepackage[english]{babel}
\usepackage{fontawesome}
\usepackage{listings}
\usepackage{fixmath}
\usepackage{xspace}
\usepackage{listings}
\usepackage{amsthm}
\usepackage{upgreek}
\usepackage{amsmath,amsfonts}
\usepackage{array}
\usepackage[caption=false,font=normalsize,labelfont=sf,textfont=sf]{subfig}
\usepackage{textcomp}
\usepackage{stfloats}
\usepackage{url}
\usepackage{verbatim}
\usepackage{cite}
\usepackage{graphicx}
\usepackage{prettyref}
\usepackage{wrapfig}
\usepackage{centernot}
\usepackage{mdframed}
\usepackage{framed}
\usepackage{tabularx}
\newcommand{\Ebb}{\mathbb{E}}

\newcommand{\Nbb}{\mathbb{N}}

\newcommand{\xv}{\mathbf{x}}
\newcommand{\yv}{\mathbf{y}}

\newcommand{\Acal}{\mathcal{A}}

\newcommand{\Ccal}{\mathcal{C}}
\newcommand{\Lcal}{\mathcal{L}}

\newcommand{\Mcal}{\mathcal{M}}
\newcommand{\Scal}{\mathcal{S}}
\newcommand{\Zcal}{\mathcal{Z}}

\renewcommand{\Gamma}{\Upgamma}
\renewcommand{\Theta}{\Uptheta}
\renewcommand{\Omega}{\Upomega}

\newcommand{\convas}{\stackrel{\text{a.s.}}{\longrightarrow}}

\newcommand{\indic}[1]{\mathbf{1}\left\{#1\right\}}

\newcommand{\simiid}{\stackrel{\text{i.i.d.}}{\sim}}

\DeclareMathOperator*{\argmin}{arg\,min}

\def\ie{\textit{i.e.}\@\xspace}
\def\eg{\textit{e.g.}\@\xspace}

\newtheorem{theorem}{Theorem}[section]

\newtheorem{lemma}[theorem]{Lemma}
\newtheorem{corollary}[theorem]{Corollary}
\newtheorem{fact}[theorem]{Fact}

\theoremstyle{definition}
\newtheorem{definition}[theorem]{Definition}

\newtheorem{example}[theorem]{Example}

\newtheorem{exercise-easy}[theorem]{Exercise}
\newtheorem{exercise-med}[theorem]{Exercise}
\newtheorem{exercise-hard}[theorem]{Exercise$^\star$}

\newtheorem*{claim*}{Claim}

\newtheorem{remark}[theorem]{Remark}
\newtheorem*{remark*}{Remark}

\newtheorem*{observation*}{Observation}

\newcommand{\savehyperref}[2]{\texorpdfstring{\hyperref[#1]{#2}}{#2}}

\newrefformat{eq}{\savehyperref{#1}{\textup{(\ref*{#1})}}}
\newrefformat{ineq}{\savehyperref{#1}{\textup{(\ref*{#1})}}}
\newrefformat{eqn}{\savehyperref{#1}{\textup{(\ref*{#1})}}}
\newrefformat{lem}{\savehyperref{#1}{Lemma~\ref*{#1}}}
\newrefformat{con}{\savehyperref{#1}{Construction~\ref*{#1}}}
\newrefformat{def}{\savehyperref{#1}{Definition~\ref*{#1}}}
\newrefformat{thm}{\savehyperref{#1}{Theorem~\ref*{#1}}}
\newrefformat{cor}{\savehyperref{#1}{Corollary~\ref*{#1}}}
\newrefformat{cha}{\savehyperref{#1}{Chapter~\ref*{#1}}}
\newrefformat{sec}{\savehyperref{#1}{Section~\ref*{#1}}}
\newrefformat{app}{\savehyperref{#1}{Appendix~\ref*{#1}}}
\newrefformat{tab}{\savehyperref{#1}{Table~\ref*{#1}}}
\newrefformat{fig}{\savehyperref{#1}{Figure~\ref*{#1}}}
\newrefformat{hyp}{\savehyperref{#1}{Hypothesis~\ref*{#1}}}
\newrefformat{alg}{\savehyperref{#1}{Algorithm~\ref*{#1}}}
\newrefformat{rem}{\savehyperref{#1}{Remark~\ref*{#1}}}
\newrefformat{item}{\savehyperref{#1}{Item~\ref*{#1}}}
\newrefformat{step}{\savehyperref{#1}{step~\ref*{#1}}}
\newrefformat{conj}{\savehyperref{#1}{Conjecture~\ref*{#1}}}
\newrefformat{fact}{\savehyperref{#1}{Fact~\ref*{#1}}}
\newrefformat{prop}{\savehyperref{#1}{Proposition~\ref*{#1}}}
\newrefformat{prob}{\savehyperref{#1}{Problem~\ref*{#1}}}
\newrefformat{claim}{\savehyperref{#1}{Claim~\ref*{#1}}}
\newrefformat{relax}{\savehyperref{#1}{Relaxation~\ref*{#1}}}
\newrefformat{rem}{\savehyperref{#1}{Remark~\ref*{#1}}}
\newrefformat{red}{\savehyperref{#1}{Reduction~\ref*{#1}}}
\newrefformat{part}{\savehyperref{#1}{Part~\ref*{#1}}}
\newrefformat{ex}{\savehyperref{#1}{Exercise~\ref*{#1}}}
\newrefformat{property}{\savehyperref{#1}{Property~\ref*{#1}}}
\newrefformat{type}{\savehyperref{#1}{Type~\ref*{#1}}}
\newrefformat{eg}{\savehyperref{#1}{Example~\ref*{#1}}}
\newrefformat{obs}{\savehyperref{#1}{Observation~\ref*{#1}}}

\definecolor{deepblue}{rgb}{0,0,0.5}
\definecolor{deepred}{rgb}{0.6,0,0}
\definecolor{deepgreen}{rgb}{0,0.5,0}

\newcommand{\lztrans}{\mathcal{T}^{\text{LZ}}}

\begin{document}

\title{Information-computation trade-offs in non-linear transforms}


\author{
Connor Ding$^{1}$, Abhiram Rao Gorle$^{1}$, Jiwon Jeong$^{1}$, Naomi Sagan$^{1}$ and Tsachy Weissman$^{1}$}

\address{$^{1}$Stanford University, Dept. of Electrical Engineering}

\subject{Information theory, data compression, machine learning}


\corres{Tsachy Weissman\\
\email{tsachy@stanford.edu}}

\begin{abstract}
In this work, we explore the interplay between information and computation in non-linear transform-based compression for broad classes of modern information-processing tasks.
We first investigate two emerging nonlinear data transformation frameworks for image compression: \textit{Implicit Neural Representations} (INRs) and \textit{2D Gaussian Splatting} (GS). We analyze their representational properties, behavior under lossy compression, and convergence dynamics.
Our results highlight key trade-offs between INR’s compact, resolution-flexible neural field representations and GS’s highly parallelizable, spatially interpretable fitting, providing insights for future hybrid and compression-aware frameworks.
Next, we introduce the textual transform 
that enables efficient compression at ultra-low bit rates-regimes, 
and simultaneously enhances human perceptual satisfaction.
When combined with the concept of denoising via lossy compression, the textual transform becomes a powerful tool for denoising tasks.
Finally, we describe a Lempel–Ziv (LZ78) transform, a universal method that, when applied to any member of a broad compressor family, produces new compressors that retain the asymptotic universality guarantees of the LZ78 algorithm.
Collectively, these three transforms illuminate the fundamental trade-offs between coding efficiency and computational cost.
We discuss how these insights extend beyond compression to tasks such as classification, denoising, and generative AI, suggesting new pathways for using non-linear transformations to balance resource constraints and performance. 
\end{abstract}



\maketitle

\section{Implicit Neural Representations and 2D Gaussian Splatting}

Data transformation lies at the core of modern lossy compression, enabling signals to be represented in domains where redundancy can be more effectively reduced. Classical image compression techniques, such as JPEG, rely on linear transformations like the Discrete Cosine Transform (DCT) to compact energy into a small set of coefficients, thereby facilitating efficient quantization and entropy coding. However, while effective for many natural images, linear transforms struggle to capture the highly nonlinear structures and spatial dependencies that often occur in complex image content, particularly for textures, fine geometric details, and sharp edges.

To address these limitations, recent research has shifted toward exploring nonlinear transformation frameworks that more closely align with the underlying statistics of natural signals. In this work, we focus on two emerging nonlinear representations for 2D images: \textit{Implicit Neural Representations} (INRs)~\cite{sitzmann2020implicitneuralrepresentationsperiodic} and \textit{Gaussian Splatting} (GS)~\cite{zhang2024gaussianimage}. Both methods replace conventional discrete pixel arrays with continuous, parametric representations, but they do so using fundamentally different inductive biases: INRs model images as continuous neural fields parameterized by neural networks, while GS models images as weighted sums of projected 2D Gaussians that explicitly encode spatial structure.

In this section of the paper, we conduct a comparative study of these two nonlinear transforms, with an emphasis on their potential for image compression. We first introduce the mathematical foundations of INRs and GS, followed by a detailed examination of their behavior under quantization and pruning. Furthermore, we investigate the effect of model capacity, convergence dynamics, and optimizer choice on both reconstruction quality and runtime efficiency. 

\begin{figure}[h!]
    \centering
    \begin{minipage}[b]{0.49\linewidth}
        \centering
        \includegraphics[width=\linewidth]{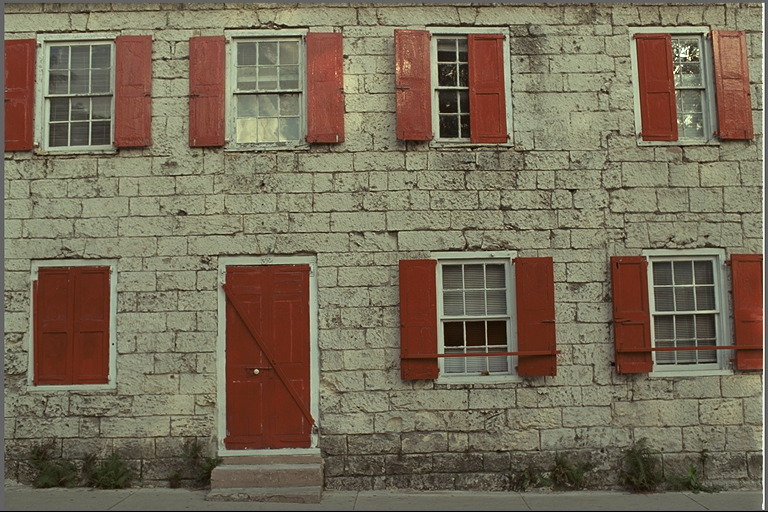}
        \caption{Example Kodak image 1 (\texttt{kodim01})}
        \label{fig:kodim01}
    \end{minipage}
    \hfill
    \begin{minipage}[b]{0.49\linewidth}
        \centering
        \includegraphics[width=\linewidth]{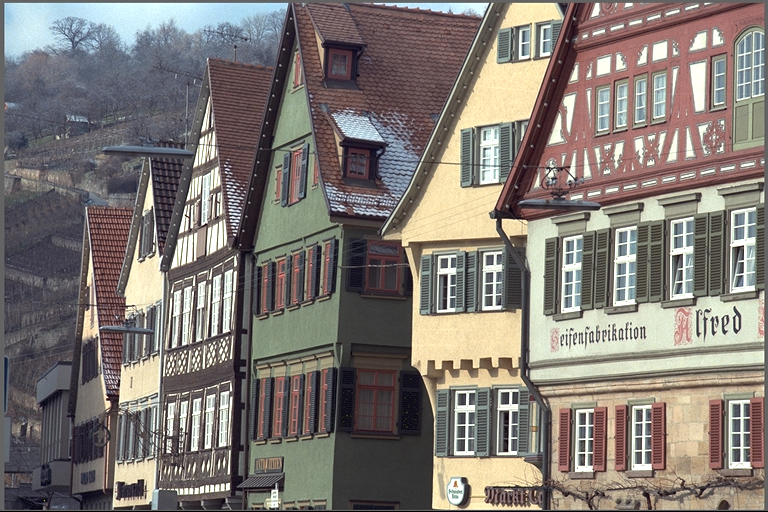}
        \caption{Example Kodak image 8 (\texttt{kodim08})}
        \label{fig:kodim08}
    \end{minipage}
\end{figure}

Our empirical evaluations are conducted on the Kodak dataset~\cite{kodak1991}, a widely used benchmark consisting of natural photographs with a resolution of $768\times512$, chosen for its relevance in image compression research. All experiments are performed on an NVIDIA A100 GPU hosted on Google Colab. Representative example images from the dataset are shown in Figures~\ref{fig:kodim01} and \ref{fig:kodim08}.

Overall, our results highlight key trade-offs between INRs and Gaussian Splatting in terms of compression effectiveness, encoding speed, interpretability, and sensitivity to quantization. These insights contribute toward the broader goal of rethinking image compression as a nonlinear transform learning problem, and point toward future hybrid frameworks that may combine the complementary strengths of neural fields and explicit splat-based representations.

\vspace{-1cm}

\subsection{Neural Transform}

In this section, we explore neural network-based nonlinear transforms for image compression, specifically those based on multi-layer perceptrons (MLPs). We first introduce the architectural design of this transformation class, discuss its formulation for image compression, and review existing compression techniques applied to this framework.

\subsubsection{Background: Implicit Neural Representations}

Neural transforms are based on \textit{Implicit Neural Representations} (INRs), which model signals as continuous functions parameterized by neural networks, rather than conventional discrete array-based formats. For image data, an INR models the mapping between continuous spatial coordinates $(x, y)$ and RGB color values $(R, G, B)$ by fitting a neural network $f_\theta$. The network is trained to minimize reconstruction error against the target image $I(x,y)$ over all pixel coordinates:
\[
\sum \|f_\theta(x,y)-I(x,y)\|_2.
\]
A common architecture for $f_\theta$ is the \textit{Sinusoidal Representation Network} (SIREN) \cite{sitzmann2020implicitneuralrepresentationsperiodic}, which employs sinusoidal activation functions to enable the network to capture high-frequency components of the signal. Specifically, for a hidden layer $h$, the activation is given by 
\[
\sin \left( w_0 (Wh + b) \right),
\]
where $W$ is the weight matrix, $b$ is the bias vector, and $w_0$ is a frequency scaling factor that controls the spectral bandwidth of the network. This choice of activation allows SIRENs to efficiently model fine-grained image features and repeating structures that are often difficult for standard activations like ReLU or tanh \cite{sitzmann2020implicitneuralrepresentationsperiodic}. In addition, due to their continuous nature, INRs provide attractive properties such as resolution-independent sampling and parameter efficiency.

\begin{figure}[h!]
    \centering
    \begin{minipage}[b]{0.49\linewidth}
        \centering
        \includegraphics[width=\linewidth]{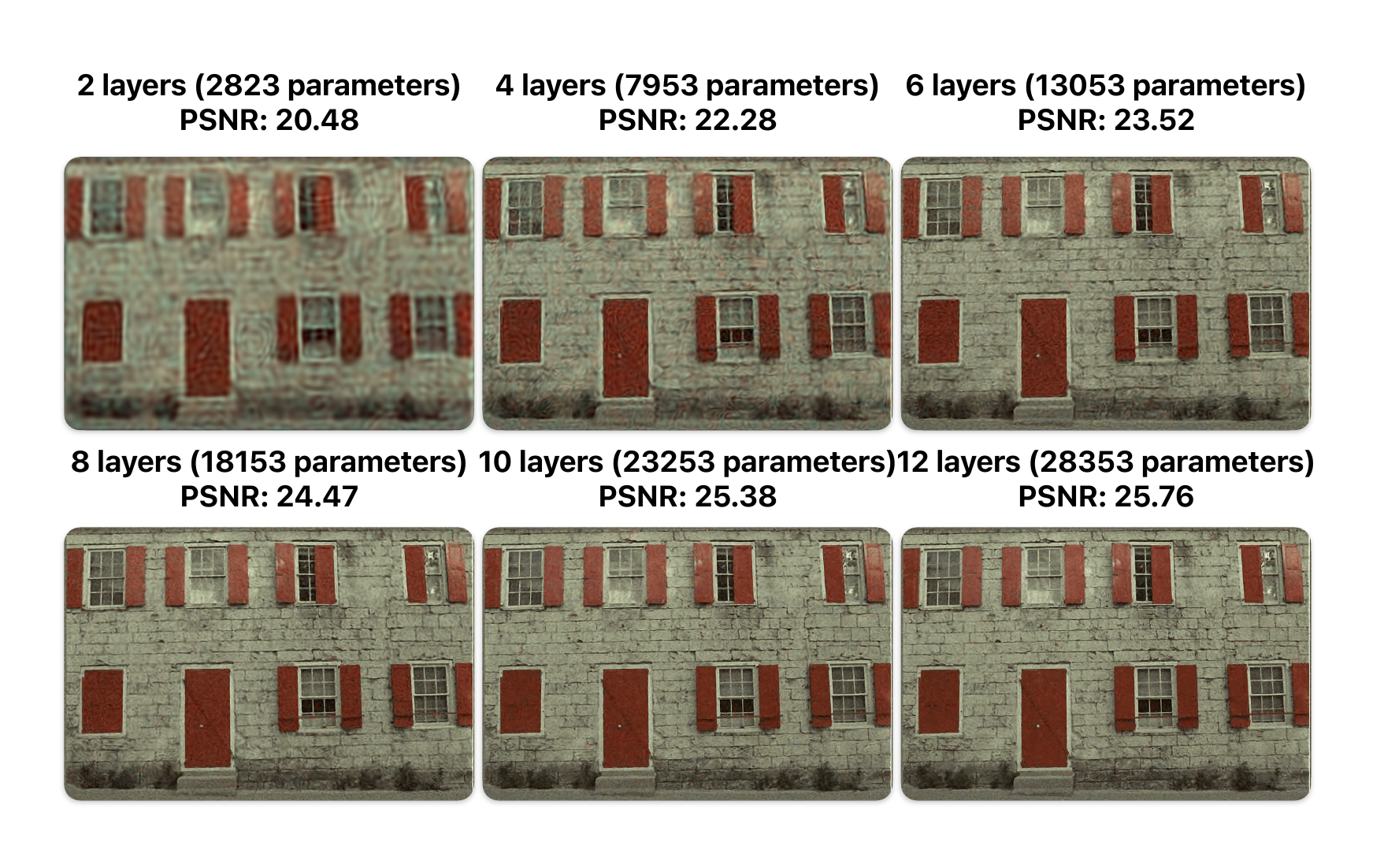}
        \caption{Reconstructed images with INR of different layer sizes}
        \label{fig:inr-all}
    \end{minipage}
    \hfill
    \begin{minipage}[b]{0.49\linewidth}
        \centering
        \includegraphics[width=\linewidth]{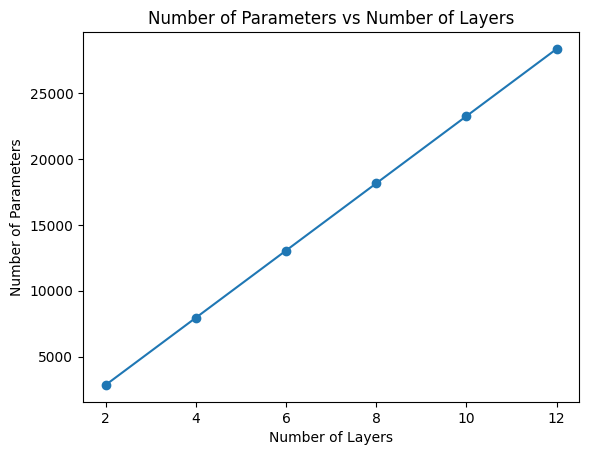}
        \caption{Parameters vs Number of Layers}
        \label{fig:param_v_layers}
    \end{minipage}
\end{figure}

To visualize the behavior of INRs, we fit a SIREN model to the first image in the Kodak dataset (\texttt{kodim01}). We fix the hidden layer width to 50, such that each layer contains weight matrices of size $50\times 50$ and bias vectors of length 50. The frequency scaling constant is set to $w_0 = 30$, following prior work \cite{dupont2021coincompressionimplicitneural}. Figure~\ref{fig:inr-all} shows reconstructed images from SIREN models with varying numbers of layers, alongside their corresponding parameter counts and PSNR values. As expected, deeper networks consistently produce better reconstructions, with sharper edges and finer detail recovery. This is because greater network depth allows the model to represent higher-frequency components of the image via multiple layers of sinusoidal modulation. 

It is important to note that these visualizations are intended for qualitative illustration rather than optimal SIREN architecture design. In particular, we do not tune for fixed total parameter counts across depths, nor optimize the tradeoff between width and depth. Rather, the experiments demonstrate the general effect that increased depth improves the capacity of the model to capture high-frequency image content.

\begin{figure}[h!]
    \centering
    \begin{minipage}[b]{0.98\linewidth}
        \centering
        \includegraphics[width=\linewidth]{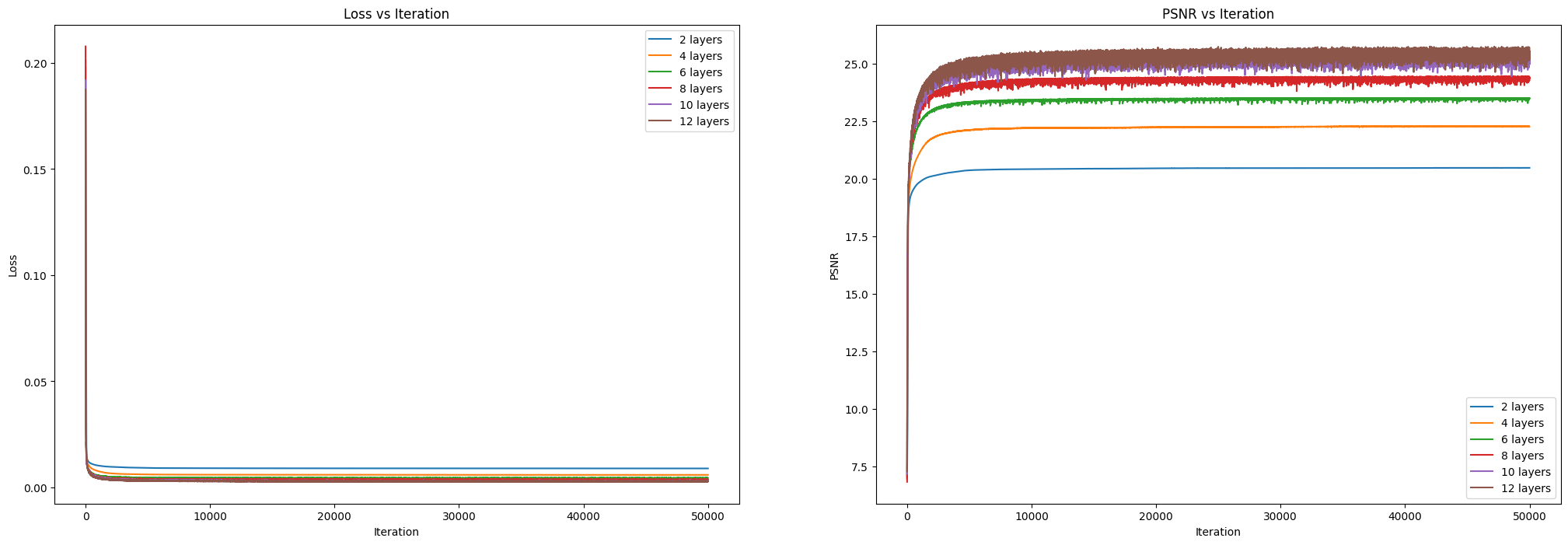}
        \caption{Convergence plot}
        \label{fig:convergence-inr}
    \end{minipage}
    \begin{minipage}[b]{0.98\linewidth}
        \centering
        \includegraphics[width=\linewidth]{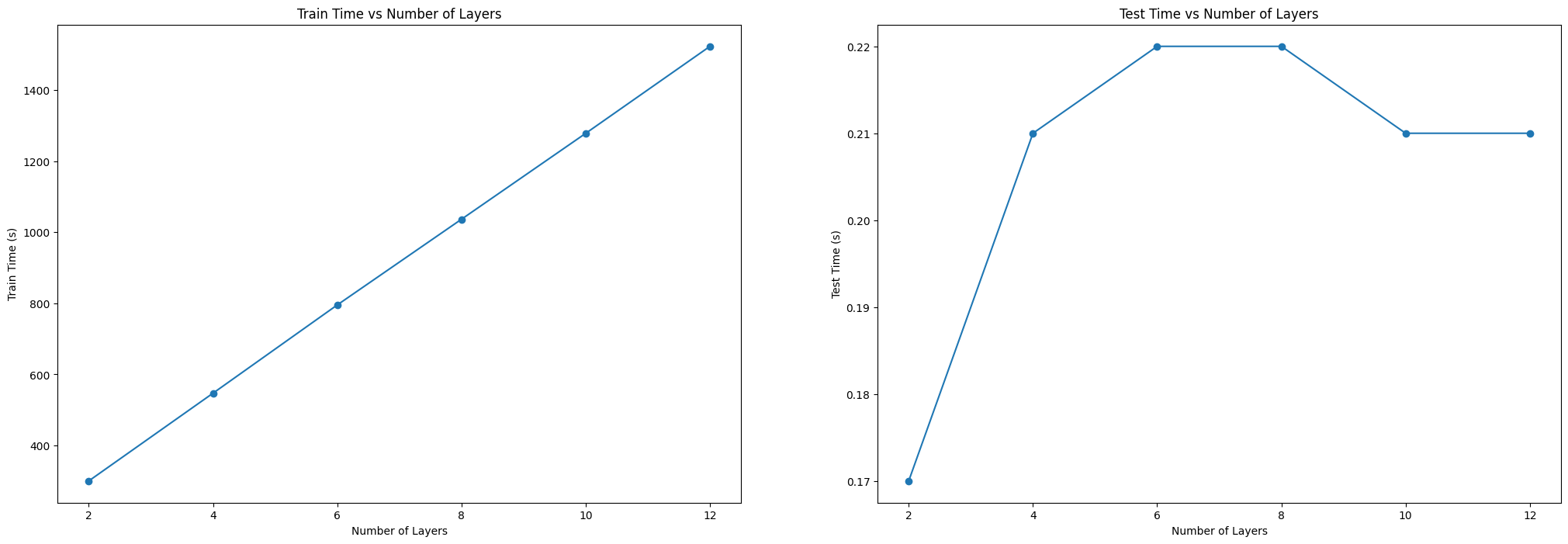}
        \caption{Training and Inference time}
        \label{fig:enc_dec_inr}
    \end{minipage}
\end{figure}

In addition to the qualitative reconstructions, we also analyze the parameter counts, convergence behavior, and computational efficiency of INRs with varying depths. Figure~\ref{fig:param_v_layers} shows that the total number of parameters increases linearly with network depth, as the hidden layer width is kept constant across all models. The training curves in Figure~\ref{fig:convergence-inr} indicate that while all models converge at comparable rates, deeper networks consistently reach lower final loss values and achieve higher PSNR. However, the convergence trajectories for deeper models tend to exhibit more fluctuations, potentially due to increased model capacity interacting with optimization dynamics. 

Importantly, these reconstruction improvements come at the cost of higher training time, which scales approximately linearly with depth as shown in Figure~\ref{fig:enc_dec_inr} (left). Interestingly, the inference time (Figure~\ref{fig:enc_dec_inr}, right) grows with depth initially but plateaus beyond 8 layers, suggesting that additional depth has limited impact on forward evaluation latency once the computational overhead saturates. Collectively, these results highlight that deeper SIREN models offer greater representational power for high-fidelity image reconstruction, but with moderately increased training cost and largely stable inference speed.

\subsubsection{Using INR as a Compressor}

As shown in Figure~\ref{fig:param_v_layers}, even the largest SIREN model (with 12 layers) requires only a few thousand parameters to represent an image, which is substantially smaller than the original uncompressed image size of $H \times W \times 3 = 1179648$ pixels (for $H = 512$, $W = 768$). This significant parameter reduction illustrates how INRs serve as compact, continuous representations of images, effectively functioning as a nonlinear form of transform coding into what we may refer to as the \textit{neural domain}. Once an image has been fit into this neural representation, conventional neural network compression techniques—such as quantization and pruning—can be directly applied to the model weights to achieve further compression.

\begin{figure}[h!]
    \centering
    \includegraphics[width=\linewidth]{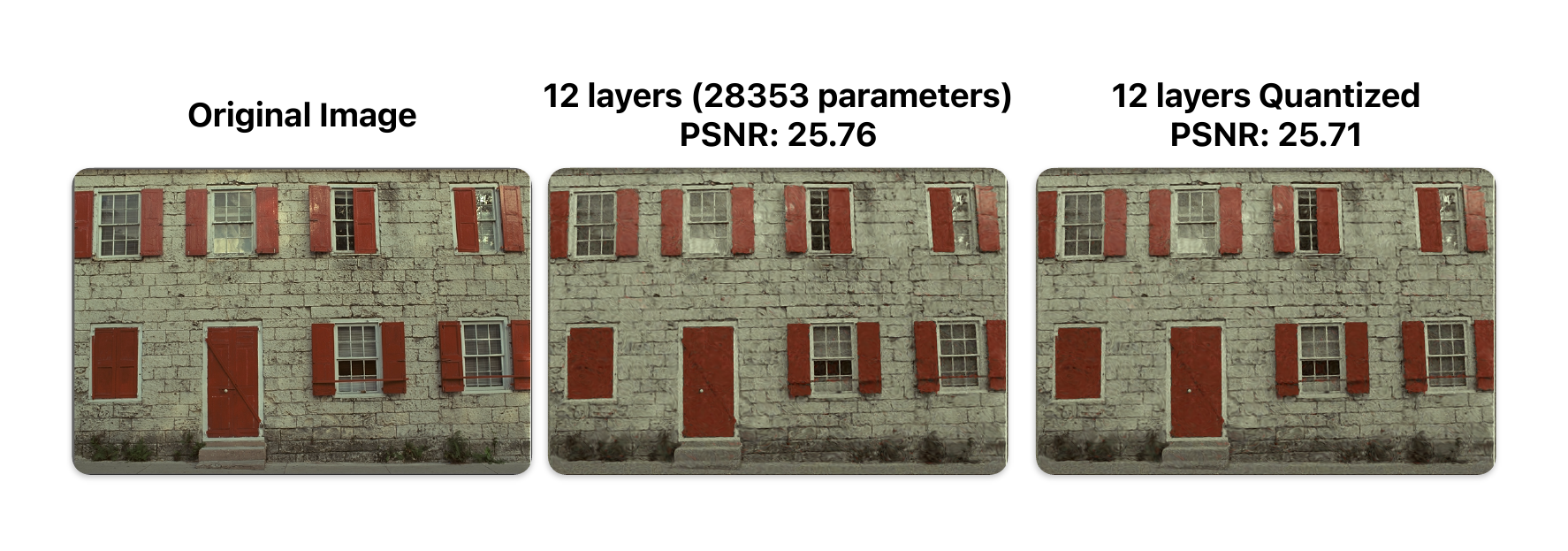}
    \caption{Replicating COIN \cite{dupont2021coincompressionimplicitneural}: Original vs. Reconstructed (32 bits) vs. Quantized (16 bits)}
    \label{fig:INR-quantization}
\end{figure}

The COIN framework, proposed by Dupont et al.~\cite{dupont2021coincompressionimplicitneural}, is one of the first works to formalize INR-based compression. After fitting an INR to the image, COIN applies uniform quantization to the network weights, reducing their precision from 32-bit floating point to 16-bit fixed-point representations. This effectively reduces storage by half while preserving high reconstruction quality with minimal degradation in PSNR. In Figure~\ref{fig:INR-quantization}, we replicate the quantization behavior from COIN and observe that aggressive weight quantization yields only marginal loss in reconstruction performance, further demonstrating the compressibility of INR representations.

\vspace{-1cm}

\vspace{-1cm}

\subsubsection{Improving COIN--Beyond Basic Quantization}

While the COIN framework~\cite{dupont2021coincompressionimplicitneural} introduced a novel paradigm for image compression by transforming data from the pixel domain into the neural domain, its original implementation remains impractical for real-world deployment. In particular, COIN underperforms conventional codecs both in terms of bitrate efficiency and encoding speed. Nevertheless, COIN opened an important new perspective: once an image is transformed into a neural representation, classical neural network compression tools such as quantization, pruning, and entropy modeling can be applied directly to the network weights. Building on this foundation, numerous extensions have been proposed to improve both the efficiency and effectiveness of INR-based compression, spanning architectural design, learning strategies, and probabilistic modeling.

One line of work builds upon COIN by introducing meta-learning frameworks. Instead of training a full neural representation from scratch for each image, a large base network is first meta-trained across a dataset. For each new image, only a lightweight modulation network is trained to adapt the base representation to the specific image content. This substantially accelerates encoding since only the modulation parameters, rather than the full network weights, need to be optimized and transmitted~\cite{dupont2022coinneuralcompressionmodalities, strümpler2022implicitneuralrepresentationsimage}.

Other approaches integrate entropy modeling directly into the INR framework. By jointly optimizing for both distortion and bitrate through a rate-distortion objective $D + \lambda R$, these methods learn entropy models conditioned on the INR parameters that enable adaptive and compact bitstream encoding. The entropy models are typically small auxiliary networks that predict probability distributions over the weights or activations, allowing for more efficient entropy coding during compression~\cite{ladune2023coolchiccoordinatebasedlowcomplexity, kim2023c3highperformancelowcomplexityneural}. More recent works adopt a probabilistic perspective using Bayesian INRs. These models place a prior distribution $p(\theta)$ over the network weights and optimize a variational posterior $q(\theta | \mathcal{D})$ for each image. Compression is then achieved by minimizing the Kullback-Leibler (KL) divergence between the posterior and prior while jointly minimizing reconstruction error. The resulting posterior samples can be encoded using relative entropy coding, allowing for principled trade-offs between uncertainty, expressiveness, and bitrate~\cite{guo2023compressionbayesianimplicitneural, he2024recombiner}.

Together, these developments demonstrate that while early INR-based compressors such as COIN primarily served as proof-of-concept, subsequent advances have pushed INR compression toward more practical, efficient, and expressive regimes, grounded in both functional representation and information-theoretic principles.

\subsubsection{Pruning INR Weights}

The aforementioned INR-based compression methods have largely focused on quantization and entropy coding, but have not fully explored one of the most classical neural network compression techniques: pruning. By selectively zeroing out small or insignificant network weights, pruning can reduce both model size and computation while preserving much of the original reconstruction quality. Since INR-based image compression operates by encoding an image into the weights of a neural network $f_\theta(x; w)$, applying pruning directly on $w$ can be interpreted as transform coding in the neural domain. Analogous to zeroing out high-frequency components in Fourier-based transform coding, pruning removes weights that contribute minimally to reconstruction, potentially yielding sparse and highly efficient neural representations.

Isik et al.~\cite{isik2022informationtheoreticjustificationmodelpruning} proposed an information-theoretic justification for pruning, showing that neural network weights are often approximately Laplacian distributed, making them successively refinable sources. Based on this property, they introduced the \textit{Successive Refinement for Pruning} (SuRP) algorithm, which iteratively refines the model by transmitting the most significant weights in a priority order. The sender starts from a fully trained model, selects weight indices according to their magnitudes, and transmits updates to the receiver, who incrementally reconstructs a sparse version of the model. This iterative process continues until the target sparsity level is reached. Isik et al. demonstrated SuRP's effectiveness primarily on classification networks such as ResNets and convolutional architectures.

\begin{figure}[h!]
    \centering
    \includegraphics[width=0.8\linewidth]{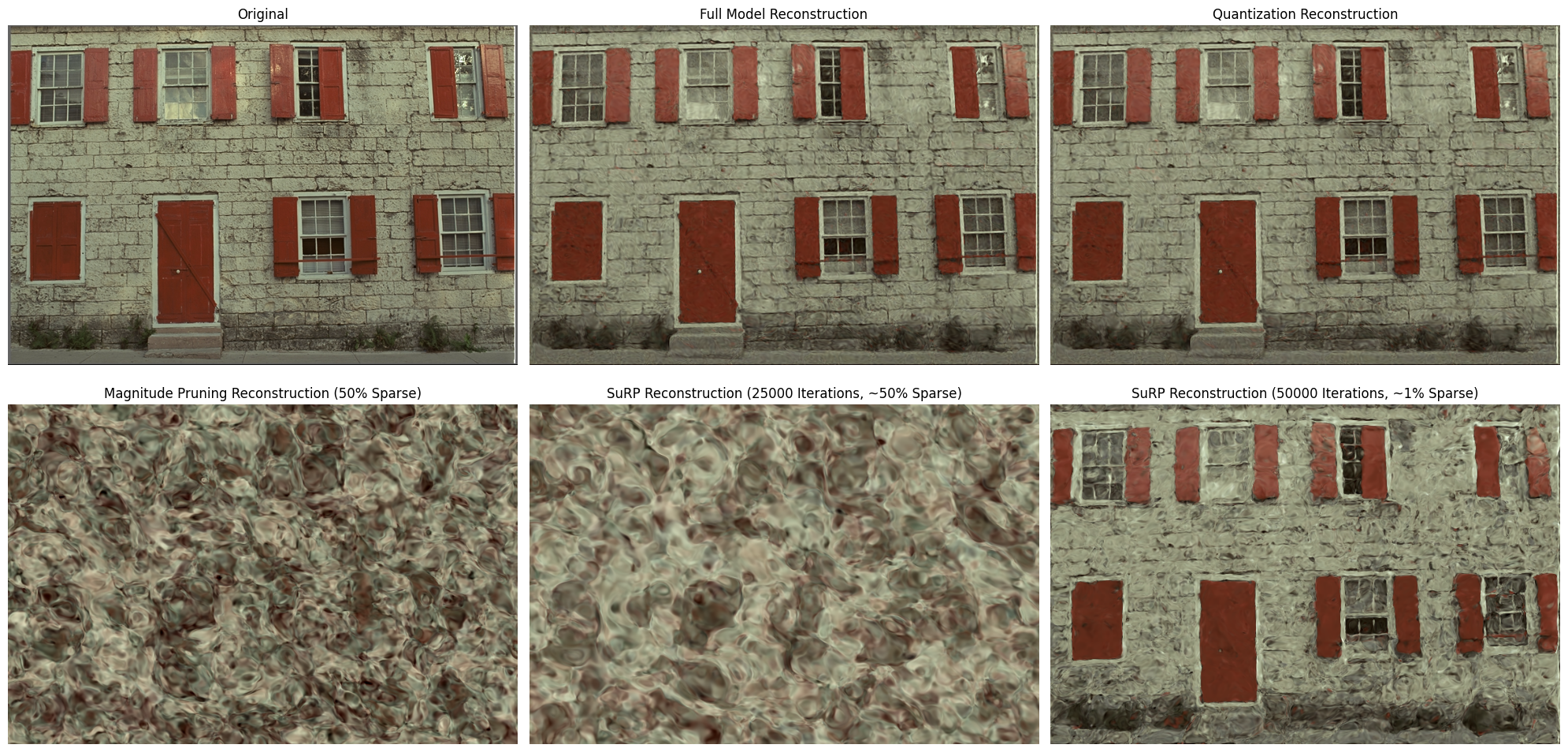}
    \caption{Original Image, Full Model Reconstruction, Quantization Magnitude Pruning, and SuRP at two different stages}
    \label{fig:SuRP_demo}
\end{figure}

Figure~\ref{fig:SuRP_demo} presents qualitative comparisons of the reconstructed images at various pruning stages. We include the original image, the full INR reconstruction, the quantized reconstruction (following the COIN approach), magnitude-based pruning at 50\% sparsity, and SuRP reconstructions at two different levels of sparsity (approximately 50\% and 99\%). Several key observations emerge from these visualizations.

While both quantization and pruning reduce model size, their effects on image quality differ significantly. Quantization with reasonable bit-depth generally preserves most of the global structure, with some expected smoothing of fine textures. In contrast, both magnitude pruning and SuRP pruning show that sparsification leads to a rapid degradation of fine-grained image details. This is especially evident for the magnitude-based pruning: even at 50\% sparsity, the image quickly collapses into severe texture distortions, losing most spatial structures.

Overall, these results suggest that while pruning can introduce sparsity into INR models, its utility for image compression remains limited. Unlike classification networks where many weights may be redundant, the SIREN weights participate more uniformly in capturing high-frequency content across the image. This makes INR weight-space much more sensitive to aggressive pruning. Therefore, naive weight sparsification may not be the most effective compression strategy for INR-based codecs, and may need to be combined with more structured sparsity, better regularization during training, or alternative parameterizations.



\subsection{2D Gaussian Splatting Transform}
While INR-based compression approaches demonstrate strong performance in compact signal representation, they remain limited by two fundamental challenges: (1) the high encoding complexity involved in training neural networks for each new image, and (2) the black-box nature of neural network weights, which offers limited interpretability or transparency. For many practical compression scenarios—such as real-time encoding, streaming, or deployment on resource-constrained edge devices—these limitations restrict the applicability of INR-based codecs.

To address these challenges, a recent line of work proposes an alternative family of transformations based on Gaussian Splatting that shows promising results in speeding up the encoding time and providing interpretability of the transformation. Building upon 3D Gaussian Splatting developed for real-time neural rendering \cite{kerbl20233dgaussiansplattingrealtime}, Zhang et al. introduced \emph{GaussianImage} \cite{zhang2024gaussianimage}, which extends the Gaussian Splatting framework to 2D images. This approach models images using collections of parameterized 2D Gaussian kernels, whose parameters can be directly optimized to fit the target image.

The 2D Gaussian Splatting (2D-GS) transform inherits several key advantages: its encoding process, while still iterative, is significantly faster than INR training; its decoding (rendering) process is highly parallelizable and extremely fast, often requiring only a simple rasterization-like procedure; and its parameters offer more interpretability, as each Gaussian corresponds to a spatially localized primitive with explicit position, shape, and color information.

In the following subsections, we present the mathematical formulation of 2D-GS, provide empirical evaluations of its fitting and compression behavior, and discuss recent extensions and open research directions for further improving its efficiency and compression capability.

\subsubsection{Background}
2D Gaussian Splatting originates from recent developments in 3D view synthesis. In particular, 3D Gaussian Splatting has demonstrated highly efficient and high-quality rendering for novel view synthesis tasks \cite{kerbl20233dgaussiansplattingrealtime}. In this framework, a 3D scene is parameterized by a collection of anisotropic 3D Gaussians, where each Gaussian is associated with a set of attributes: 3D mean (position), covariance matrix (shape and orientation), color coefficients, opacity, and spherical harmonics for view-dependent effects. Rendering is performed via differentiable $\alpha$-blending, where the scene is projected and accumulated using the Gaussian parameters. These parameters are optimized through gradient descent to minimize the reconstruction loss between the rendered and ground truth views.

Since its introduction, 3D Gaussian Splatting has rapidly grown into an active area of research, with work focusing on accelerating both training and inference, improving memory efficiency, developing compression algorithms for Gaussian splats, and applying the technique to broader vision and graphics problems. We refer interested readers to the recent comprehensive surveys \cite{chen2025survey3dgaussiansplatting, 10870258} for a detailed overview of the state of 3D Gaussian Splatting research.

\subsubsection{Image Representation with 2D Gaussian Splatting}
Inspired by 3D Gaussian Splatting, Zhang et al. \cite{zhang2024gaussianimage} introduced Gaussian Splatting for 2D images, where an image is represented as a mixture of $N$ Gaussians. Each Gaussian is described by a set of parameters: position $\mu \in \mathbb{R}^2$, covariance matrix $\Sigma \in \mathbb{R}^{2\times 2}$, and color coefficients $c \in \mathbb{R}^3$. Due to the symmetry of the covariance matrix, each Gaussian can be fully specified using 8 parameters: 2 for position, 3 for covariance (via Cholesky factorization), and 3 for color.

The rendered color of a pixel located at $\begin{bmatrix} x \\ y \end{bmatrix}$ is computed by aggregating the contributions from nearby Gaussians through:

\begin{equation}
    c(x, y) = \sum_{n \in \mathcal{N}} c_n \exp{\left\{ -\left( \mu_n - \begin{bmatrix} x \\ y \end{bmatrix} \right)^{T} \Sigma_n^{-1} \left( \mu_n - \begin{bmatrix} x \\ y \end{bmatrix} \right) \right\}},
    \label{eq:render-equation}
\end{equation}

where $\mathcal{N}$ denotes the set of Gaussians whose influence overlaps with the current pixel.

To fit an image, the Gaussian parameters are initialized randomly and optimized via gradient descent by minimizing the standard $\ell_2$ loss between the reconstructed and target image:

\begin{equation}
    \sum_{(x,y)} \|\hat{I}(x,y) - I(x,y)\|_2,
\end{equation}

where $\hat{I}$ is the image reconstructed via Eq.~\ref{eq:render-equation}, and $I$ is the ground truth image.

\begin{figure}[h!]
    \centering
    \includegraphics[width=0.75\linewidth]{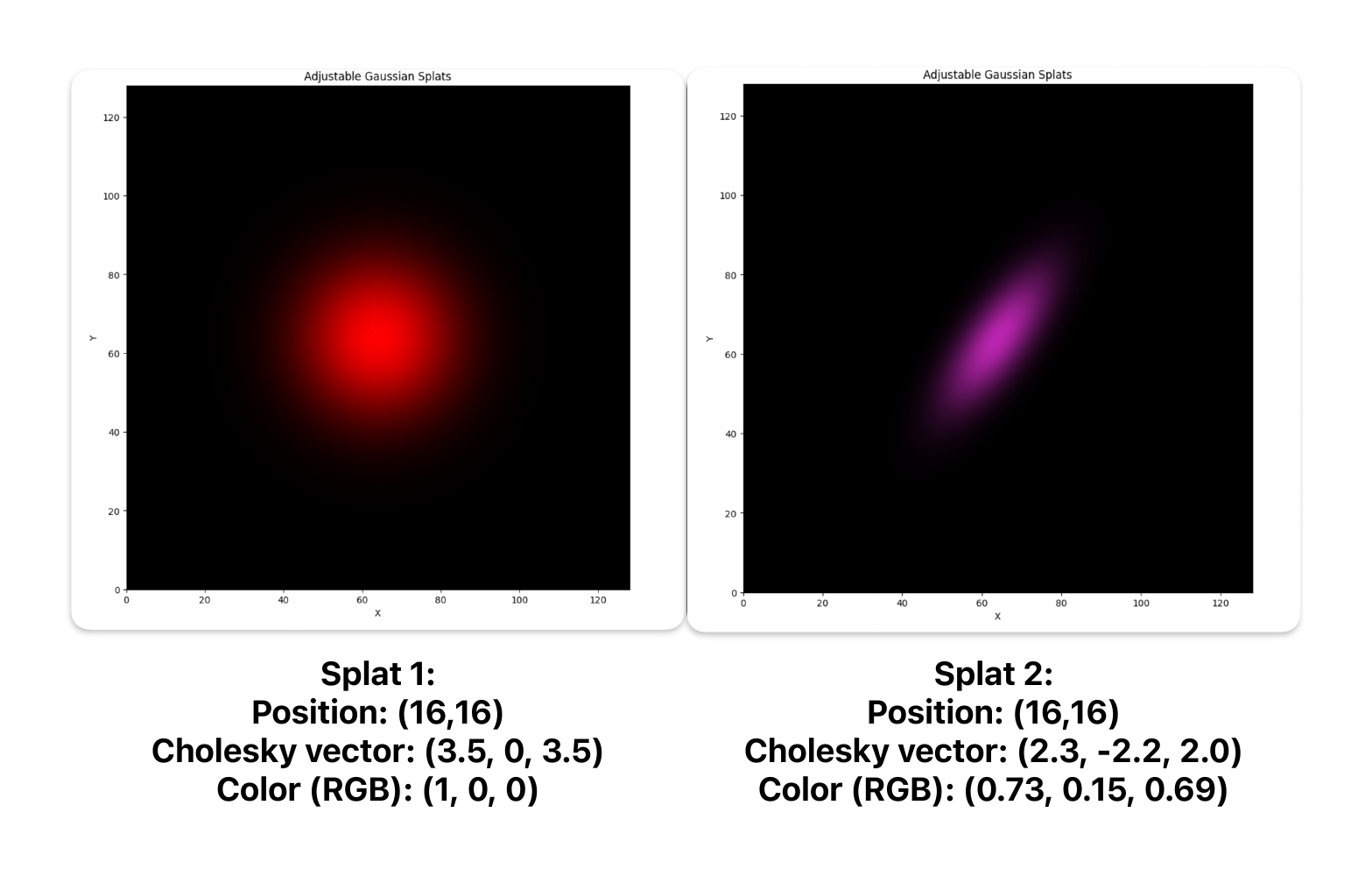}
    \caption{Example Splats with the same position but different covariance matrices and color coefficients}
    \label{fig:example-splats}
\end{figure}

Compared to INR-based representations, the rendering function for 2D-GS is much simpler, as it does not involve any neural network evaluations or activations. Crucially, the gradients of the rendering function can be analytically derived \cite{zhang2024gaussianimage}, which allows efficient implementation of backpropagation without invoking PyTorch’s automatic differentiation engine. This leads to substantially faster optimization and evaluation. In Figure~\ref{fig:example-splats}, we illustrate two example splats that share the same spatial location but differ in their covariance structures and color coefficients. These primitives form the building blocks for image reconstruction in 2D Gaussian Splatting.

\begin{figure}[h!]
    \centering
    \begin{minipage}[b]{0.49\linewidth}
        \centering
        \includegraphics[width=\linewidth]{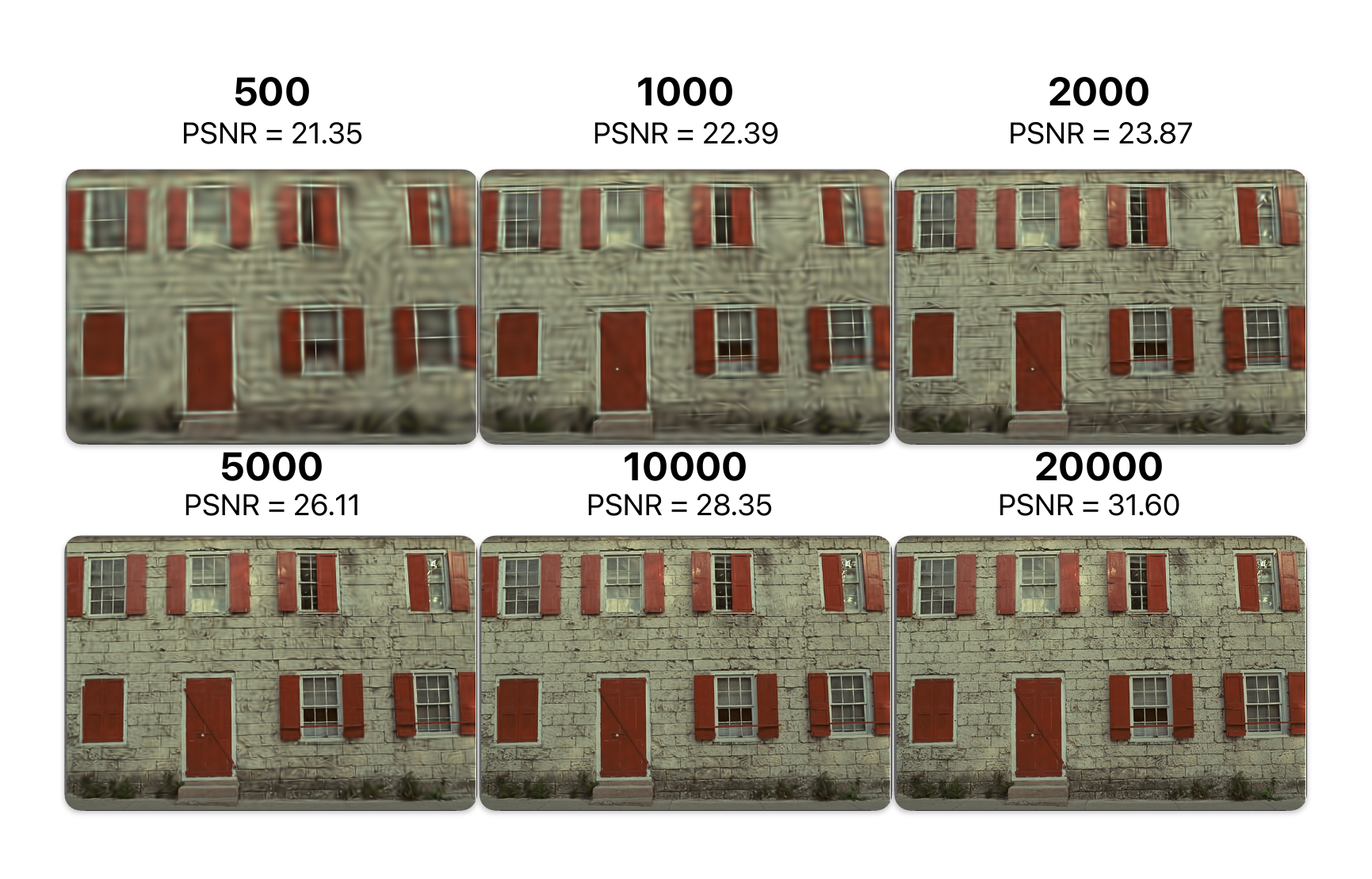}
        \caption{Fitted image with different numbers of Gaussian Splats}
        \label{fig:fitted_gs}
    \end{minipage}
    \hfill
    \begin{minipage}[b]{0.49\linewidth}
        \centering
        \includegraphics[width=\linewidth]{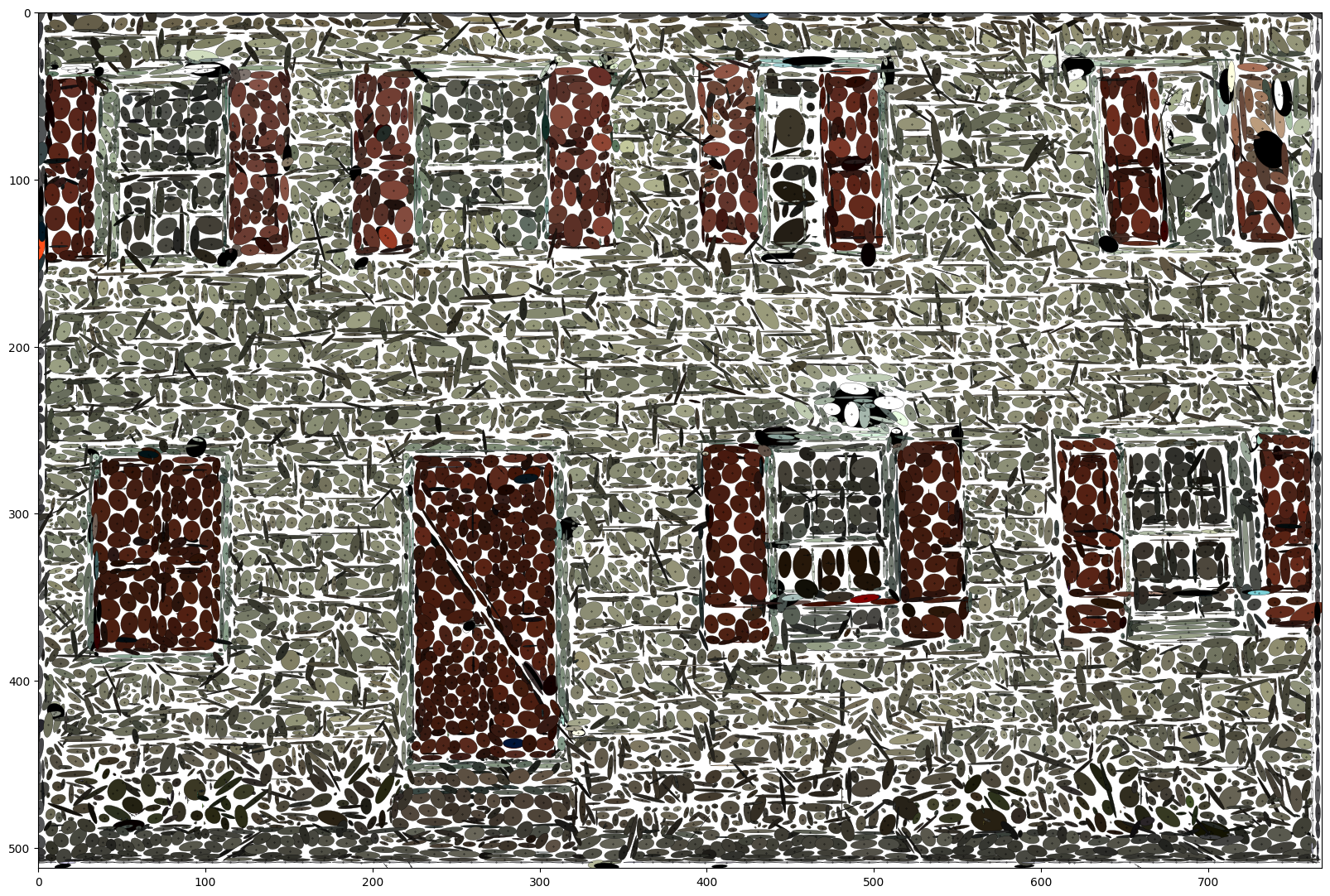}
        \caption{Ellipsoids $(N=10000)$}
        \label{fig:ellipsoids}
    \end{minipage}
\end{figure}

We executed the scripts from \cite{zhang2024gaussianimage} on an example image in the Kodak dataset and observe the following. Figure~\ref{fig:fitted_gs} shows the reconstruction with different numbers of Gaussian splats. One can see that the PSNR values and visual quality get better as the number of Gaussians used increases. At low capacity (e.g., 500 splats), the reconstruction is visibly blurry and lacks high-frequency structure. As the number of splats increases to 10,000 and beyond, the reconstructed image becomes significantly sharper and visually more faithful to the ground truth. The PSNR rises from 21.35 at 500 splats to 31.60 at 20,000 splats, confirming the benefit of greater capacity in capturing spatial and color detail. To further understand the structure behind the reconstruction, we visualize the fitted Gaussians in Figure~\ref{fig:ellipsoids}. Each ellipse represents a 2D Gaussian's position, shape, and color. We observe that ellipsoids show anisotropic shapes, suggesting that the fitting process adapts both the scale and orientation of each splat to local image geometry. These results demonstrate how 2D Gaussian splatting can compactly and adaptively encode image content using a relatively small number of continuous primitives.

\vspace{-1cm}

\subsubsection{Convergence and Runtime Analysis}
\begin{figure}[h!]
    \centering
    \begin{minipage}[b]{0.98\linewidth}
        \centering
        \includegraphics[width=\linewidth]{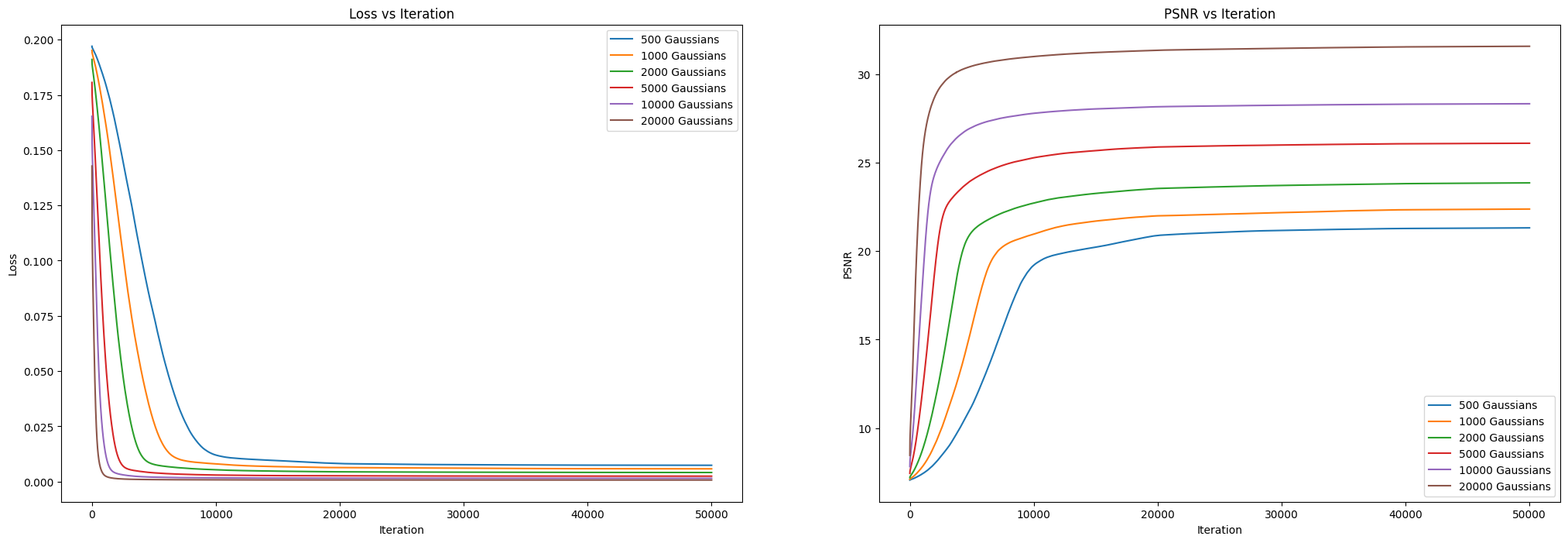}
        \caption{Convergence plot across different iterations of Gaussians}
        \label{fig:convergence}
    \end{minipage}
    \begin{minipage}[b]{0.98\linewidth}
        \centering
        \includegraphics[width=\linewidth]{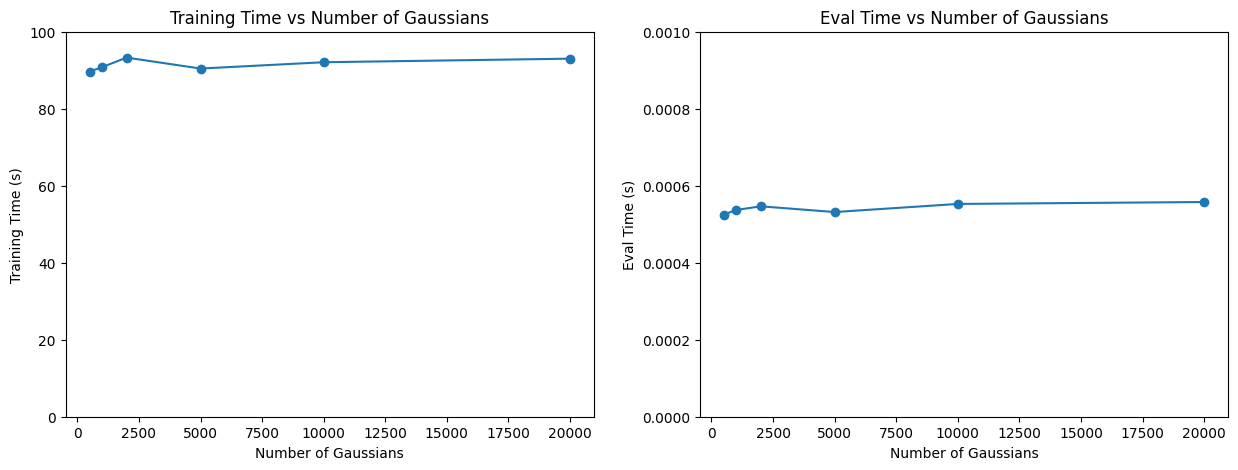}
        \caption{Encoding and Decoding time of Gaussian splats, as a function of number of Gaussians.}
        \label{fig:enc_dec_gs}
    \end{minipage}
\end{figure}

We further analyze the convergence behavior and computational performance of the 2D Gaussian splatting method. As shown in Figure~\ref{fig:convergence}, increasing the number of Gaussians leads to faster and smoother convergence of both the loss and PSNR curves. Larger models exhibit not only higher final PSNR values but also reach low-error regimes in fewer iterations. This confirms that overparameterization helps in fitting the image more effectively, likely due to the ability to model local structures with more spatially specific splats.

Despite increasing the number of splats, the runtime results in Figure~\ref{fig:enc_dec_gs} show a strikingly flat trend. Both training time and evaluation time remain nearly constant across all configurations, indicating that the implementation scales well and fully utilizes GPU parallelism. The training time is dominated by fixed overhead and GPU kernel launch costs, rather than per-splat computation, while the evaluation time per image remains in the sub-millisecond range even with 20,000 Gaussians. These findings highlight one of the major advantages of Gaussian splatting over implicit neural representations: fast, parallelizable rendering and optimization, even at high capacity.

\subsubsection{Use in Compression}
Analogous to implicit neural representations (INRs), the image is transformed from the standard RGB domain $\mathbb{R}^{H\times W\times 3}$ to the Gaussian Splatting (GS) domain $\mathbb{R}^{N\times 8}$, where $N$ is the number of Gaussians used to parameterize the image. This transformation provides a compact and continuous representation of the image, enabling the application of various compression techniques directly in the GS domain. For example, one could quantize the parameters, apply entropy coding, or exploit spatial correlations among the Gaussians to reduce the bitrate. The original GaussianImage \cite{zhang2024gaussianimage} framework, in addition to introducing the image fitting paradigm, attempted at applying it to image compression by separately quantizing the Gaussian attributes (positions, covariances and colors). The biggest advantage of the compressor is its decoding speed, while in terms of rate-distortion performances, it shows some advantages over JPEG and COIN at certain bit-rate levels. 

\begin{figure}[h!]
    \centering
    \begin{minipage}[b]{0.32\linewidth}
        \centering
        \includegraphics[width=\linewidth]{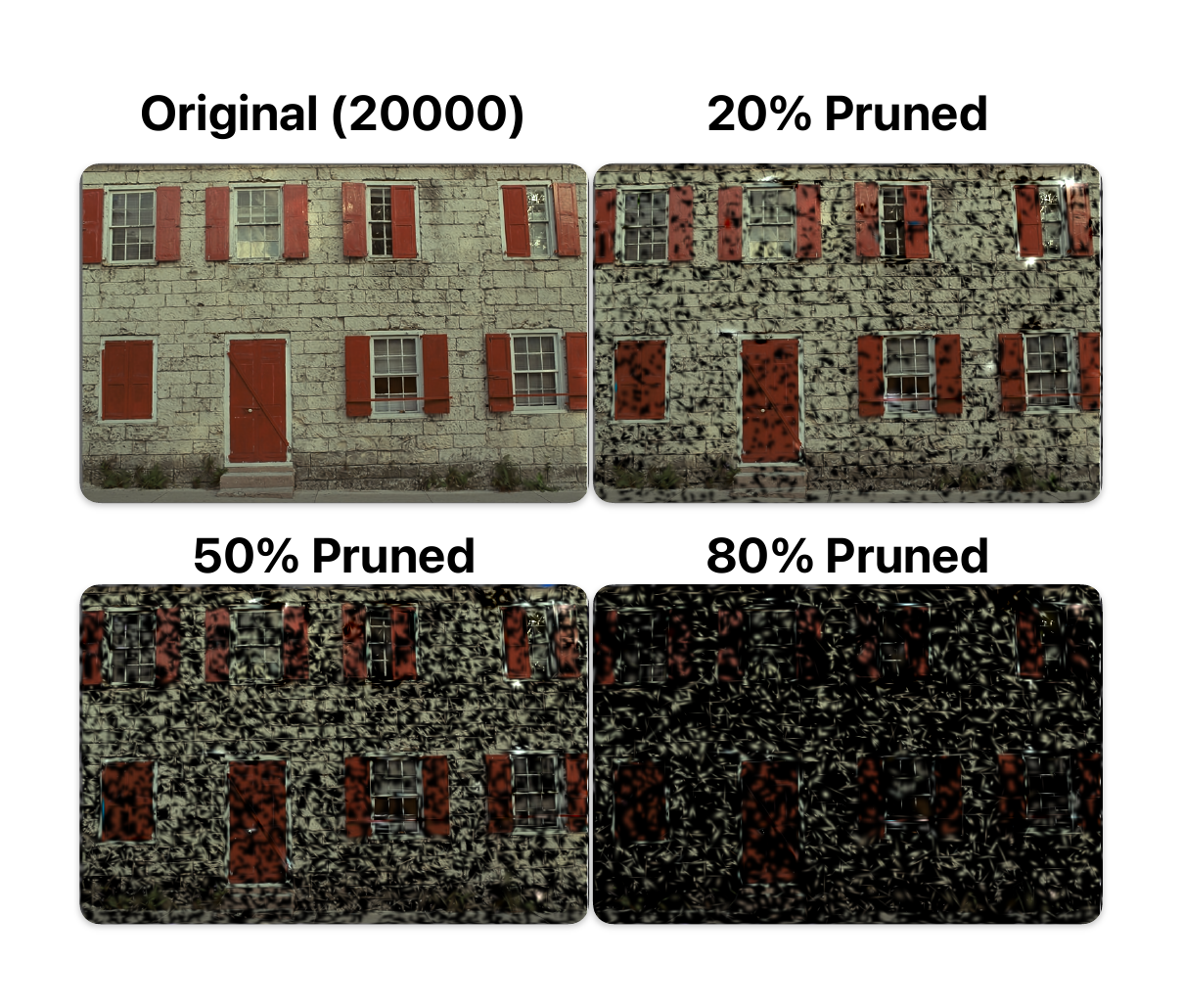}
        \caption{Pruning Different \% of Gaussians}
        \label{fig:pruned}
    \end{minipage}
    \hfill
    \begin{minipage}[b]{0.64\linewidth}
        \centering
        \includegraphics[width=\linewidth]{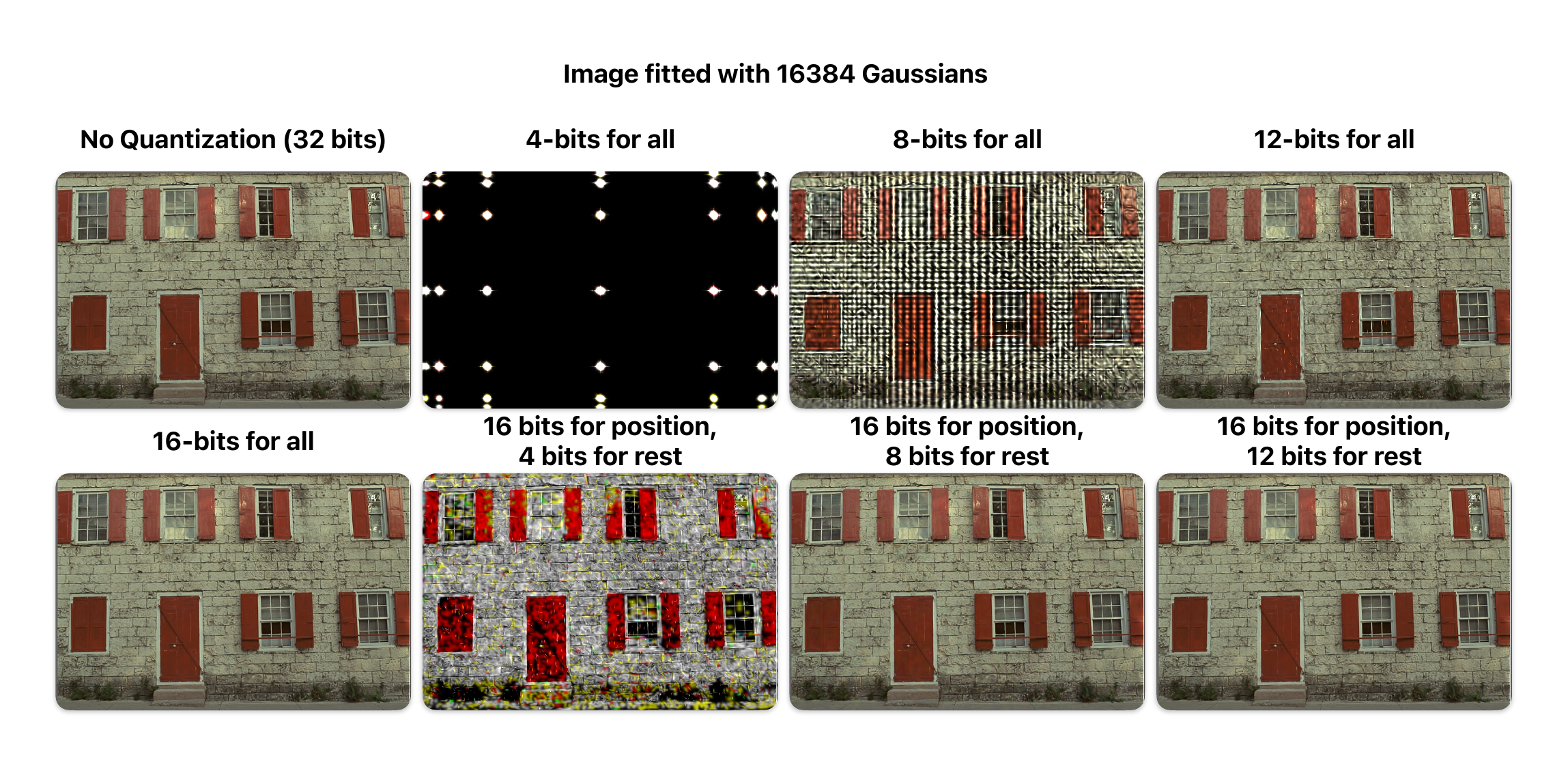}
        \caption{Quantizing Guassian splat parameters to different numbers of bits.}
        \label{fig:quantized}
    \end{minipage}
\end{figure}

To evaluate the compressibility and robustness of 2D Gaussian Splatting, we investigate two post-encoding techniques: pruning and quantization. For pruning, we randomly remove a certain percentage of the 20000 Gaussians fitted to an image with no further fine-tuning or adjustment, as shown in Figure~\ref{fig:pruned}. We can see that removing some Gaussians directly cause degradation to the images, with the level of degradation directly proportional to the number of Gaussians removed. Unlike network pruning, each Gaussian is directly responsible for rendering a particular region of the image, so removing that Gaussian will remove the corresponding region as well. This shows that further optimization of Gaussian parameters is needed to preserve image fidelity. The question of whether fitting a large amount of Gaussian primitives followed by pruning then refining results in a better fitted image than simply initializing the desired number of Gaussians from the start has yet to be investigated. 

Figure~\ref{fig:quantized} shows the results of applying various quantization strategies to the Gaussian parameters. A key insight is that the position component of each Gaussian is highly sensitive to quantization: using fewer than 16 bits for positions (especially 4 or 8 bits) leads to visible artifacts such as blockiness, loss of alignment, and even total collapse of the representation in extreme cases. In contrast, the other parameters (e.g., scale, color) are more robust to quantization, and can be compressed more aggressively without major visual degradation. The observation directly aligns with the findings in GaussianImage \cite{zhang2024gaussianimage}. Besides keeping the positions as 16 bits, optimized the number of bits for uniform quantization for the covariance parameters and applied residual vector quantization for the color coefficients. The findings suggest that future compression schemes should treat Gaussian position parameters as higher-precision channels, while more aggressively quantizing others to optimize for bitrate without sacrificing perceptual quality.

\subsubsection{Related Work and Future Directions}
Since 2D Gaussian Splatting is a very recent development, only a handful of works have emerged, primarily focusing on enhancing the representational power, improving encoding speed, and extending the approach to video compression. In \cite{zhu2025large}, the authors identified limitations of the original 2D-GS framework \cite{zhang2024gaussianimage} in fitting high-resolution images, and proposed a multi-level architecture that allocates a small number of Gaussians to capture the coarse structure and the rest to finer details. This hierarchical approach significantly improves reconstruction quality at lower computational cost. To accelerate the encoding process, \cite{tai2025eigengsrepresentationeigenspacegaussian} proposed using a learned initialization based on the eigenimage of a dataset. They fit a 2D-GS representation to this eigenimage and used it to initialize the fitting of individual images, which results in faster convergence and higher final PSNR compared to the random initialization used in \cite{zhang2024gaussianimage}.

The GS framework has also been extended to video representation and compression. Since video can be viewed as a temporal sequence of images, it is natural to model it with a series of Gaussian splat representations. \cite{gupta2025neuralvideocompressionusing, wang2025gsvcefficientvideorepresentation} explore strategies such as treating I-frames and P-frames differently. They propose initializing the splat parameters of P-frames using those from nearby I-frames, which leads to significantly faster convergence and better temporal coherence than fitting each frame independently from scratch.

Given the nascent stage of 2D Gaussian Splatting, there are numerous promising directions for future research to enhance its practical utility and theoretical foundations:
\begin{itemize}
    \item \textbf{Faster encoding}: Current encoding relies on iterative optimization, which is time-consuming. Developing learning-based encoders or meta-learned initialization strategies could significantly speed up the fitting process.
    \item \textbf{Compression-aware training}: Jointly optimizing splat parameters with compression constraints (e.g., bitrate control, quantization robustness) would make GS more suitable for real-world codecs.
    \item \textbf{Downstream tasks}: Beyond compression, GS representations may be used in downstream vision tasks such as segmentation, object detection, or style transfer, offering compact representations with strong inductive biases.
    \item \textbf{Video representation}: More principled temporal modeling, such as learning motion-aware splat transformations or hierarchical keyframe structures, could improve the efficiency of video encoding using GS.
    \item \textbf{Hardware acceleration}: Given the rasterization-like structure of GS rendering, specialized GPU kernels or hardware-aware optimization pipelines could further accelerate training and inference.
\end{itemize}

\subsection{Discussion}


\begin{table}[h!]
\small
\centering
\renewcommand{\arraystretch}{1.2}
\begin{tabular}{|p{3cm}|p{4.5cm}|p{4.5cm}|}
\hline
\textbf{Aspect} & \textbf{INR (Neural Fields)} & \textbf{Gaussian Splatting (GS)} \\
\hline
\textbf{Representation} & Continuous neural function & Parametric 2D Gaussians \\
\hline
\textbf{Encoding Time} & Slow (iterative optimization) & Fast (parallelizable fitting) \\
\hline
\textbf{Rendering Time} & Fast (MLP inference) & Very fast (rasterization) \\
\hline
\textbf{Interpretability} & Low (latent parameters) & High (spatially explicit) \\
\hline
\textbf{Compression Tools} & Quantization, pruning, variational modeling, entropy coding & Quantization, entropy coding \\
\hline
\textbf{Quantization Sensitivity} & Moderate & High (especially positions) \\
\hline
\textbf{Best Use Cases} & Continuous signal encoding & Fast, interpretable image/video encoding \\
\hline
\end{tabular}
\caption{Comparison between INR and Gaussian Splatting (GS) as data transformations for image compression.}
\label{tab:inr-vs-gs}
\end{table}

\vspace{-1cm}

Data transformation lies at the heart of effective lossy compression, enabling high-dimensional signals to be represented in domains where redundancy can be more efficiently reduced. While classical linear transforms such as the Discrete Cosine Transform (DCT) underpin legacy codecs like JPEG, their limited expressiveness for complex image structures has motivated the development of more powerful nonlinear alternatives. In this work, we investigated two emerging nonlinear transformation frameworks for images: \textit{Implicit Neural Representations} (INRs) and \textit{Gaussian Splatting} (GS). As summarized in Table~\ref{tab:inr-vs-gs}, these two approaches offer distinct advantages and trade-offs in the context of image compression.

INRs encode images as continuous neural fields, allowing compact parameterization and flexible resolution sampling. They are well-suited for modeling fine-grained structures and high-frequency details. However, they require expensive iterative optimization for each image, exhibit sensitivity to weight quantization and pruning, and lack interpretability due to the abstract nature of neural network weights. While techniques such as quantization, meta-learning, and variational modeling have been developed to improve INR-based compression, practical deployment remains limited by their encoding complexity.

In contrast, 2D Gaussian Splatting directly models images as a sum of spatially explicit Gaussian primitives. The rasterization-based rendering process enables extremely fast decoding, and the optimization procedure is highly parallelizable and substantially faster than INR training. The explicit spatial structure of GS makes its parameters more interpretable and more naturally suited for manipulation during compression. However, unlike INRs, GS does not offer flexible resolution sampling, as the rendering is tied to the discrete pixel grid. Our experiments demonstrated that GS achieves strong reconstruction performance across varying model sizes, with stable convergence behavior and flat runtime scaling. Nevertheless, we observed that GS representations are particularly sensitive to the quantization of position parameters, whereas other parameters such as covariance and color can tolerate aggressive compression.

Broadly, both INR and GS reflect a shift from classical array-based representations toward \textit{function-based} compression paradigms, where the signal representation itself is learned or optimized. These approaches open new opportunities for integrating representation learning with compression objectives. Future directions may include hybrid models that combine the flexibility of INRs with the fast, interpretable structure of GS, compression-aware training objectives that directly optimize for bitrate, and extensions to downstream tasks such as generation, perception, and video representation. As both paradigms continue to evolve, they offer promising pathways toward compact, expressive, and efficient image compression frameworks.

\section{Textual Transform Domain}


\subsection{Motivation}
The textual transform, introduced in \cite{weissman2023toward}, is grounded in the idea that text naturally aligns with how we perceive and describe concepts. Analogous to how Fourier transform represents a signal in terms of its frequency components, the textual transform describes a signal using words. For example, the textual transform of an image would consist of words corresponding to objects within the image, along with their attributes including size and location. 

Based on the fact that words can be interpreted as a code optimized in many years of evolution, the textual transform has a potential for achieving improved compression performance while maintaining high levels of human satisfaction from the reconstructions. It also exhibits some useful characteristics of an efficient transform. Words perform as meaningful basis functions in the transform domain as each word encapsulates rich semantic content, and it's possible to preserve proximity since similar inputs often yield similar text representation. 

An additional advantage is that the representation in the transform domain is human-readable and queryable. This not only enhances the interpretability of the signal and its transformed version, but also can be leveraged in denoising problem by identifying and filtering the words related to noise in the transform domain, as explored in later subsection.

\subsection{Lossy Compression via the Textual Transform}
A number of recent papers have explored the use of textual transforms to achieve ultra-low-rate compression, especially in image and video domain. In the image domain, \cite{lei2023text} showed how text can serve as a core component for compression by representing an image using an extremely low bitrate (below 0.003 bits-per-pixel) while still preserving meaningful semantics. They also addressed how additional sketch information can further improve the semantic quality. \cite{careil2023towards} presented how latent diffusion model combined with textual description can yield even higher image quality in terms of semantic-level metrics. \cite{arikan2024semantic} used human semantic satisfaction scores to show that textual transforms can often achieve a similar level of human satisfaction as images compressed by JPEG, despite operating at significantly lower bitrates. In the video domain, \cite{tandon2022txt2vid} proposed a video compression framework that extracts a text transcript to reduce the bitrate by two to three orders of magnitude, while maintaining a comparable Quality-of-Experience to the standard audio-video codecs.

\subsection{Denoising via the  Textual Transform}
We propose a new application domain for the textual transform: the denoising problem. This idea is motivated by the concept of ``denoising via lossy compression''. It was introduced in \cite{natarajan1995filtering} based on the fact that random noise is harder to compress compared to ordered information. The core idea is that, for a signal corrupted by additive white noise, effective denoising can be achieved by applying a lossy compression algorithm with an appropriately chosen distortion level. This technique has been theoretically justified in several papers that establish performance bounds \cite{natarajan1995filtering, weissman2005empirical, donoho2002kolmogorov}, and its practical effectiveness has also been demonstrated in various settings \cite{jalali2012denoising, ochoa2016effect}. 

Building on the fact that the textual transform offers strong compression capabilities and enables easy identification and removal of the components related to noise in transform domain thanks to its human-readability, we explore two text-based denoising approaches. We focus here  on denoising problems in the image domain for concreteness with the implication that the ideas carry over to other types of data. 

\subsubsection{Experimental Setup}
We use 512$\times$512 RGB or grayscale images, obtained by cropping the center of images from the Kodak dataset \cite{kodak1991}. Each pixel value is represented using an 8-bit integer. We consider three types of noise on image - Gaussian, Poisson, and salt-and-pepper noise. For Gaussian noise, which is a typical and widely studied type in image denoising literature, is modeled as additive white Gaussian noise with variance 2000. Poisson noise arises in imaging systems due to the discrete nature of photon detection, and we implement it using the following equation:
\[\Tilde{I}_\text{Poisson}(x,y) = \frac{1}{2} \text{Poisson} \left(2 I(x,y)\right),\]
where $I(x,y)$ is the pixel intensity at location $(x,y)$ and $\Tilde{I}_\text{Poisson}(x,y)$ is the pixel value at the same location in the image corrupted by Poisson noise. The scaling factor of 2 is introduced to simulate a low-light imaging condition. Salt-and-pepper noise for grayscale image randomly assigns some pixels to either 0 (pepper noise) or 255 (salt noise), and it's implemented by the equation below.
\[
\Tilde{I}_\text{Salt-and-pepper}(x, y) =
\begin{cases}
0, & \text{with probability } \frac{25}{510} \\
255, & \text{with probability } \frac{25}{510} \\
I(x, y), & \text{with probability } 1 - \frac{25}{255}
\end{cases}
\]
where $\Tilde{I}_\text{Salt-and-pepper}(x, y)$ is the pixel intensity at location $(x,y)$ of the image corrupted by salt-and-pepper noise.

\subsubsection{Denoising Guided by Text}

\begin{figure}
    \centering
    \includegraphics[width=1\linewidth]{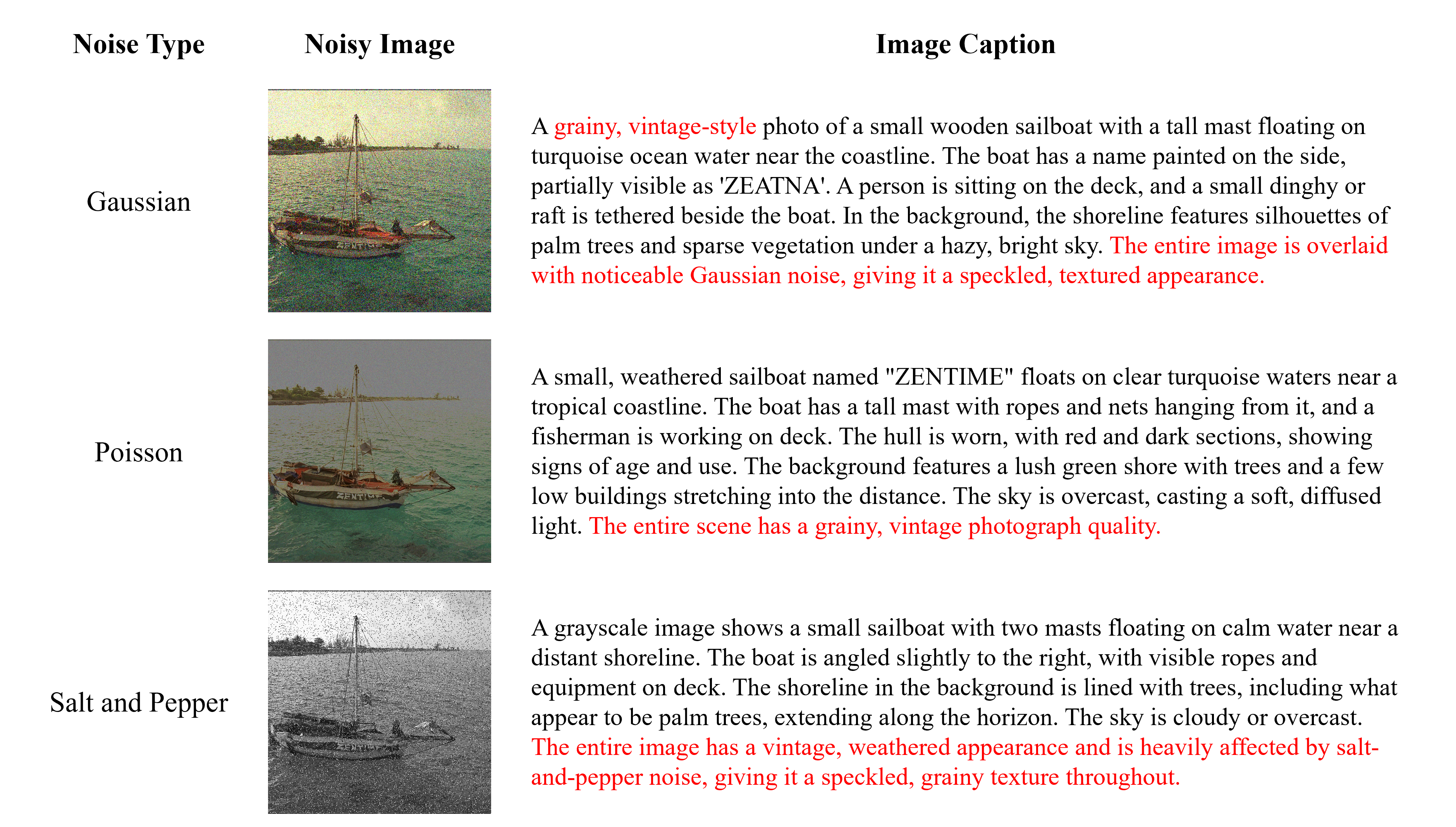}
    \caption{Image captions for noisy images generated by ChatGPT 4o. Red colored text, marked by human, describes the noise of the images.}
    \label{fig:denoise_text_caption}
\end{figure}

\begin{figure}
    \centering
    \includegraphics[width=1\linewidth]{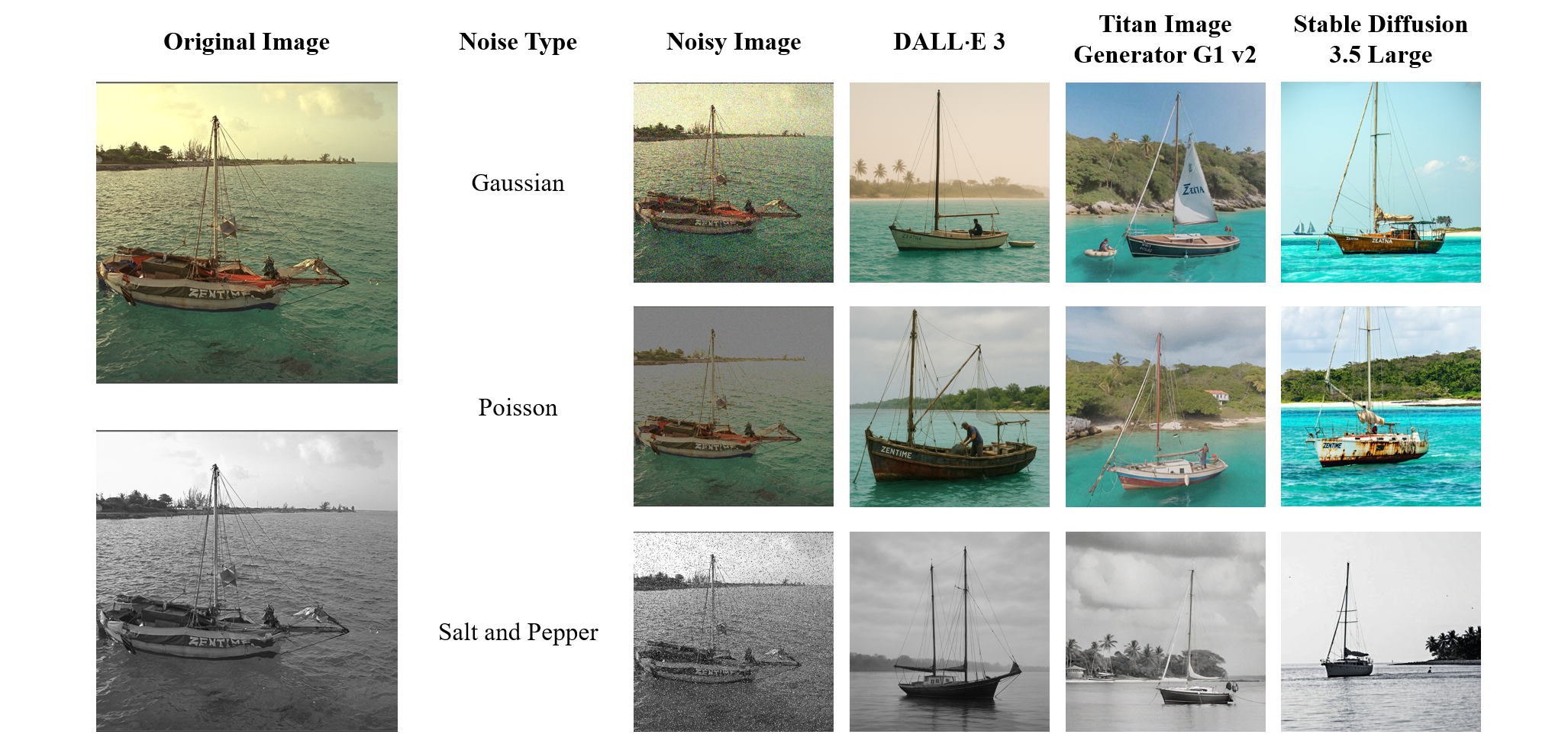}
    \caption{Results of denoising guided by text. Three columns on the right shows the denoised image generated by different text-to-image generation models. }
    \label{fig:denoise_text_result}
\end{figure}

Given a noisy image, we first extract its textual description using either a human annotator or a large language model (LLM). As illustrated in Figure \ref{fig:denoise_text_caption}, these image captions capture meaningful information about the input image including background, objects, orientation, color, and so on. Also, they describe the noise of the images using terms such as `grainy', `speckled', or `textured'. Using either human judgement or an LLM, we can filter out those words related to noise from the caption. Now we can use this filtered text as a prompt for a text-to-image generation model to retrieve the underlying noise-free image.

Figure \ref{fig:denoise_text_result} presents the denoised images, generated by three different text-to-image generation model using the filtered text from Figure \ref{fig:denoise_text_caption}. For Titan Image Generator G1 v2, we use a prompt strength of 8 to keep the image realistic while maintaining the semantics from the image caption. All images from Titan Image Generator G1 v2 are generated using the same random seed for consistency. We  observe that all three models generate highly realistic images that are free from the injected noie, while still preserving the essential semantics of the original image.

\subsubsection{Denoising Guided by Text and Lossily Compressed Sketch}
Building on the Text+Sketch framework \cite{lei2023text}, which incorporates an additional sketch input to guide text-to-image generation, we propose using a ``conditioning image'' to better steer the generation process. Since the Holistically-nested Edge Detection (HED) map extracted from noisy image, which was used in \cite{lei2023text}, often fails to capture the structure of the noise-free image, and the compression rate is not the primary objective of our denoising task, we instead introduce a lossily compressed version of the noisy image as a sketch input. Particularly, we employ the JPEG compression standard, which is based on DCT \cite{wallace1992jpeg}. Prior work has shown that shrinkage of empirical wavelet coefficients can effectively denoise signals corrupted by additive white Gaussian noise \cite{donoho1994ideal, donoho1995adapting, donoho1995denoising}. Although JPEG 2000 \cite{taubman2002jpeg2000} uses wavelet-based compression, we choose the original JPEG format for standardized compatibility. To assess the generality of our approach, we apply this scheme not only to images corrupted by Gaussian noise, but also to those affected by Poisson and salt-and-pepper noise, evaluating its effectiveness across diverse practical noise settings. 

\begin{figure}
    \centering
    \includegraphics[width=1\linewidth]{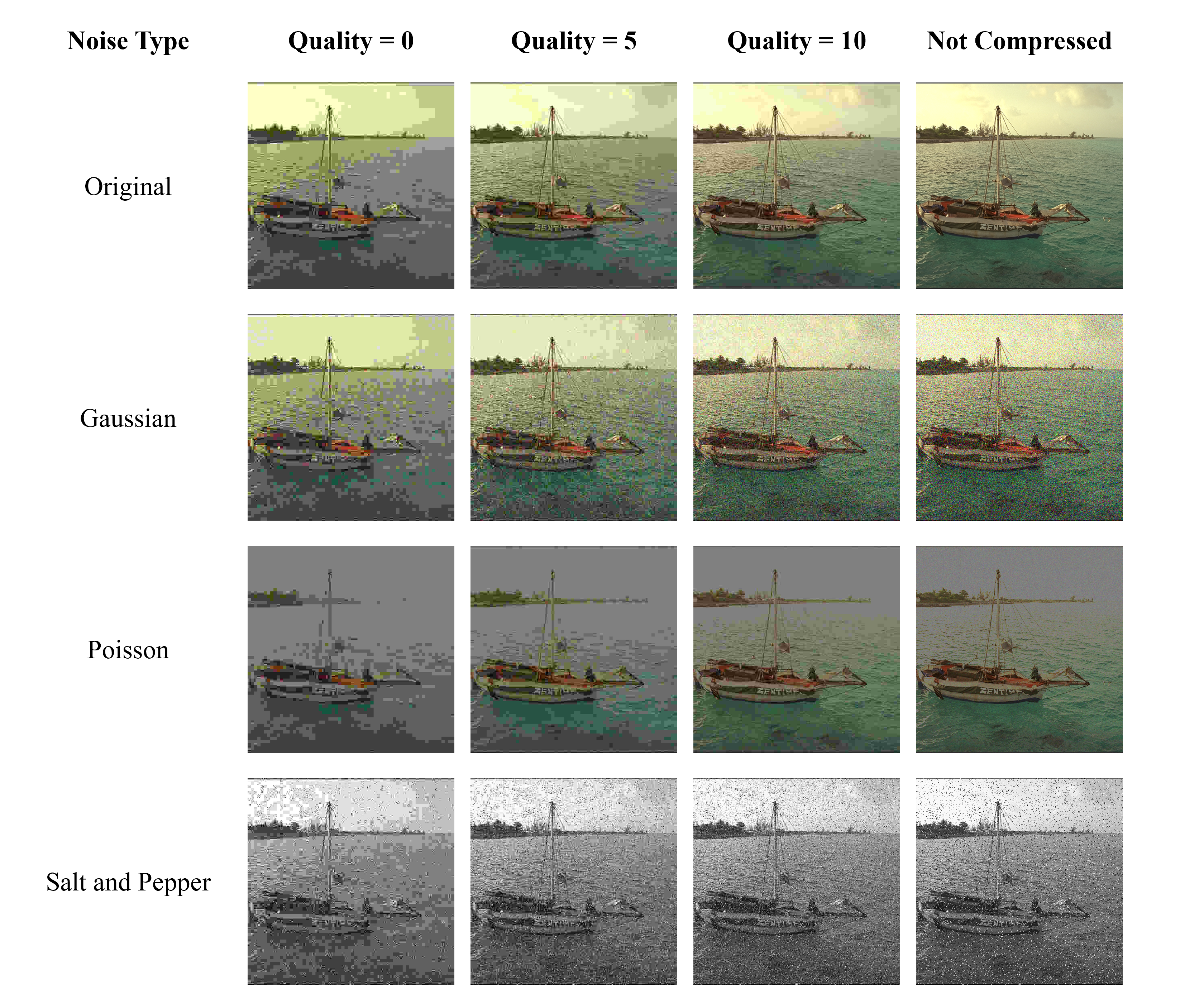}
    \caption{Images compressed using JPEG with different image quality.}
    \label{fig:jpeg_quality}
\end{figure}

Figure \ref{fig:jpeg_quality} illustrates the example of images compressed using the JPEG format. We use the Python Pillow library for compression, where the quality parameter (ranging from 0 to 95) controls the trade-off between compression rate and image fidelity. 0 represents the lowest quality (highest compression), and 95 represents the highest quality (lowest compression).

Comparing original image and the one corrupted by Gaussian noise, the level of Gaussian noise gets lower as the quality decreases. Two images become perceptually similar when quality is equal to zero, indicating that JPEG compression effectively suppresses high-frequency noise. Similarly for salt-and-pepper noise, we can observe the reduction of white and black outlier pixels for the image compressed with quality zero. For the image corrupted by Poisson noise, much of the grainy texture is smoothened out in the compressed images. In all cases, while some color information is altered or lost due to compression, the overall structural content of the original image is preserved. This preservation of global structure suggests that JPEG-compressed images can serve as effective conditioning images for text-to-image generation models.

\begin{figure}
    \centering
    \includegraphics[width=1\linewidth]{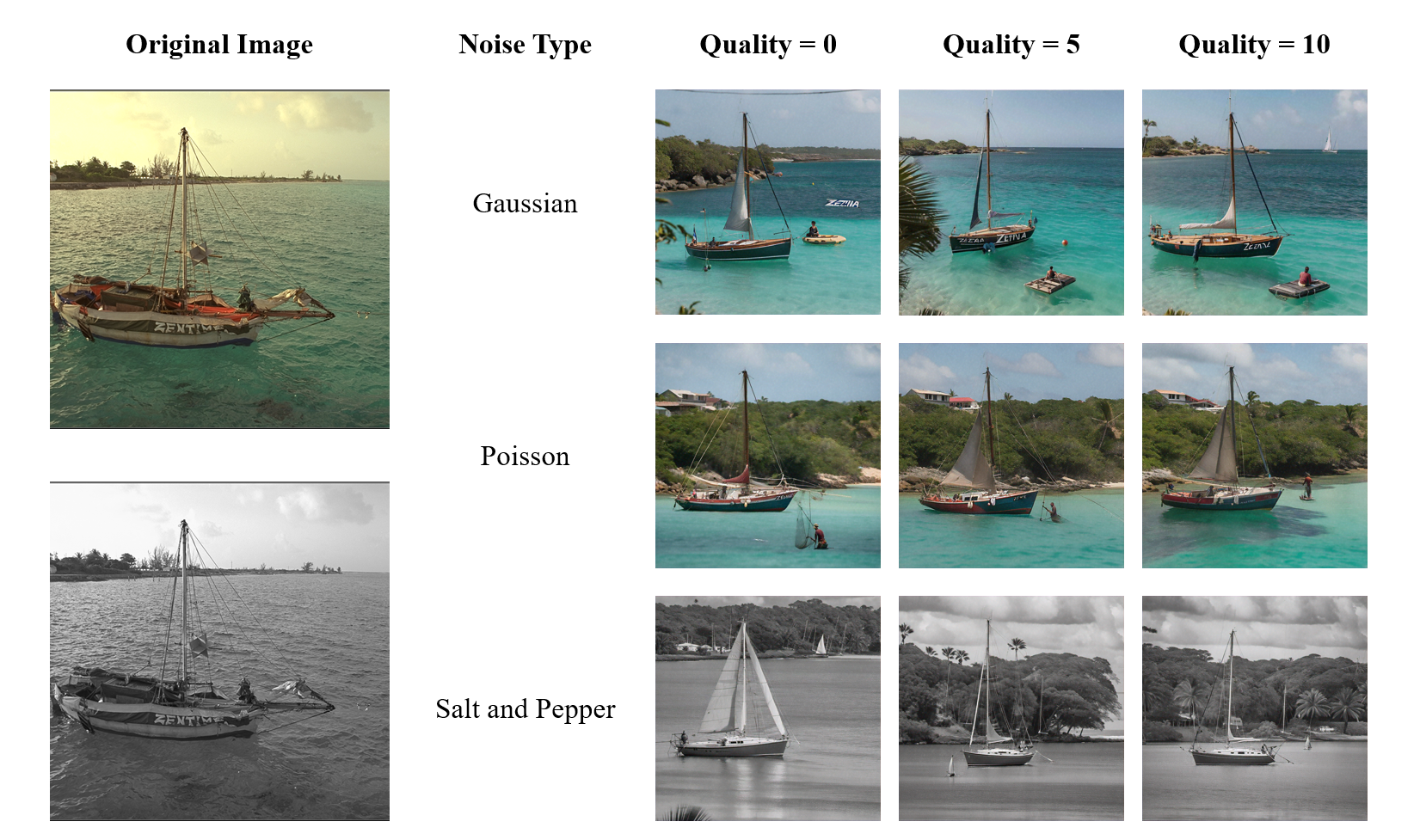}
    \caption{Results of denoising guided by text and lossily compressed sketch. Three columns on the right show the denoised image generated with different conditioning images, which are the noisy images compressed using JPEG with the indicated image quality.}
    \label{fig:denoise_text_jpeg_result}
\end{figure}

Building on our previous experiment using text alone, we conduct additional experiments that incorporate both text prompts and conditioning images. We use Titan Image Generator G1 v2, which supports both text prompt and a conditioning image as inputs. For conditioning, we set the control mode to canny edge and the control strength to 0.5. The prompt strength remains at 8, consistent with the previous experiment, and we use the same random seed across all cases for consistency. The conditioning images are derived from the JPEG-compressed noisy images shown in Figure \ref{fig:jpeg_quality}, and the text prompts are the filtered image captions from Figure \ref{fig:denoise_text_caption}. 

The resulting denoised images are presented in Figure \ref{fig:denoise_text_jpeg_result}. The generated images appear visually realistic according to human perception and maintain the semantics of the original image. More notably, compared to the results in Figure \ref{fig:denoise_text_result}, the spatial alignment of the main object (e.g., the boat) in the generated images more consistently matches that of the original image. This highlights how using JPEG-compressed image as an additional input can be helpful in terms of providing structural information that may not be fully captured by text alone.

\subsubsection{Discussion}
Our scheme offers several compelling advantages. First, it achieves high perceptual quality by leveraging state-of-the-art text-to-image generation models on the decoder side. Traditional denoising methods are often evaluated using metrics such as Mean Squared Error (MSE) or Peak Signal-to-Noise Ratio (PSNR), and they tend to generate blurry images which fall outside of the desired manifold of images and appear less realistic \cite{elad2023image}. This tradeoff between distortion (measured by metrics such as MSE) and perceptual quality has been theoretically established in \cite{blau2018perception}, showing that optimizing distortion often comes at the cost of reduced perceptual fidelity. In contrast, our approach prioritizes perceptual quality. Although our denoiser outputs images with high pixel-wise distortion compared to the original image, it maintains strong semantic and visual realism, thanks to the generative power of text-to-image models. Furthermore, our scheme can perform universally on different types of noise. Since multimodal LLMs can inherently identify the noise from the input image, it eliminates the need for model-specific training or fine-tuning across different noise types. This also leads to a simple pipeline, as it relies entirely on off-the-shelf models without any task-specific adaptation. Finally, our scheme enhances explainability in the denoising process. Since the intermediate representation is in human-readable text form, it offers interpretable insights into how and why denoising actions are taken.

For future work, we plan to perform a quantitative comparison of our method against existing denoisers using a range of evaluation metrics. These include distortion metrics such as MSE and PSNR, perception metrics such as NIQE \cite{mittal2013making}, human satisfication scores, as well as computational efficiency. Also, this paper focused on three of the typical image noises, but we may consider more complex forms of degradation, including structured noise or noise that is not spatially homogeneous. Finally, we see potential in extending our framework to broader image restoration problems, such as deblurring, inpainting and colorization, leveraging the flexibility of text-to-image generation.

\section{Lempel-Ziv (LZ) Transform}\label{sec:lz-transform}
In the realm of sequential probability modeling, we introduce the Lempel-Ziv (LZ) \cite{originalLZ77paper,originalLZ78paper} transform.
This is a transformation from one probability model to another, and can be thought of as a ``wrapper'' around a probability model that, in many circumstances, enhances its accuracy.
Specifically, the LZ78 transform takes as input any probability model that captures at least zero-order statistics of the data (e.g., for English text, this would be the average proportion of each letter in the data).
The output is a probability model that is \textbf{universal} in the sense that, as the length of the input data goes to infinity, it is at least as accurate as the best finite-memory model. 

\vspace{-3pt}

\subsection{Background}
\subsubsection{Sequential Probability Assignment (SPA)}
We consider one-dimensional discrete data sequences over a finite alphabet (e.g., each element in the sequence comes from a fixed, finite set). 
A \textit{sequential probability assignment (SPA)} is a function that, given a \textit{prefix} of a sequence, or all of the past elements, produces an estimated probability distribution for the next symbol.

A very simple sequential probability assignment is one that estimates the probability of the next symbol according to the proportion of times each symbol has appeared previously in the sequence.
This SPA only uses \textit{zero-order} information, so it is simple to compute but only accurate in limited cases, e.g., when the data is independent and identically distributed (i.i.d.).
For real-world data, it fails to capture the correlations between symbols in the sequence.
For instance, if the prefix of a sequence over the English alphabet is ``\texttt{informatio}'', it is highly likely that the next symbol is ``\texttt{n}'', even if ``\texttt{n}'' by itself is fairly uncommon in the rest of the sequence.

At the other extreme, large language models (LLMs) are highly accurate, but at a high computational cost in both training and inference.
The LZ78 transform we present provides a framework for constructing provably universal sequential probability assignments in a more computationally-efficient regime.

\vspace{-3pt}

\paragraph{Notation and Definitions}
Let $\Acal$ is a finite alphabet with size $A \triangleq |\Acal|$.
For instance, a \textit{binary sequence} has alphabet $\Acal_\text{bin} = \{0, 1\}$, and DNA nucleotide sequences have alphabet $\Acal_\text{DNA} = \{\texttt{A}, \texttt{C}, \texttt{G}, \texttt{T}\}$.

$\Mcal(\Acal)$ represents the simplex of probability distributions over $\Acal$.
In other words, elements of $\Mcal(\Acal)$ are all length-$A$ vectors that are valid probability mass functions (PMFs) over $\Acal$.
For the binary alphabet, 
\[\Mcal(\Acal_\text{bin}) = \left\{\begin{pmatrix}
    1-\theta & \theta
\end{pmatrix} \,:\, \theta \in [0, 1]\right\}.\]

An \textit{individual sequence} is a deterministic infinite sequence $\xv = \left(x_1,\, x_2,\, \cdots\right)$, where $x_i \in \Acal$.
$x^n$ denotes the first $n$ symbols of $\xv$, and $x_k^\ell$ is the segment between indices $k$ and $\ell$ (inclusive).
$\Acal^n$ represents the set of all length-$n$ sequences over alphabet $\Acal$, and $\Acal^* \triangleq \bigcup_{k \geq 0} \Acal^k$ is the set of all finite sequences over $\Acal$, including the empty sequence.
For a pair of sequences $x^n \in \Acal^n$ and $y^m \in \Acal^m$, $(x^n \,^\frown y^m) \in \Acal^{m+n}$ is their concatenation, with $x^n$ first and then $y^n$.

A SPA is denoted as follows:
\begin{definition}[Sequential Probability Assignment]
    A SPA is a set of functions
    \[q \triangleq \{q_t\}_{t \geq 1},\quad\text{where}\quad q_t(\cdot|x^{t-1}) \in \Mcal(\Acal)\]
    is a PMF estimating the probability distribution of $x_t$.
    Where unambiguous, we omit, the subscript $t$ in $q_t(\cdot)$.    
\end{definition}

The accuracy of a SPA is often based on \textit{cross-entropy log loss},
\begin{definition}[Log Loss]
    For sequence $x^n$, the log loss incurred by SPA $q$ is
    \[\Lcal(q, x^n) \triangleq \frac{1}{n} \log \frac{1}{q(x^n)} = \frac{1}{n} \sum_{t=1}^n \log \frac{1}{q(x_t|x^{t-1})}.\]
    For infinite sequence $\xv$, the asymptotic log loss is computed via the limit supremum as $n\to\infty$,
    \[\Lcal(q, \xv) \triangleq \limsup_{n\to\infty} \Lcal(q, x^n).\]
    At every timestep, $t$, the marginal log loss is low if the true symbol, $x_t$ is assigned a high probability according to SPA $q$, and vice versa.
\end{definition}

\subsubsection{Compression and Sequential Probability Modeling}

Compression serves as a proxy for learning because an algorithm’s ability to compress data reflects its capacity to capture underlying structure. 
At a high-level, lossless compression is achieved by allocating fewer bits in the compressed representation to higher-probability (more common) patterns and more bits to less common patterns.
Even though the less common patterns require more bits to compress, they appear infrequently and therefore contribute relatively few overall bits to the full compressed representation.
To identify which patterns in the data are of higher probability, a good compressor must, implicitly or explicitly, include a good probability model of the data it compresses.
In fact, with data generated from a (stationary) stochastic process, the fundamental limits of compression and sequential probability modeling are both equal to the entropy rate of the process; see the textbook \cite{coverThomasBook} for more details. Via a process called \textit{arithmetic coding} \cite{rissanenLangdon1979}, any sequential probability model can be directly transformed into a compressor, where the compression ratio is proportional to the log loss incurred by the SPA.

In addition, there are techniques to induce a SPA from a universal compressor.
\cite{frank2000textcategorization} proposed a probability model based on the  normalized compression distance (NCD) \cite{cilibrasi2004clusteringcompression} metric, which measures sequence similarity based on the joint compression ratio between two sequences.
Given any compressor, this method produces $q_t(\cdot|x^{t-1})$ based on the compression ratio of $x^{t-1} \,^\frown a$, $\forall a \in \Acal$.
Recently, \cite{dele2024language} explored use of this SPA, with the GZIP compressor, for autoregressive generation.
This form of SPA, though generalizable to any compressor, has room for improvement in both efficiency (computing $\{q_t\}_{t=1}^n$ is generally quadratic complexity) and accuracy (the results from \cite{dele2024language} do not indicate high generative quality).
Less generalizable but more scalable are SPAs proposed in \cite{langdon1983LZNote,feder1991gambling,federMerhavGutman1992,weissman2007Filtering}, which, instead of using a compressor as a black box, produce a SPA via the internal structure of the LZ78 universal compressor \cite{originalLZ78paper}. 
These methods distill from the LZ78 compressor the features that would make it a good probability model.
The LZ78 transform we propose in this section is a broad generalization of such LZ78-based SPAs that has the potential to be fairly accurate across many tasks while maintaining computational efficiency.

\subsubsection{LZ78 Compression}
The LZ78 transform is heavily based on the LZ78 universal compressor \cite{originalLZ78paper}.
This compressor, like the popular LZ77 compression \cite{originalLZ77paper} (seen in industry-standard compressors such as GZIP and ZStandard), takes advantage of repeated sections in the input data.
As an example, consider compressing the sequence ``\texttt{Hello world. Hello world!}''.
The second time ``\texttt{Hello world}'' appears, we need not use many bits to represent this subsequence: it is sufficient to tell the decoder that it is a repetition of the first 11 symbols.
This is exactly what LZ77 does, and LZ78 follows a similar intuition.\footnote{We base our LZ78 transform off of LZ78 instead of LZ77 because the former is more conducive to sequential probability modeling, and has broader theoretical guarantees.}

LZ78 takes advantage of repeating patterns by parsing the input sequence into a list of \textit{phrases}, which are consecutive subsequences such that each phrase is equal to a previously-seen phrase, plus one new symbol at the end.
This so-called \textit{incremental parsing} can be visualized via a prefix tree, as described via example below:\footnote{cf. \cite{sagan2024familylz78baseduniversalsequential} for a formal algorithmic description.}
\begin{example}[LZ78 Prefix Tree]
    We demonstrate the LZ78 incremental parsing procedure, and corresponding prefix tree, for the binary sequence $x^n = \texttt{00011001}$.
    
    The LZ78 tree begins as a single root node, which we say corresponds to the ``empty phrase.''
    \begin{center}
        \begin{tikzpicture}[
        level distance=1cm,
          level 1/.style={sibling distance=6cm},
          level 2/.style={sibling distance=2cm},
          level 3/.style={sibling distance=1.5cm},
          level 4/.style={sibling distance=1cm}
          ]
          \node[rectangle, fill=yellow!15, rounded corners, draw]{\parbox[t]{3em}{\centering\small root}};
        \end{tikzpicture}
    \end{center}
    
    The first symbol, \texttt{0}, constitutes the first LZ78 phrase.
    This is added to the tree as a new leaf.
    \begin{center}
        \begin{tikzpicture}[
        level distance=1cm,
          level 1/.style={sibling distance=6cm},
          level 2/.style={sibling distance=2cm},
          level 3/.style={sibling distance=1.5cm},
          level 4/.style={sibling distance=1cm}
          ]
          \node[rectangle, fill=yellow!15, rounded corners, draw]{\parbox[t]{4em}{\centering\small root}}
            child {
                node[rectangle, fill=cyan!15, rounded corners, draw]{\parbox[t]{4em}{\centering\small node \texttt{0} \\{\scriptsize(phrase 1)}}}
            };
        \end{tikzpicture}
    \end{center}

    Starting from the root, we parse the second symbol.
    As there is already a \texttt{0} node branching from the root, we traverse to that node and continue parsing.
    In parsing the next \texttt{0}, we notice there is no node labeled \texttt{00}, add a corresponding branch off of the \texttt{0} node and then return to the root.
    \texttt{00} is then the second LZ78 phrase for $x^n$. 

    \begin{center}
        \begin{tikzpicture}[
        level distance=1cm,
          level 1/.style={sibling distance=6cm},
          level 2/.style={sibling distance=2cm},
          level 3/.style={sibling distance=1.5cm},
          level 4/.style={sibling distance=1cm}
          ]
          \node[rectangle, fill=yellow!15, rounded corners, draw]{\parbox[t]{4em}{\centering\small root}}
            child {
                node[rectangle, fill=gray!5, rounded corners, draw]{\parbox[t]{4em}{\centering\small node \texttt{0} \\{\scriptsize(phrase 1)}}}
                child {
                    node[rectangle, fill=cyan!15, rounded corners, draw]{\parbox[t]{4em}{\centering\small node \texttt{10} \\{\scriptsize(phrase 2)}}}
                }
            };
        \end{tikzpicture}
    \end{center}
    
    At the end of the LZ78 parsing process, $x^n$ is split into the phrases \texttt{0,00,1,10,01}, with corresponding prefix tree

    \begin{center}
        \begin{tikzpicture}[
        level distance=1.2cm,
          level 1/.style={sibling distance=6cm},
          level 2/.style={sibling distance=2cm},
          level 3/.style={sibling distance=1.5cm},
          level 4/.style={sibling distance=1cm}
          ]
          \node[rectangle, fill=yellow!15, rounded corners, draw]{\parbox[t]{4em}{\centering\small root}}
            child {
                node[rectangle, fill=gray!5, rounded corners, draw]{\parbox[t]{4em}{\centering\small node \texttt{0} \\{\scriptsize(phrase 1)}}}
                child {
                    node[rectangle, fill=cyan!15, rounded corners, draw]{\parbox[t]{4em}{\centering\small node \texttt{00} \\{\scriptsize(phrase 2)}}}
                }
                child {
                    node[rectangle, fill=cyan!15, rounded corners, draw]{\parbox[t]{4em}{\centering\small node \texttt{01} \\{\scriptsize(phrase 5)}}}
                }
            }
            child {
            node[rectangle, fill=gray!5, rounded corners, draw]{\parbox[t]{4em}{\centering\small node \texttt{1} \\{\scriptsize(phrase 3)}}}
                child {
                    node[rectangle, fill=cyan!15, rounded corners, draw]{\parbox[t]{4em}{\centering\small node \texttt{10} \\{\scriptsize(phrase 4)}}}
                }
            };
        \end{tikzpicture}
    \end{center}
\end{example}

\begin{definition}[LZ-based Notation]
    From the LZ78 parsing algorithm, we define some notation that will be useful for discussing the LZ78 transform:
    \begin{center}
        \small
        \renewcommand{\arraystretch}{1.3}
        \begin{tabularx}{\linewidth}{
          >{\raggedright\arraybackslash}p{1.8cm}
          >{\raggedright\arraybackslash}X}\toprule
            \textbf{Notation} & \textbf{Description} \\
            \midrule
            $\Zcal(x^n)$ & Set of \textbf{all LZ78 phrases} in the parsing of $x^n$, i.e., all nodes of the LZ78 prefix tree, including the root. \\
            $C(x^n)$ & \textbf{Number of LZ78 phrases} or nodes in the LZ78 tree, i.e., $C(x^n) = |\Zcal(x^n)|$. \\
            $z_c(x^{t-1})$ & The \textbf{LZ78 context} of $x_t$, i.e., the prefix of the phrase that $x_t$ belongs to (not including $x_t$)
            This is the current node of the prefix tree being traversed when parsing $x_t$. \\
             $\mathcal{Y}\{x^m, z\}$; & The \textbf{subsequence} of $x^m$ that has LZ78 context $z$, or the ordered sequence of symbols parsed while at node $z \in \Zcal(x^n)$ of the LZ78 tree. E.g., in the example above, $\mathcal{Y}\{x^n, \texttt{0}\} = (0, 1)$ because, from node \texttt{0}, we first 
             parse a \texttt{0} and then a \texttt{1}. Likewise, $\mathcal{Y}\{x^n, \texttt{root}\} = (0, 0, 1, 1, 0)$, or the first symbol of each phrase. \\
             $\mathcal{N}(a|x^n)$ & The number of times that symbol $a$ appears in $x^n$. \\
             $\mathcal{N}_\text{LZ}(a|x^n, z)$ & The number of times that symbol $a$ appears in $\mathcal{Y}\{x^n, z\}$. \\
            \bottomrule
        \end{tabularx}
    \end{center}    
\end{definition}

\subsection{Details of the LZ78 Transform}
Using the correspondence between compression and probability modeling, we use the structure of LZ78 to transform any ``good zero-order'' SPA into a universal SPA (i.e., one that works well on any input data, as the sequence length goes to infinity).
We first define the LZ78 transform and discuss universality over individual sequences.
Then, we present the main theoretical results pertaining to the LZ78 transform.

\begin{definition}[LZ78 transform]\label{def:lz-transform}
    Given any SPA $q = \{q_t\}_{t\geq 1}$, the \textit{LZ78 transform} of the SPA is
    \[\lztrans\{q\} = \hat{q} \triangleq \{\hat{q}_t(x_t|x^{t-1}\}_{t \geq 1},\text{ where } \hat{q}(x_t|x^{t-1}) = q\left(x_t\,\big|\,\mathcal{Y}\{x^{t-1}, z_c(x^{t-1})\}\right).\]
\end{definition}
The LZ78 transform can be thought of as making $C(x^n)$ copies of the original SPA, one for each node of the LZ78 tree.
Each such copy operates on a disjoint subset of the data, namely the symbols parsed when traversing the corresponding node of the tree.
This procedure can transform a zero-order SPA into a one that is competitive against any finite-memory SPA by conditioning on the current symbol's LZ context.
By \cite{originalLZ78paper}, length of the average LZ context grows unbounded with the sequence length, so the LZ78 SPA conditions on increasingly-longer context lengths.
I.e., it uses increasingly more information from the past to make a probability estimate. 

\subsubsection{Universality of SPAs in the Individual Sequence Setting}
To discuss universality of SPAs in the individual sequence setting (i.e., evaluating a SPA on a deterministic, infinite sequence), we define the following fundamental limit:
\begin{definition}[Optimal Finite-State SPA Log Loss]
    Let $\mathcal{F}_M$ be the set of all SPAs such that the current prediction depends solely on the current state of an $M$-state finite state machine, with state transition at time $t$ being a function of $x_t$ (cf. \cite{sagan2024familylz78baseduniversalsequential} for a formal mathematical description thereof).
    These are called \textit{finite-state} or \textit{finite-memory} SPAs.

    For infinite sequence $\xv$, we define the \textit{optimal finite-state log loss} as
    \[\lambda(\xv) \triangleq \lim_{M\to\infty} \lambda_M(\xv) \triangleq \lim_{M\to\infty} \limsup_{n\to\infty} \min_{q \in \mathcal{F}_M} \frac{1}{n}\log \frac{1}{q(x^n)},\]
    where the outer limit exists because $\lambda_M(\xv)$ is decreasing and bounded below by $0$.
\end{definition}
As proven in \cite{sagan2024familylz78baseduniversalsequential}, $\lambda(\xv)$ is equivalent to the following quantity, which is easier to conceptualize:
\begin{definition}[Optimal Markov SPA Log Loss]
    Let $\Mcal_k$ be the set of all $k$-order Markov SPAs, i.e., ones where the probability assignment only depends on a length-$k$ context before the current symbol.
    Specifically, a $k$-order Markov SPA satisfies $q_t(x_t|x^{t-1}) = q_t(x_t|x_{t-k}^{t-1})$ for $t > k$.
    For $ t \leq k$, there are no restrictions on $q_t$, which can be set to attain $0$ log loss.

    The optimal log loss over this class is defined as
    \[\mu(\xv) = \lim_{k\to\infty} \mu_k(\xv) \triangleq \lim_{k\to\infty} \limsup_{n\to\infty} \mu_k(x^n),\]
    where
    \[\mu_k(x^n) \triangleq \min_{q \in \Mcal_k} \frac{1}{n}\log \frac{1}{q(x^n)} = \min_{q \in \Mcal_k} \frac{1}{n} \sum_{t={k+1}}^n \log \frac{1}{q(x_t|x_{t-k}^{t-1})}.\]
\end{definition}

\begin{remark}\label{rem:k-order-markov-loss-sum-zero-order-loss}
    For any sequence $\xv$,
    \[n \cdot \mu_k(x^n) = \sum_{z\in\Acal^k} \left|\{ k+1 \leq t \leq n : x_{t-k}^{t-1}=z\}\right| 
    \mu_0\left(\{y_t : k+1 \leq t \leq n,  y_{t-k}^{t-1}=z\}\right),\]
    where $\{y_t : k+1 \leq t \leq n,  y_{t-k}^{t-1}=z\}$ represents an \textit{ordered} subsequence of $x^n$.
    \begin{proof}
        As a $k$-order Markov SPA is allowed to attain a log loss of zero over the first $k$ symbols,
        \begin{align*}
            n\mu_k(x^n) &= \min_{q \in \Mcal_k} \sum_{t=k+1}^n \log \frac{1}{q(x_t|x_{t-k}^{t-1})} = \sum_{z\in\Acal^k} \min_{q_z \in \Mcal(\Acal)} \log \frac{1}{q_z\left(\{x_t : x_{t-k}^{t-1}=z\}\right)} \\
            &= \sum_{z\in\Acal^k} \left|\{x_t : x_{t-k}^{t-1}=z\}\right| \mu_0\left(\{x_t : x_{t-k}^{t-1}=z\}\right).
        \end{align*}
    \end{proof}
\end{remark}

\begin{remark}
    $\lambda(\xv)$ and $\mu(\xv)$ are also equivalent to the finite-state compressibility of $\xv$ (the corresponding fundamental limit for individual sequence compression), as is described in \cite{sagan2024familylz78baseduniversalsequential}.
\end{remark}

Now, we can define the notion of universality we use to evaluate the LZ78 transform:
\begin{definition}[Universal SPA]\label{def:universal}
    SPA $q$ over alphabet $\Acal$ is universal if, for any individual sequence $\xv$,
    \[\limsup_{n\to\infty} \frac{1}{n}\log \frac{1}{q(x^n)} \leq \lambda(\xv) = \mu(\xv).\]
\end{definition}


\subsubsection{Main Theoretical Results}
\begin{theorem}\label{thm:lz-transform}
    Suppose SPA $q$ satisfies
    \[\limsup_{n\to\infty} \max_{x^n \in \Acal^n} \left(\frac{1}{n}\log \frac{1}{q(x^n)} - \mu_0(x^n)\right) \leq 0,\]
    then its LZ-transformed version, $\hat{q} = \lztrans\{q\}$ is universal as per \prettyref{def:universal}.
\end{theorem}

The proof of this result can be broken into two primary components.
First,
\begin{lemma}\label{lem:lz-transform-part-1}
    If SPA $q$ satisfies
    \[\limsup_{n\to\infty} \max_{x^n \in \Acal^n} \left(\frac{1}{n}\log \frac{1}{q(x^n)} - \mu_0(x^n)\right) \leq 0,\]
    then, for $\hat{q}= \lztrans\{q\}$,
    \[\limsup_{n\to\infty} \max_{x^n \in \Acal^n} \left(\frac{1}{n}\log \frac{1}{\hat{q}(x^n)} - \frac{1}{n} \sum_{z \in \Zcal(x^n)} \left|\mathcal{Y}\{x^n, z)\} \right| \mu_0\left( \mathcal{Y}\{x^n, z)\} \right)\right) \leq 0\]

    Proof sketch.
    \textup{First, we divide the log loss of $\hat{q}$ into the log loss incurred at each node of the LZ78 tree, which allows us to directly apply the definition of the LZ78 transform from \prettyref{def:lz-transform}.
    By the condition on SPA $q$ from the theorem statement, the difference between the achieved and desired log loss at any node of the LZ78 tree decays uniformly as the length of the corresponding subsequence, $\mathcal{Y}\{x^n, z\}$, grows.
    Using basic properties of the LZ78 tree, we show that, as $n\to\infty$, most timesteps are at nodes such that $\mathcal{Y}\{x^n, z\}$ is long (loosely speaking).
    From there, the result of this lemma directly follows.
    The full proof is in \prettyref{app:lz-transform-proofs}}.
\end{lemma}
Then, it is left to show that:
\begin{lemma}\label{lem:lz-transform-part-2}
    For any individual sequence $\xv$,
    \[\limsup_{n\to\infty} \frac{1}{n} \sum_{z \in \Zcal(x^n)} \left|\mathcal{Y}\{x^n, z)\} \right| \mu_0\left( \mathcal{Y}\{x^n, z)\} \right) \leq \mu(\xv).\]

    Proof sketch.
    \textup{Let the argument of the limit supremum the theorem statement be denoted $(\ast)$.
    We first fix an arbitrary $k$ and show that $\limsup_{n\to\infty} (\ast) \leq \mu_k(\xv)$.
    Applying \prettyref{rem:k-order-markov-loss-sum-zero-order-loss}, the fact that $\mu_k$ is non-increasing in $k$, and Jensen's inequality, we upper-bound $(\ast)$ by the average of $\mu_k$ over the subsequence of $x^n$ with LZ context length $\geq k$ and $\mu_0$ over the remaining timepoints, weighted by the length of the respective subsequence.
    By \cite{originalLZ78paper}, the number of LZ78 phrases in $x^n$ is sublinear in $n$.
    This, along with the fact that exactly $k$ timesteps in each phrase have a LZ context length $< k$, means that the $\mu_0$ term in the weighted average vanishes, resulting in $(\ast) \leq \mu_k(x^n) + o(1)$.
    Taking $n\to\infty$ and then $k\to\infty$ completes the proof.
    }
\end{lemma}

\subsubsection{Example LZ78 Transforms}\label{sec:example-lz-transforms}
Perhaps the simplest SPA that satisfies the condition of \prettyref{thm:lz-transform} is an additive perturbation of the empirical distribution:
\begin{equation}
    q_t(x_t=a|x^{t-1}) = \frac{\mathcal{N}(a|x^{t-1}) + \gamma}{t - 1  + A\gamma},\quad \gamma > 0. \label{eqn:bayesian-mixture}
\end{equation}
$\gamma$ must be strictly positive, or else the SPA can incur infinite log loss (specifically, if the symbol corresponding to $x_t$ was never seen in $x^{t-1}$).
Such a SPA is simple to compute, as it only requires knowledge the number of times each symbol $a \in \Acal$ appears in the sequence.

As per \cite{cover1972admissibility}, this SPA is a Bayesian mixture under a Dirichlet prior.
\cite{sagan2024familylz78baseduniversalsequential} shows that any such Bayesian mixture, under regularity conditions on the prior,\footnote{Specifically, that the prior has full support over $\Mcal(\Acal)$.}, satisfies the condition of \prettyref{thm:lz-transform}.
In fact, applying the LZ78 transform to such SPAs forms the SPA family of \cite{sagan2024familylz78baseduniversalsequential}.

The LZ78 transform can also be applied to more sophisticated SPAs, for instance n-gram SPAs or context tree weighting (CTW) \cite{willems1995CTW}, which is universal over inputs from fixed-depth tree sources.
Applying the LZ78 transform to such SPAs is left for future exploration.

\subsubsection{Computational Complexity}
To analyze the computational complexity of the LZ78 transform, we define $\mathcal{C}_T(q, n)$ as the time complexity of computing the SPA $q$ up until timestep $n$, and $\mathcal{C}_M(q, n)$ as the corresponding memory complexity.
Take $\mathcal{C}_X(q, n)$ to be an arbitrary complexity, i.e., either time or memory.

If $\hat{q} = \lztrans\{q\}$, then the complexity (time or memory) of $\hat{q}$ is
\begin{equation}
    \Ccal_X(\hat{q}, n) = \sum_{z \in \Zcal(x^n)} \Ccal_X\left(q, \left| \mathcal{Y}\{x^n, z)\} \right|\right). \label{eqn:lz-transform-complexity}
\end{equation}
This follows directly from the definition of the LZ78 transform: each node of the LZ78 tree contains a different SPA, which is evaluated at the subsequence $\mathcal{Y}\{x^n, z)\}$.

This complexity is in general non-trivial to compute, and a full analysis of \eqref{eqn:lz-transform-complexity} is reserved for ongoing work.
Here, we consider an illustrative example and simple bound.
First, consider a base SPA $q$ where $\Ccal_T(q, n) = O(n)$ and $\Ccal_M(q, n) = O(1)$.
This is a fairly common case, covering, e.g., all of the SPAs discussed \prettyref{sec:example-lz-transforms}.
In that case,
\begin{align*}
    \mathcal{C}_T(\hat{q}) &= \sum_{z \in \Zcal(x^n)} O\left(\left| \mathcal{Y}\{x^n, z)\} \right|\right) = O\left(\sum_{z \in \Zcal(x^n)} \left| \mathcal{Y}\{x^n, z)\} \right| \right) = O(n), \\
     \mathcal{C}_M(\hat{q}) &= \sum_{z \in \Zcal(x^n)} O(1) = O(C(x^n)) \stackrel{(a)}{=} O\left(\frac{n}{\log n}\right),
\end{align*}
where (a) is due to a result from \cite{originalLZ78paper}.

For a general $\mathcal{C}_X(q, n)$ that is increasing in $n$, a loose upper-bound on the complexity of $\hat{q}$ is
\[\mathcal{C}_X(\hat{q}, n) \leq C(x^n) \max_{z \in \Zcal(x^n)} \Ccal_X\left(q, \left| \mathcal{Y}\{x^n, z)\} \right|\right) =  C(x^n) \,\,\Ccal_X\!\!\left(q, C(x^n)\right),\]
as the maximum value of $\left| \mathcal{Y}\{x^n, z)\} \right|$ is achieved at the root, which has been visited $C(x^n)$ times.
This bound is tight for $\Ccal_X(q, n) = O(1)$, and is looser the more that $\Ccal_X(q, n)$ grows with $n$.


\section{Applications of the Lempel-Ziv Transform}
As we have discussed from a theoretical perspective in \prettyref{sec:lz-transform}, compression-based learning offers a principled and computationally efficient alternative to conventional deep learning approaches, particularly for modeling structured sequential data.
Rooted in classic information theory, algorithms such the LZ78 transform enable universal, model-free representations that naturally capture regularities in discrete sequences without requiring extensive parameter tuning or large-scale training.
In this section, we explore applications for which these methods can achieve competitive performance while operating with a significantly lower computational overhead compared to modern deep generative model.

The applications of the LZ78 transform (primarily with the inner SPA being the Dirichlet mixture SPA discussed in \prettyref{sec:example-lz-transforms}) to classification, discrete filtering, and sequence generation have been explored in \cite{omri2025genomic,yan2025filt,ding2025lzmidicompressionbasedsymbolicmusic}.

\subsection{Classification}
An LZ-Transformed version of a SPA can be used for classification by building $c$ SPAs, where $c$ is the number of distinct labels: we build one SPA on the subset of the training data corresponding to each class.
Then, any new sample can be classified by computing the corresponding log loss using each of the $c$ SPAs (without modifying the parameters of the SPA, \eg, not adding any new leaves), and choosing the class with the  \textit{smallest} log loss.

\begin{figure}[htbp]
    \centering
    \includegraphics[width=0.75\linewidth]{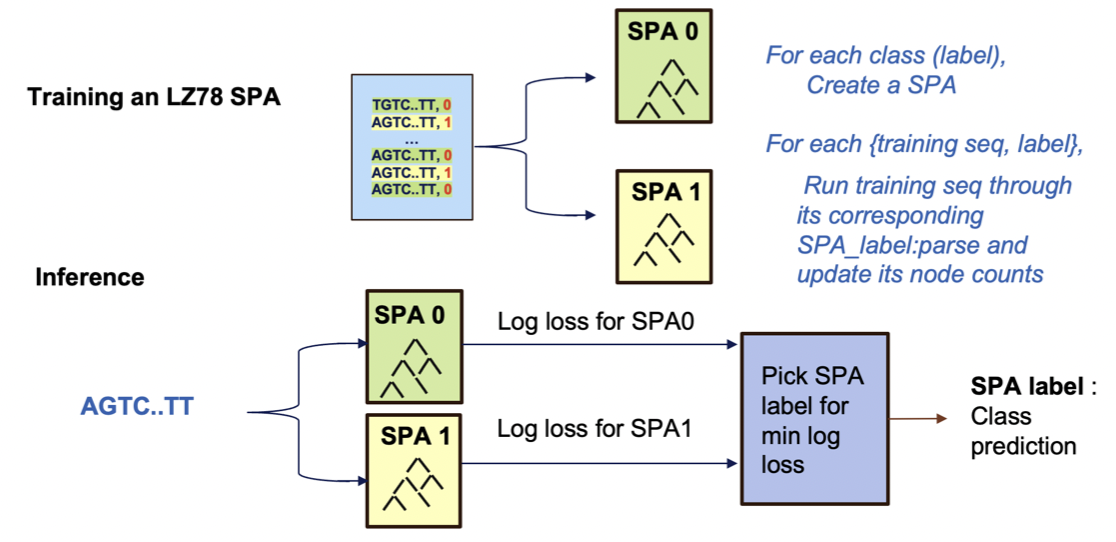}
    \caption{LZ78-based Classifier via training an LZ78-SPA on Labeled Data (diagram from \cite{omri2025genomic})}
    \label{fig:enter-label}
\end{figure}

In~\cite{sagan2024familylz78baseduniversalsequential}, preliminary experiments are conducted across both image and text domains, using the LZ78 transform of a Bayesian mixture SPA under the Dirichlet$(0.1, \dots, 0.1)$ prior (see \cite{sagan2024familylz78baseduniversalsequential} for more details on datasets and implementation).
As the LZ78 transform of a SPA tends to be more accurate as the length of the input sequence increases, we loop through the ``training data'' of each SPA 20 times for image datsets and 5 times for text datasets.
The results, also summarized in Table~\ref{tab:class} below, demonstrate that despite the simplicity and non-parametric nature of the model, reasonable classification performance can be achieved with minimal preprocessing and efficient tree construction.

\begin{table}[h]
\small
\centering
\caption{Results of classification experiments using the LZ78 SPA.}
\begin{tabular}{l|cccc}
\toprule
\textbf{Dataset} & MNIST & Fashion-MNIST & IMDB & Spam Emails \\
\midrule
Accuracy (\%) & 75.36 & 72.16 & 75.62 & 98.12 \\
Training Time (s) & 14 & 15 & 16 & 14 \\
\bottomrule
\end{tabular}
\label{tab:class}
\end{table}

\subsubsection{Genomic Data Classification}
One domain where the LZ78 transform performs especially well as a classifier is genomics.
In this section, we briefly summarize how the LZ78 transform can be applied to genomic data classification, as outlined in~\cite{omri2025genomic}. 

\textbf{Motivation and background:}
In genomic data analysis, \emph{sequence classification} assigns a label to a DNA (or protein) string drawn from the four-letter alphabet \(\{A,C,G,T\}\) (or the 20 amino acids).  Accurate, efficient classification underpins biodiversity studies, evolutionary inference, pathogen surveillance, and precision medicine, where it aids in detecting functional regions, pathogenic variants, or drug–response markers~\cite{dna1,dna2}. A rich ecosystem of methods exists, ranging from information-theoretic heuristics and classical machine-learning pipelines to billion-parameter genomic language models (gLMs) such as DNABERT-2~\cite{dna3}.  While deep transformers set state-of-the-art accuracy, their training costs are significantly large: DNABERT-2, for example, consumes tens of gigabytes of raw genomes and \(\sim\)14 GPU-days.
This \emph{scale gap} where modern gLMs achieve high accuracy but demand resources unattainable in many clinical or field settings, whereas LZ78 transform training (with a linear-time-computable internal SPA) is linear-time and CPU-friendly.

This aspect, along with the theoretical results on the LZ78 transform presented earlier provide further motivation for using the LZ78-SPA as a competitive, off-the-shelf classifier. ~\cite{omri2025genomic} delves deeper into the construction of a lightweight DNA classifier based on the LZ78 transform of the prior from \eqref{eqn:bayesian-mixture}.
This streamlined LZ78 pipeline not only narrows (and often closes) the accuracy gap with DNABERT-2 on the 28-task GUE benchmark~\cite{omri2025genomic}, but does so with \(\mathbf{\sim400\times}\) less training compute and \(\mathbf{>100\times}\) faster inference, underscoring the latent power of compression-based models for genomic analytics.

Prior related work, such as Ziv–Merhav cross-parsing for relative entropy~\cite{lzdna1}, normalized relative compression for DNA taxonomy~\cite{lzdna3,lzdna4}, and variable-order Markov tests for protein families~\cite{lzdna2}, has already hinted at the promise of such compressors for bio-sequences, often matching or surpassing classical ML baselines at a fraction of the computational cost.


\textbf{Methodology:}
In general, LZ-Transform-based classification is performed via the basic procedure outlined earlier.
A large component of the accuracy achieved in \cite{omri2025genomic}, however, relies on a hyperparameter sweep to choose the inner SPA on which the LZ78 transform is performed (among other implementation-specific hyperparameters).
The internal SPA is chosen from the set of Bayesian Mixture SPAs under Dirichlet priors (\ie, \eqref{eqn:bayesian-mixture}) with $\gamma \in\{0.1,0.33,0.5,0.75,1,3,5\}$.
These SPAs are simple to compute and sufficiently rich for the hyperparameter sweep to yield competitive accuracy.
For each hyper-parameter combination, one SPA per class is trained, mean validation log-loss is evaluated, and the
\((\gamma^\star,E^\star)\) that minimizes this loss is selected.
Since tree structures are reused across settings, the average sweep finishes in seconds on a single CPU core, retaining the \(O(n)\) training and \(O(|\mathcal{A}|)\) per-symbol inference costs that make LZ78 attractive.
See \cite{omri2025genomic} for full details of the hyperparameter sweep.

\begin{figure}[tbp]
    \centering
    \includegraphics[width=0.98\linewidth]{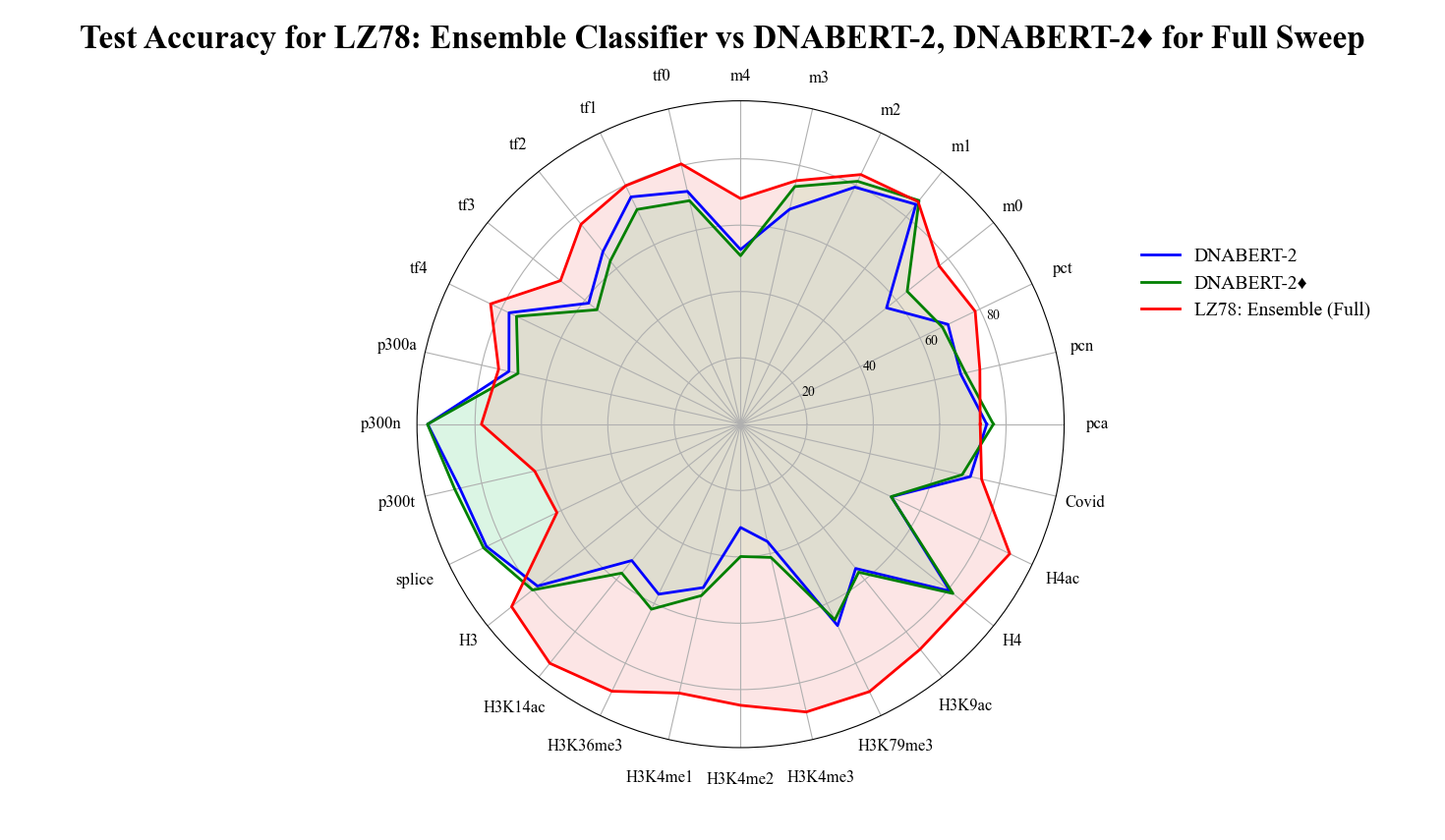}
    \caption{Radar Plot of LZ-Transform-based Genomics classification, compared with DNABERT-2 (figure from \cite{omri2025genomic})}
    \label{fig:genomics-results}
\end{figure}


\textbf{Competitiveness with transformer baselines:}
The same tuned LZ78 classifier was benchmarked on the 28-task \textsc{Genomic-Understanding Evaluation} (GUE) suite \cite{zhou2024dnabert2efficientfoundationmodel}, where DNABERT-2 \cite{zhou2024dnabert2efficientfoundationmodel}, an efficient 117 M parameter transformer—serves as the standing neural baseline.  
As noted in~\cite{omri2025genomic} (reproduced in \prettyref{fig:genomics-results}), LZ78 classifier attains \textbf{higher or parity accuracy on 25/28 datasets}  
(89.2\% of the benchmark), including gains of \(\!\ge\!47\) percentage points on several challenging epigenetic-mark prediction tasks,  
while requiring \(\sim\!30\times\) less training data, \(\sim\!400\times\) less training compute  
($<$ 1 hour on a modern CPU versus 14 days on 8 GPUs), and up to \(\mathbf{128\times}\) faster inference. Even on the three tasks where DNABERT-2 retains an edge (two promoter-detection variants and \textit{splice} site detection), LZ78’s accuracy deficit (13–24 pp) is often offset by its dramatically lower hardware footprint, making it a compelling alternative for resource-constrained genomic pipelines. These results underscore that sequence-compression primitives, when coupled with a modest hyper-parameter search, can rival or surpass state-of-the-art deep models in real-world classification without compromising on efficiency.


\subsection{Discrete Filtering}
The LZ78 transform also induces a universal discrete filter, as discussed in \cite{yan2025filt}.
In universal \textit{discrete filtering}, the task is to recover an underlying, noise–free sequence after it has been distorted by an \textit{unknown} discrete-memoryless channel (DMC).
A DMC is a function that takes as input the noise-free sequence outputs a random sequence where each symbol is drawn independently, conditional on the corresponding input symbol.
For instance, if the input is denoted $x^n$, then the output, $Y^n$ is drawn as $Y_i \sim P_{Y|X=x_i}$, where $P_{Y|X}$ is some conditional distribution.

Classical Bayesian filters succeed when both the clean-source statistics and the channel law are known a priori, but such knowledge is rarely available in practice. Our focus is therefore on \textit{universal filters} that must perform well for \textit{any} finite-alphabet source without assuming a model, while also respecting strict look-ahead or delay constraints that frequently arise in low-latency sensing and communication systems. In~\cite{yan2025filt}, it is shown that a broad family of \textit{Sequential Probability Assignments (SPAs)}, including those induced by the LZ78 transform, are \textit{universal} for this setting: for every source–channel pair they attain, up to vanishing redundancy, the minimum possible expected loss among all filters that are limited to the same look-ahead.
For the LZ78 transform with the internal SPA from \eqref{eqn:bayesian-mixture}, e.g., the result delivers both a performance guarantee and a practical algorithm that is readily implementable on resource-constrained hardware.

Here, we summarize key results on LZ78-based universal filtering from \cite{yan2025filt}; refer to \cite{yan2025filt} for full technical details and proofs of the results outlined here.
%

\subsubsection{Preliminaries}\label{sec:denoising-prelim}
\begin{itemize}
    \item Let the clean signal at any timepoint be denoted $X$ with finite alphabet $\mathcal{A}_X$.
    $X$ is random and distributed according to PMF vector $P_X$.
    A length-$n$ clean sequence is denoted $X^n$.
    \item The noisy observation $Z$ takes values in $\mathcal{A}_Z$ and is produced by a DMC with \textit{invertible} transition matrix $\Pi$, i.e., $\Pr\{Z = z \mid X = x\} = \Pi(z,x)$, where $\Pi^{-1}$ exists.
    The full noisy sequence is denoted $Z^n$.
    \item A filter outputs $\hat{X} \in \mathcal{A}_{\hat{X}}$ after observing a window of past (and possibly future) noisy symbols whose length is capped by a prescribed look-ahead $d$ (or, equivalently, delay $d$).
    \item Accuracy is measured with a loss function $\Lambda: \mathcal{A}_X \times \mathcal{A}_{\hat{X}} \to [0,\infty)$, with matrix representation is $A(i,j) = \Lambda(x_i, \hat{x}_j)$, and $A_{\max} = \max_{i,j} A(i,j)$.
\end{itemize}

Given knowledge of the distribution of the noisy sequence $Z^n$ and the channel matrix $\Pi$, we can derive (via Bayes' rule and some algebra) the distribution of $X_t$, conditioned on different subsequences of $Z^n$:

\begin{align*}
    P_{X_t \mid Z_t=z}
   &= \frac{\Pi(\!\cdot,z)\,\odot
          \bigl(\Pi^{-\!\top} P_{Z_t}\bigr)(z)}
          {P(Z_t=z)}
   \triangleq F\!\bigl(P_{Z_t},\Pi,z\bigr), \\
   P_{X_t \mid Z_1^t = z_1^t} &=
        F\!\bigl(P_{Z_t \mid Z_1^{t-1}}, \Pi, z_t\bigr), \\
        P_{X_t \mid Z_{t+1}^{\,n} = z_{t+1}^{\,n}} &= F\!\bigl(P_{Z_t \mid Z_{t+1}^{\,n}}, \Pi, z_t\bigr),
\end{align*}
where $\odot$ denotes the Hadamard (element-wise) product.
This result relies on the fact that the channel is memoryless with an invertible channel matrix.




\subsubsection{Bounding Excess Loss}

In an estimation problem specified by a loss $\Lambda:\mathcal{A}_X\times\mathcal{A}_{\hat X}\!\to\![0,\infty)$ and a
distribution $P_X$, the \textbf{Bayes response} minimises the expected loss:
\begin{align}
\hat X_B(P_X)
      &\triangleq \arg\min_{\hat x}
         \mathbb{E}\!\bigl[\Lambda(X,\hat x)\bigr]
        = \arg\min_{\hat x}\!
          \sum_{x\in\mathcal{A}_X}
          \Lambda(x,\hat x)\,P_X(x)
\label{eq:bayes}
\end{align}

In addition, if we are given side information in the form of random variable $W$ and conditional distribution $P_{X|W}$, the conditional Bayes response is:
\begin{lemma}[Optimal estimator with side information]
\label{lem:opt-with-side-info}
Let $W$ be an auxiliary random variable.  The estimator that minimizes
$\mathbb{E}\bigl[\Lambda(X,\hat X(W))\bigr]$ is
\[
  \hat X^\star(W)=\hat X_B\bigl(P_{X\mid W}\bigr).
\]
\end{lemma}
For our purposes, the side information is some subsequence of the noisy signal $Z^n$, meaning the optimal estimator relies $P_{X|Z_k^\ell}$ (for some $k \leq \ell$).
As described above, this can be derived using the distribution of $Z^n$.

The distribution $P_{Z_t}$ is generally not known \textit{a priori}.
It, however, can be estimated via a sequential probability assignment (SPA).
If we replace $P_{Z_t}$ with a SPAs with universality guarantees, such as the LZ78 Transform of ``good zero-order SPAs,'' we can then develop a universal estimator for the Bayes response.
In the remainder of this section, state tight upper bounds on the excess estimation loss and present LZ78-transform-based filters that achieve them with minimal computational overhead. 

\textbf{SPA-Based Estimators:}
\cite{yan2025filt} considers three key scenarios: causal estimation, estimation with delay and estimation with lookahead.
For each scenario, a denoiser is produced by deriving the Bayes optimal estimator and replacing the appropriate distributional information of $Z^n$ with a SPA, $Q_{Z^n}$.
The resulting estimator is called the \textit{mismatched estimator}.
\begin{enumerate}
    \item \textbf{Causal estimation}: When only past and current observations $Z^t$ are available, the side information of \prettyref{lem:opt-with-side-info} is $Z^{t-1}$, in which case the Bayes optimal causal estimator and corresponding mismatched estimator under SPA $Q$ are:
    \begin{align*}
      \hat X_t^{\mathrm{opt}}(Z^t) &=
         \hat X_B\!\Bigl(
           F\bigl(P_{Z_t\mid Z_1^{t-1}},\Pi,Z_t\bigr)
         \Bigr) \\
         &\approx \hat X_B\bigl(\tilde P_{X_t\mid Z^t}\bigr) \triangleq \hat X_B\bigl(F\bigl(Q_{Z_t\mid Z_1^{t-1}},\Pi,Z_t\bigr)\bigr). 
    \end{align*}

    \item \textbf{Estimation with delay}:
    Practical systems frequently suffer a fixed lag, so that at time~$t$
    one can only access the \emph{delayed} noisy sequence
    $Z^{t-d}\!=\!(Z_1,\dots,Z_{t-d})$ for some $d\!\ge\!1$.
    Here, the Bayes optimal estimator and estimator under SPA $Q$ are:
    \begin{align*}
      \hat X_t^{\mathrm{opt}}(Z^{t-d}) &=
         \hat X_B\!\Bigl(
           F\bigl(P_{Z_t\mid Z_1^{t-d}},\Pi,Z_t\bigr)
         \Bigr) \\
         &\approx   \hat X_B\!\bigl(
          \Pi^{-\!\top} Q_{Z_t\mid Z^{t-d}}.
      \bigr). 
    \end{align*}

    \item \textbf{Estimation with lookahead}:
    When limited future samples are available, the filter may exploit a look-ahead window of length $\ell\!\ge\!0$.  At time~$t$ the observable data consist of
    $\bigl(Z^{t-1},Z_t,Z_{t+1}^{\,t+\ell}\bigr)$. The Bayes-optimal rule and SPA-estimated equivalent are now:
    \begin{align*}
      \hat X_t^{\mathrm{opt}}\!\bigl(Z^{t+\ell}\bigr)
        &=
        \hat X_B\!\bigl(
            P_{X_t\mid Z^{t-1},\,Z_{t+1}^{\,t+\ell}}
        \bigr) \\
        &\approx \hat X_B\!\Bigl(
           \tilde P_{X_t\mid Z^{t-1},\,Z_{t+1}^{\,t+\ell}}\Bigr)
           \triangleq \hat X_B\!\Bigl(
        F\!\bigl(
           Q_{Z_t\mid Z^{t-1},\,Z_{t+1}^{\,t+\ell}},\,
           \Pi,\,Z_t
        \bigr)
      \Bigr).
    \end{align*}
    
\end{enumerate}

\textbf{Excess loss bounds:}
In each of the above cases, the loss of the mismatched estimator can be bounded with respect to the relative entropy between $P_{Z^n}$ and $Q_{Z^n}$.
\begin{theorem}[Excess loss bound for causal estimation]
\label{thm:excess-causal}
For any sequence $Z^n$ and SPA~$Q$, the normalised expected excess
loss satisfies
\begin{align*}
\mathbb{E}\!\left[
    \frac1n\!\sum_{t=1}^n \Lambda\bigl(X_t,
        \hat X_B^{Q_{Z^{t-1}}}(Z_t)\bigr)
      - \frac1n\!\sum_{t=1}^n \Lambda\bigl(X_t,
        \hat X_t^{\mathrm{opt}}(Z^t)\bigr)
\right]
     \;\le\;
     \sqrt{2\,C_1(\Pi)\,\Lambda_{\max}}
     \,\sqrt{\frac1n\,D\bigl(P_{Z^n}\Vert Q_{Z^n}\bigr)},
\end{align*}
where\; $C_1(\Pi) =\|\Pi^{-\!\top}\|_{\max}\,\,|\mathcal{A}_X|$,\;
$\Lambda_{\max}=\max_{x,\hat x}\Lambda(x,\hat x)$,  and
$D(\cdot\Vert\cdot)$ denotes relative entropy.
\end{theorem}

This inequality is \textbf{distribution-agnostic}: it controls the regret caused by any mismatched SPA solely via its cumulative KL divergence from the true law of the observations. The tightness of the bound is discussed further in~\cite{yan2025filt}.

\begin{theorem}[Excess loss bound for delay]
\label{thm:excess-delay}
For any fixed delay $d\!\ge\!1$,
\begin{align*}
\mathbb{E}\!\left[
      \frac1n\!\sum_{t=1}^n
        \Lambda\!\bigl(X_t,\hat X_{B}^{Q_{Z^{t-d}}}(Z_t)\bigr)
     -\frac1n\!\sum_{t=1}^n
        \Lambda\!\bigl(X_t,\hat X_t^{\mathrm{opt}}(Z^{t-d})\bigr)
\right]
  \;\le\;
  \sqrt{2\,\Lambda_{\max} C_1(\Pi)}
  \;\sqrt{\frac{d}{n}\;
          D\!\bigl(P_{Z^n}\,\|\,Q_{Z^n}\bigr)}.
\end{align*}
\end{theorem}

\begin{theorem}[Excess loss bound for look-ahead]
\label{thm:excess-lookahead}
For any $\ell\!\ge\!0$,
\begin{align*}
\mathbb{E}\!\left[
      \frac1n\!\sum_{t=1}^n
        \Lambda\!\bigl(
           X_t,
           \hat X_{B}^{Q_{Z^{t-1},Z_{t+1}^{\,t+\ell}}}(Z_t)
        \bigr)
     -\frac1n\!\sum_{t=1}^n
        \Lambda\!\bigl(
           X_t,
           \hat X_t^{\mathrm{opt}}(Z^{t+\ell})
        \bigr)
\right]
  \;\le\;
  \sqrt{2\,C_1(\Pi)\,\Lambda_{\max}} \sqrt{\frac{1+\ell}{n}\cdot \mathcal{D}}. 
\label{eq:lookahead-bound}
\end{align*}
where $\mathcal{D} = D(P_{Z^{n+\ell}}|| Q_{Z^{n+\ell}})$.
\end{theorem}


In both the delay and lookhead cases, the penalty grows with the square root of the corresponding delay or lookhead length.
Hence, for any \emph{fixed} $\ell$ or $d$, the excess loss still vanishes provided $\frac1n D(P_{Z^{\,n+\ell}}\!\parallel Q_{Z^{\,n+\ell}})\!\to\!0$. 
Together, Theorems~\ref{thm:excess-causal}, \ref{thm:excess-delay}, and \ref{thm:excess-lookahead}
characterize the robustness of SPA-based universal filters across strictly causal, delayed, and look-ahead scenarios.

\subsubsection{Universal Filtering Schemes}
We now explore universal filtering schemes for the causal, delayed, and lookahead settings. The bounds on the excess loss due to the mismatch, introduced in the previous section, are in terms of the relative entropy. If we have universal SPAs, which satisfy the equation below,
\begin{equation}
    \lim_{n \to \infty} \frac{1}{n} D \left( P_{Z^n} \Vert Q_{Z^n} \right) = 0,\quad \forall P \text{ stationary},
\label{eq:univspa}
\end{equation}
we can implement a scheme whose excess loss converges to zero. Thus, we can ensure the universality of the induced estimators under universal SPAs.
For such a SPA, we can compute the following universal filters (in the sense that, for stationary sources, their normalized expected loss converges to the corresponding loss of the optimal filter):

\textbf{Causal filtering:}
Given $Q$ satisfying~\ref{eq:univspa},
\begin{align}
    \hat{X}_t^{\text{ao}}(Z^t) &\triangleq \hat{X}_B \left( F\left( Q_{Z_t | Z^{t-1}}, \Pi, Z_t \right) \right) \\
    &= \argmin_{\hat{X}(\cdot): \mathcal{A}_{Z^t} \to \mathcal{A}_{\hat{X}}} \sum_{x \in \mathcal{A}_X} \Lambda \left( x, \hat{X}(Z^t) \right) \frac{ \Pi(x, Z_t) \left( \Pi^{-T}(x, \cdot) Q_{Z_t | Z^{t-1}} \right) }{ Q_{Z_t | Z^{t-1}}(Z_t | Z^{t-1})}.
\label{eq:ao-filter}
\end{align}

\textbf{Filtering with delay:}
For delay $d \geq 1$,
\begin{align}
    \hat{X}_t^{\text{ao}}(Z^{t-d}) &\triangleq \hat{X}_B \left( Q_{X_t | Z^{t-d}} \right) \nonumber
    = \hat{X}_B \left( \Pi^{-T} Q_{Z_t | Z^{t-d}} \right) \nonumber \\
    &= \arg\min_{\hat{X}(\cdot): \mathcal{A}_{Z^{t-d}} \to \mathcal{A}_{\hat{X}}} \sum_{x \in \mathcal{A}_X} \Lambda \left( x, \hat{X}(Z^{t-d}) \right) \Pi^{-T} Q_{Z_t | Z^{t-d}}(x).
    \label{eq:ao-univ}
\end{align}

The complexity of calculating $Q_{Z_t | Z^{t-d}}$ from the SPA increases exponentially with $d$, requiring marginalization over $Z_{t-d+1}^{t-1}$ from $Q_{Z_{t-d+1}^t | Z^{t-d}}$. This makes the computation infeasible for large $d$ or $\mathcal{A}_Z$, even when the original SPA
has a per-time-step complexity of $O(1)$. A practical alternative to exact marginalization in delayed filtering is to use \textit{Monte Carlo simulations} to approximate $Q_{Z_t \mid Z^{t-d}}$. This involves generating $M$ sample sequences from $Q_{Z_{t-d+1}^t \mid Z^{t-d}}$ and observing the outcomes for $Z_t$. This reduces the computational complexity from exponential in $d$ to linear in $M$ and $d$, i.e., $O(Md)$.

By adjusting $M$, one can balance between computational cost and approximation accuracy, with even modest $M$ values often yielding effective results.  The the precise theorem and proof is presented in~\cite{yan2025filt}, which quantifies the approximation error and establishes bounds on the excess loss due to sampling.
The result characterizes the trade-off between approximation quality and computational complexity.

\textbf{Filtering with lookahead:}
With a lookahead of $\ell \geq 0$,
\begin{align}
\hat{X}_t^{\text{ao}} \left(Z^{t+\ell} \right) &\triangleq \hat{X}_B \left( F\left( Q_{Z_t \mid Z^{t-1}, Z_{t+1}^{t+\ell}}, \Pi, Z_t \right) \right) \nonumber \\
&= \argmin_{\hat{X}(\cdot): \mathcal{A}_{Z^{t+\ell}} \to \mathcal{A}_{\hat{X}}}
\sum_{x \in \mathcal{A}_X} \Lambda\left(x, \hat{X}(Z^{t+\ell}) \right) \nonumber \times \frac{
\Pi(x, Z_t) \left( \Pi^{-T}(x, \cdot) Q_{Z_t \mid Z^{t-1}, Z_{t+1}^{t+\ell}} \right)
}{
Q_{Z_t \mid Z^{t-1}, Z_{t+1}^{t+\ell}}(Z_t \mid Z^{t-1}, Z_{t+1}^{t+\ell})
}.
\label{eq:ao-la}
\end{align}

When implementing such a filter with lookahead $\ell \geq 0$, the conditional probability $Q_{Z_t \mid Z^{t-1}, Z_{t+1}^{t+\ell}}$ can be efficiently computed using Bayes' rule and the chain rule of probability. This allows future observations to be incorporated sequentially, with per-time-step complexity that scales linearly in $\ell$.

\subsubsection{Empirical Results}
To assess the performance of the universal filter described above, \cite{yan2025filt} applies it to a Markov process setting. The performance is then compared with the Wiener filter, a traditional linear filtering technique, as well as with the theoretical optimum of non-linear filtering for this setting.

The noise-free data $X_t \in \{-1, 1\}$ is generated from a first-order symmetric binary Markov source with transition probability $p$. It is then corrupted by an i.i.d. additive noise process $N_t$, where $N_t$ takes on values of $+1$ and $-1$ with equal probability.
This defines a noisy observation model $Z_t = X_t + N_t$.
See \cite{yan2025filt} for more details, including the resulting channel matrix.

\begin{figure}[h]
    \centering
    \includegraphics[width=0.5\linewidth]{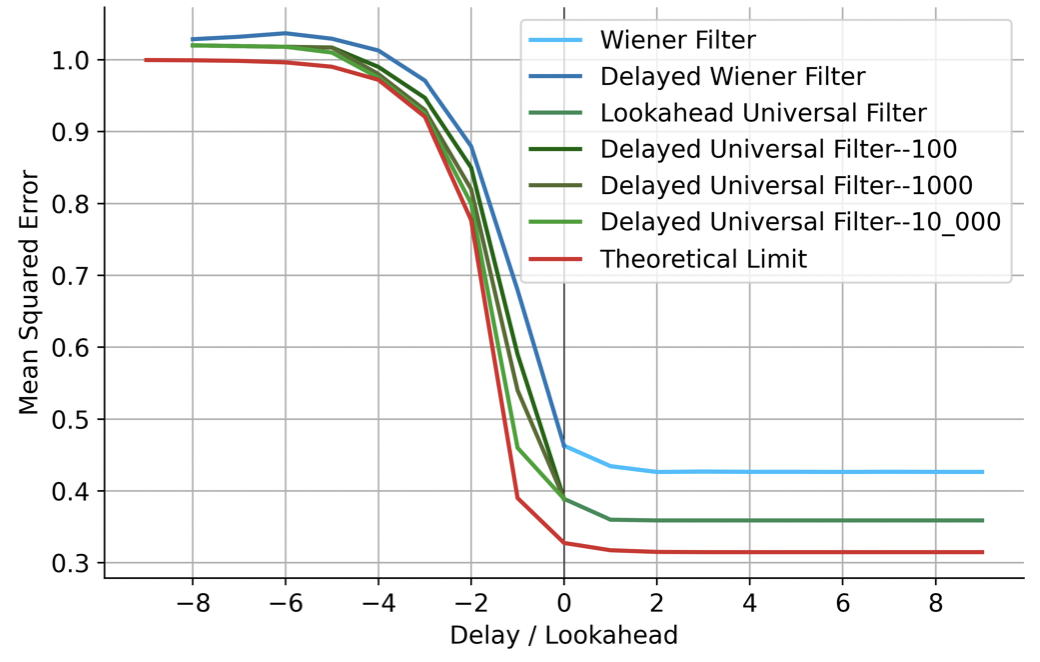}
    \caption{MSE loss of universal filter, increasing the number of Monte Carlo steps (figure from \cite{yan2025filt})}
    \label{fig:markov-filt}
\end{figure}

Figure~\ref{fig:markov-filt} plots the mean–squared–error (MSE) attained by the proposed universal filter across a continuum of \emph{delay/look-ahead indices} (negative values denote delay $d$, positive values denote look-ahead~$\ell$).  The solid black curve is the theoretical MSE limit for the controlled Markov process, obtained via dynamic-programming
evaluation of the Bayes estimator.

\textbf{Accuracy versus Monte-Carlo budget:}
For the \textit{delayed} regime, three Monte-Carlo budgets are reported $(M\!=\!100,\;1{,}000,\;10{,}000)$ when approximating
$Q_{Z_t\!\mid Z^{t-d}}$. With $M=100$ the universal filter already tracks the optimal curve closely, deviating by at most
$1.2\!\times\!10^{-3}$ in high-delay regions; increasing $M$ to
$10{,}000$ closes the gap almost completely, confirming the
$\sqrt{\log M / M}$ error decay predicted in~\cite{yan2025filt}. The look-ahead results (right half of the plot) match the theoretical bound without sampling, since $Q_{Z_t\!\mid Z^{t-1},Z_{t+1}^{t+\ell}}$ is computable in closed form.

\textbf{Computation/accuracy trade-off:}
The above experiment highlights a practical knob for real-time systems: larger $M$ yields tighter performance at the cost of $O(Md)$ arithmetic operations per step. In latency-critical environments one can cap $M$ or even revert to the exact $O(1)$ causal filter and still achieve graceful degradation.

The LZ78-SPA filter requires only: a) $O(1)$ updates for the causal predictor and b) $O(|\mathcal{A}_X|)$ arithmetic for the Bayes response, making the overall per-timestep complexity linear in the alphabet size.
In contrast, neural network-based filters incur at least $\Omega(n^2)$ multiply-accumulate operations per time step and demand GPU-class memory for weights and activations, rendering them impractical for low-compute edge devices.
Across all described regimes—causal, delayed, and look-ahead: the universal LZ78-based filter attains near-Bayes MSE with orders-of magnitude lower compute and memory footprint than neural alternatives, confirming its suitability for resource-constrained, low-latency applications.

\subsection{Sequence Generation}
The LZ78 transform lends itself naturally to causal generation of individual sequences.
\cite{sagan2024familylz78baseduniversalsequential} outlines a procedure for using the LZ78 transform of the Dirichlet mixture SPA of \eqref{eqn:bayesian-mixture} for causal unconditional sequence generation.
This includes two phases: a ``training'' phase and a ``generation'' phase.
During the training phase, an LZ78 prefix tree is formed using the available training sequences.
As described in \cite{sagan2024familylz78baseduniversalsequential}, this prefix tree includes the number of times that each node was traversed, which is sufficient information to compute the LZ78-transformed SPA.
After the training phase, the tree is frozen (as in classification): no new nodes are added and the traversal counts for each node are no longer updated.

Then, the generation phase involves repeating two steps: (1) generating a symbol according to the SPA at the current node of the LZ78 tree, (2) traversing the tree according to the newly-generated symbol (returning to the root if necessary).
Optionally, before performing causal generation, we can traverse the tree using a pre-determined sequence known as ``seeding data.''
To improve the quality of generated sequences, \cite{sagan2024familylz78baseduniversalsequential} also employes several heuristics, including temperature, top-$k$ sampling, and \textit{backshift parsing}.
In backshift parsing, upon reaching the leaf of the LZ78 prefix tree (where the corresponding SPA has not seen any symbols), we return to the root and traverse with a suffix of the generated data before generating the next symbol.
Refer to \cite{sagan2024familylz78baseduniversalsequential} for more details.

Preliminary experiments for byte-level English text generation (on the \texttt{tiny-shakespeare} dataset~\cite{tinyshakespeare}, \texttt{realnewslike} segment of C4~\cite{c4}, and \texttt{tinystories} dataset~\cite{tinystories}), as presented in \cite{sagan2024familylz78baseduniversalsequential} are promising---training took a mere 6 minutes for 500~MB of data, and even for the 1~MB \texttt{tiny-shakespeare} dataset, the LZ78 SPA qualitatively captured the grammar and style of the training data.
This preliminary experiment becomes a natural inspiration for the task of \cite{ding2025lzmidicompressionbasedsymbolicmusic}: Symbolic Music Generation.

\subsubsection{Theoretical Results}

Define the \textit{LZ78-SPA} as the LZ transform of a Dirichlet mixture SPA, i.e.,
\begin{equation}
    q^{LZ,\gamma}(a|x^{t-1})\triangleq \frac{N_{LZ}(a|x^{t-1})+\gamma}{\sum_{a'\in \mathcal{X}}N_{LZ}(a'|x^{t-1})+\gamma|\mathcal{X}|}.
    \label{eq:spa}    
\end{equation}
To justify our use of the LZ tree model, we present the following theorem, which establishes that an LZ tree trained on a sufficiently large dataset will closely approximate the true data distribution: 
\begin{theorem}[Universal Convergence of LZ78-SPA]
\label{thm:LZ78-main}
Let \(P\) be the law of a process with components taking values in a finite alphabet \(\mathcal{X}\), and let \(Q^m\) be the LZ78-based sequential probability assignment (SPA) constructed using \(m\) i.i.d training sequences from \(P_{X^n}\). Then, for any fixed \(n\),
\[
D\bigl(P_{X^n} \,\big\|\, Q^m_{X^n}\bigr) \;\xrightarrow[m\to\infty]{\text{a.s.}}\; 0,
\]
where \(D(\cdot\|\cdot)\) denotes the relative entropy, or Kullback--Leibler divergence.
\end{theorem}

\emph{Proof Sketch.} By construction, each node in the LZ78 tree tracks the empirical frequency of symbols following a particular context, and since every context with nonzero probability is visited infinitely often, the frequency with which a symbol \(a\) appears converges to the true conditional probability \(P(a | \text{context})\). Consequently, the LZ78-SPA, which returns $q(a|\textrm{context})$, an estimate of $P$ via empirical frequencies, assigns probabilities that  approximate $P$ increasingly accurately with more training data. As a result, when the assigned probabilities agree with the source distribution for all positively-probable contexts, the relative entropy \(D(P_{X^n}\|Q^m_{X^n})\) converges to 0 almost surely as \(m\to\infty\). A full, detailed proof is provided in Appendix~\ref{sec:generation-proofs}. Here, we emphasize that the key idea relies on the law of large numbers, which can be invoked because each relevant context is visited infinitely often. This allows us to conclude that local (node-wise) empirical distributions converge to the true underlying source probabilities.

\subsubsection{Symbolic Music Generation}

Deep learning–based generative models have achieved remarkable success in text, image, and audio synthesis. However, their substantial computational demands, particularly in sampling procedures employed in diffusion-based models, pose significant challenges for practical deployment because of high latency and the reliance on specialized hardware. Recent research has explored more computationally tractable alternatives that maintain competitive output quality. As we saw above, \cite{sagan2024familylz78baseduniversalsequential} introduces a learning framework based on universal sequential probability assignments (SPAs) derived from the celebrated Lempel-Ziv (LZ78) \cite{originalLZ78paper} compression. Using LZ-parsing under \textit{stationary} assumptions, the approach proposed next represents sequences efficiently within a tree-based structure. This methodology is backed by strong theoretical guarantees on runtime, memory usage as presented earlier.

\textbf{LZMidi}\cite{ding2025lzmidicompressionbasedsymbolicmusic} focuses on the generation of symbolic music, which refers to the task of generating music in a structured, discrete format, typically represented as MIDI or other symbolic encodings rather than raw audio waveforms. With its discrete structure and finite alphabet, symbolic music is well-suited to LZ-based SPAs. \textbf{LZMidi} induces an LZ78-based SPA on symbolic music, using this as a tool for symbolic music generation. Empirical evaluations indicate that LZ78-based SPA produces music of excellent perceptual quality, quantified using various metrics like Fréchet Audio Distance (FAD), Wasserstein Distance (WD) and KL-divergence (KL), while significantly reducing both training time and sampling overhead. \textbf{LZMidi} is directly motivated by foundational concepts stated above: it leverages universal compression (via LZ78-based SPA) to efficiently approximate the underlying statistical structure of symbolic music. This allows \textbf{LZMidi} to effectively capture the intrinsic redundancy and repetitive structure in musical sequences, providing a resource-efficient alternative for symbolic music generation. 

\vspace{2pt}

\begin{figure}[htbp]
    \centering
    \includegraphics[width=0.9\linewidth]{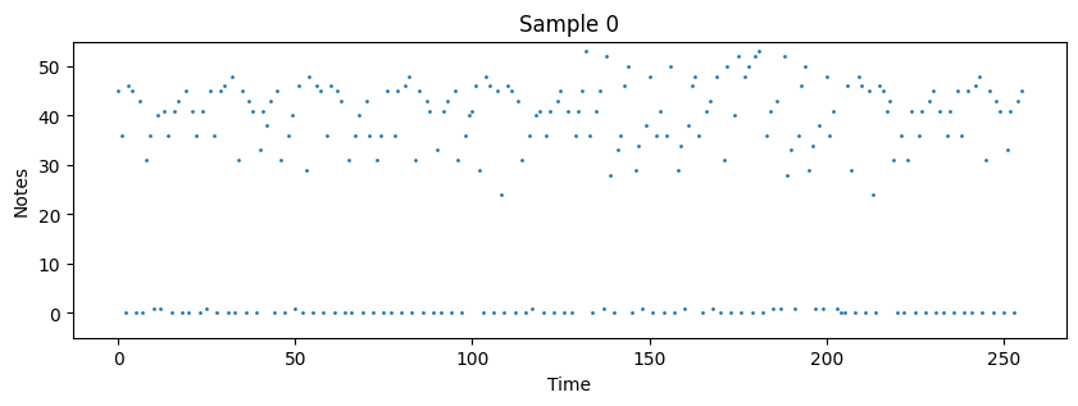}
    \caption{Sample Midi File from the Lakh MIDI Dataset}
    \label{fig:sample}
\end{figure}

\textbf{Methodology:}
The \textbf{Lakh MIDI Dataset (LMD)}, containing 648,574 samples, each with 256 notes drawn from the alphabet \( \mathcal{X} = \{0, 1, \dots, 89\} \), is used to train the LZ-based model for symbolic music generation. Here, 0 represents a rest, 1 denotes consecutive note continuation, and 2–89 correspond to actual pitch values. Figure \ref{fig:sample} illustrates a sample MIDI sequence. To build the alphabet for the LZ model, we simply treat each individual note as a symbol within the alphabet $\mathcal{X} = \{0, 1, 2, \dots, 89\}$, allowing us to traverse the tree and update the SPA sequentially for each note in the dataset. 

\vspace{-0.5cm}

\vspace{-1cm}

\textbf{Baseline Model:}
The \textbf{ASD3PM} (Absorbing State Denoising Diffusion Probabilistic Model) from \cite{plasser2023discretediffusionprobabilisticmodels}, a recent state-of-the-art model for generation of symbolic music is adopted as the benchmark. ASD3PM has been shown to outperform transformer-based autoregressive models~\cite{choi2020encodingmusicalstyletransformer} and prior continuous diffusion approaches~\cite{mittal2021symbolicmusicgenerationdiffusion}.
Through an iterative denoising process, ASD3PM directly denoises discrete MIDI tokens (instead of relying on latent representations): a forward chain stochastically masks tokens into an absorbing ``blank'' state, while a neural reverse network combining convolutional and transformer layers predicts the original sequence, say $x_0$. The loss function minimizes the evidence lower bound (ELBO) on the likelihood of $x_0$:
\begin{equation}
\mathcal{L}_{\text{ELBO}} = \mathbb{E}_{q(x_0)}\left[\sum_{t=1}^T \frac{T - t - 1}{T} \mathbb{E}_{q(x_t|x_0)} \left[ \log p_\theta(x_0|x_t) \right]\right]
\end{equation}

Training therefore optimizes an ELBO that directly conditions on $x_0$, enabling fewer diffusion steps and faster sampling than continuous-latent or autoregressive alternatives. This design helps ASD3PM outperform other baselines, especially on polyphonic infilling, yet it remains a heavyweight neural baseline against which a simpler, lightweight, unconditional \textbf{LZMidi} baseline is evaluated.

\medskip

\textbf{Evaluation Metrics:}
The quality of the generated music is evaluated through the following set of metrics:

\textbf{1) Framewise Self-Similarity Metrics:}
To evaluate statistical similarity between generated and original sequences, the overlapping area (OA) metrics from \cite{mittal2021symbolicmusicgenerationdiffusion} are adopted, which quantifies local pitch and duration distributions. Using a sliding 4-measure window (with a 2-measure hop), we fit Gaussian PDFs to pitch (\( p(k) \)) and duration (\( d(k) \)) distributions. The overlapping area (OA) between adjacent windows is defined as:
\begin{equation}
\text{OA}(k, k+1) = 1 - \text{erf}\left( \frac{c - \mu_1}{\sqrt{2}\,\sigma_1} \right) + \text{erf}\left( \frac{c - \mu_2}{\sqrt{2}\,\sigma_2} \right)
\end{equation}
where \( c \) is the intersection of the two Gaussian PDFs, and \( (\mu_1,\sigma_1), (\mu_2,\sigma_2) \) are the respective means and standard deviations. From the overlapping areas for pitch (\(\text{OA}_P\)) and duration (\(\text{OA}_D\)), the \textit{consistency} (C) and \textit{variance} (Var) are computed as follows:
\begin{equation}
\text{C} = \max\left(0, 1 - \frac{|\mu_\text{OA} - \mu_\text{GT}|}{\mu_\text{GT}}\right), \quad \text{Var} = \max\left(0, 1 - \frac{|\sigma_\text{OA}^2 - \sigma_\text{GT}^2|}{\sigma_\text{GT}^2}\right).
\end{equation}
where \( \mu_\text{OA}, \sigma_\text{OA}^2 \) and \( \mu_\text{GT}, \sigma_\text{GT}^2 \) are the means and variances of generated and ground-truth samples, respectively. \textit{Consistency} measures alignment with ground truth, while \textit{variance} reflects diversity. Higher consistency suggests realistic sequence structure, while balanced variance prevents mode collapse. However, strong OA scores alone \textbf{do not} guarantee perceptual quality, necessitating complementary evaluation (e.g., FAD). This motivates further evaluation with alternative qualitative metrics described below.

\textbf{2) Fréchet Audio Distance (FAD):}
Fréchet Audio Distance (FAD), inspired by the Fréchet Inception Distance (FID) \cite{heusel2018ganstrainedtimescaleupdate}, quantifies how closely the statistical distribution of generated audio aligns with real data. It computes the Fréchet distance between feature embeddings extracted from a pre-trained model (e.g., VGGish \cite{7952261}). Lower FAD scores indicate better perceptual similarity, making it a robust metric for evaluating generative quality.

\textbf{3) KL-Divergence (KL):}
The well-known Kullback-Leibler (KL) divergence measures the distance between the probability distributions of real and generated data. While lower KL values suggest better alignment, it primarily favours distributional similarity over perceptual quality. Knowing that KL can favor models producing mode-collapsed outputs, FAD is used as the primary metric for evaluating generation fidelity.

\textbf{4) Wasserstein Distance (WD):}
In addition to audio-based metrics like consistency, variance, FAD, and KL divergence, the Wasserstein Distance (WD) \cite{arjovsky2017wasserstein, gulrajani2017improved} is also used to evaluate numerical sequence distributions—particularly useful for hyperparameter tuning without costly neural network inference. WD measures the minimal cost of transforming one distribution into another, with lower values indicating a closer match between generated and real data distributions.

\medskip

\textbf{Training and Generation Setup:}
The LZMidi model is trained using the provided implementation of the LZ78-transformed Dirichlet micture SPA from \cite{sagan2024familylz78baseduniversalsequential}.
The \textit{Lakh MIDI Dataset} is split into an \textbf{80/20} train-test split. 
or training, LZMidi iterates through all samples in the training set, updating the LZ tree with each sequence.
For evaluation, 1,000 samples are generated for each value of block length.
All experiments are conducted on a CPU (Apple M1 Chip, 2021 MacBook).

\begin{figure}[htbp]
    \centering
    \includegraphics[width=0.9\linewidth]{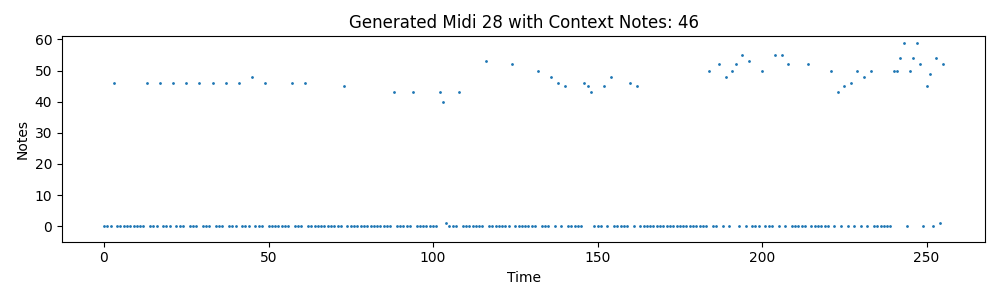}
    \vspace{2mm} 
    \includegraphics[width=0.9\linewidth]{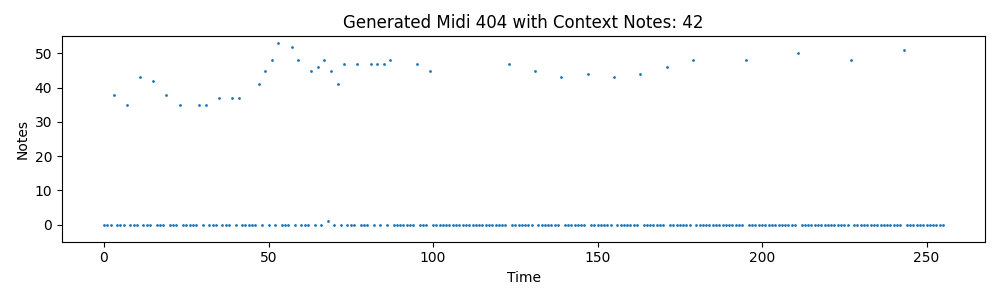}
    \vspace{2mm}
    \includegraphics[width=0.9\linewidth]{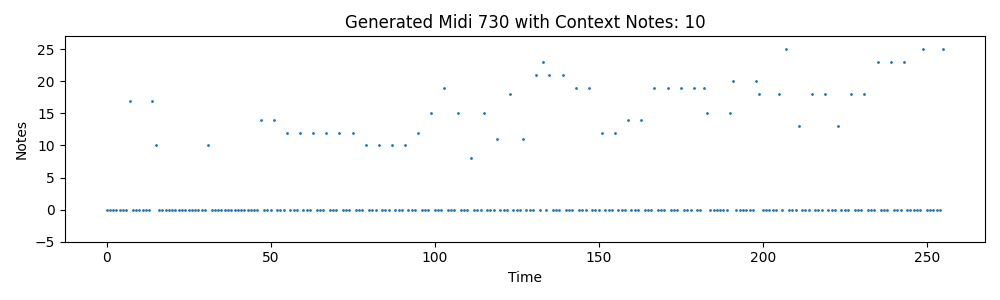}
    \caption{MIDI plots for Generated Samples using the LZMidi approach.}
    \label{fig:sample_all}
\end{figure}

To find the best configuration of the LZ78-transformed SPA (including Dirichlet parameter), we perform a hyperparameter sweep on the following parameters:

\begin{center}
        \small
        \renewcommand{\arraystretch}{1.3}
        \begin{tabularx}{\linewidth}{
          >{\raggedright\arraybackslash}p{3cm}
          >{\raggedright\arraybackslash}X}\toprule
            \textbf{Hyperparameter} & \textbf{Description} \\
            \midrule
            \textbf{Dirichlet Param.} ($\gamma$) & Determines the proximity of the SPA to the empirical distribution. A smaller $\gamma$ results in the SPA being closer to the empirical distribution. \\
            \textbf{Top-$K$} & A generation heuristic parameter: the model is only allowed to select the symbols with the $K$ highest probability. \\
            \textbf{Temperature} ($T$) & This parameter controls the randomness of the generated output by adjusting the probabilities of predicted symbols.
            $0$ means deterministic sampling, $1$ means sampling directly from the SPA, and higher values push the sampling distribution closer to uniform. \\
            \textbf{Minimum context} & The length of the suffix used for backshift parsing.
            \textit{This value is not swept, but rather set to a constant $64$.}\\
            \bottomrule
        \end{tabularx}
    \end{center}   
The sweep is also performed using Optuna \cite{akiba2019optunanextgenerationhyperparameteroptimization} to find the parameters optimizing Wasserstein distance (WD). ~\cite{ding2025lzmidicompressionbasedsymbolicmusic} suggests a categorical selection of the hyperparameters to Optuna for the hyperparameter sweep with the Wasserstein distance as the objective to minimize. It is observed that $\gamma$ and temperature affect the generation quality the most, with a smaller $\gamma\approx 5\times 10^{-5}$ and a larger temperature $T\approx 0.8$ being optimal in terms of the Wasserstein distance. For the final generation, $\gamma = 5\times 10^{-5}, T = 0.8, $ and $K=8$ are the chosen hyperparameters. MIDI plots of some of the generated samples are shown in Fig.~\ref{fig:sample_all}. Some of our generated music samples (corresponding to the plots in \ref{fig:sample_all}) are also attached \href{https://drive.google.com/drive/folders/1YglJn_KnBWZnze5xDQgHP5ufKqmZsaHn?usp=share_link}{here} for the reader's listening. 

Note that the ASD3PM baseline is also independently trained due to significant differences in sequence length, since \textbf{LZMidi} uses sequences of 256 tokens, whereas~ \cite{plasser2023discretediffusionprobabilisticmodels} trained on 1024 token sequences making direct use of these results unsuitable for a fair comparison. \cite{plasser2023discretediffusionprobabilisticmodels} reports a 24-hour training duration on $4\times$ NVIDIA 2080 Ti GPUs. Due to computational constraints and for a fair comparison, training time hers is fixed to approximately one hour.

\medskip

\textbf{Quantitative Results:}
The consistency and variance metrics are computed by comparing generated samples with 1000 randomly sampled sequences from both the training and test sets (in Table~\ref{tab:c_var_metrics}). LZMidi exhibits high consistency and variance for both pitch and duration, closely matching the dataset statistics. In fact, variance is \textbf{higher} in all cases for LZMidi.

\vspace{-0.5cm}
\begin{table}[ht]
\caption{Consistency and Variance}
    \centering
\label{tab:c_var_metrics}
    \begin{tabular}{|c|c|c|c|c|c|c|c|c|} \hline 
         &  \multicolumn{4}{|c|}{Training Set}& \multicolumn{4}{|c|}{Test Set}\\ \hline 
 & \multicolumn{2}{|c|}{Pitch}& \multicolumn{2}{|c|}{Duration}& \multicolumn{2}{|c|}{Pitch}& \multicolumn{2}{|c|}{Duration}\\ \hline 
 & C& Var&  C&Var&  C&Var&  C&Var\\\hline
         LZMidi&  $0.97$&  $\textbf{0.92}$&  $0.97$&$\textbf{0.93}$&  $0.97$&$\textbf{0.93}$& $0.97$&$\textbf{0.94}$\\ \hline 
 ASD3PM& $\textbf{0.98}$& $0.85$& $\textbf{0.99}$& $0.87$& $\textbf{0.98}$& $0.86$& $\textbf{0.99}$&$0.87$\\\hline
    \end{tabular}
\end{table}

The quality of the generated MIDI samples is evaluated using the three aforementioned metrics: Wasserstein Distance (WD), Fréchet Audio Distance (FAD), and Kullback-Leibler (KL) Divergence.  As shown in Table \ref{tab:fad_kl_metrics}, the  generated data is compared to both the training and testing datasets. \textbf{LZMidi} achieves a much lower WD and FAD in both the training and test sets, indicating superior fidelity and distributional alignment.

\vspace{-5pt}

\begin{table}[htbp]
    \centering
\caption{WD, FAD and KL Divergence metrics}
\label{tab:fad_kl_metrics}
    \begin{tabular}{|c|c|c|c|c|c|c|}\hline
 & \multicolumn{3}{|c|}{Training}&  \multicolumn{3}{|c|}{Test}\\\hline  
         &   WD&FAD&  KL &  WD&FAD&KL\\ \hline 
         LZMidi&   \textbf{8.57}&$\textbf{0.69}$& \textbf{1.42} &  \textbf{8.39}& \textbf{0.64}&\textbf{1.37}\\ \hline 
 ASD3PM&  27.91&4.22&$2.29$ &  27.96&4.05&2.26\\\hline
    \end{tabular}
  
\end{table}

\textbf{Training, Generation Time, and Memory Usage:} The  training time, generation time (per sample), and memory usage of \textbf{LZMidi} are also evaluated to underscore its computational advantages over deep learning–based methods. Table~\ref{tab:train_gen_time} summarizes these metrics for different \(L\) settings. The metrics for our models are substantially lower than those of D3PM. For instance, while our training time is fixed at approximately one hour (reflecting our computational constraints), \cite{mittal2021symbolicmusicgenerationdiffusion} report a 6.5-hour training duration on an Nvidia Tesla V100 GPU. 

\vspace{-5pt}

\begin{table}[htbp]
    \centering
\caption{Training Time, Generation Time, Memory Usage}
\label{tab:train_gen_time}
    \begin{tabular}{|>{\centering\arraybackslash}p{0.1\linewidth}|>{\centering\arraybackslash}p{0.20\linewidth}|>{\centering\arraybackslash}p{0.25\linewidth}|>{\centering\arraybackslash}p{0.15\linewidth}|} \hline 
         &  Training Time (s)&  Generation Time (s/sample)& Model Size (MB)\\ \hline 
         LZMidi&  \textbf{107.7}&  \textbf{0.016}& \textbf{287.1}\\ \hline 
 ASD3PM& 3480& 5.4 &306.2\\\hline
    \end{tabular}  
\end{table}

\textbf{Floating point operations (FLOPS) comparison:}

Next, we analyze the computation (FLOPS) required for sequential MIDI generation using the \textbf{ASD3PM} baseline. We set the sequence length for the generated MIDI files to 256 timesteps as in the experiments. To evaluate the computational efficiency of our diffusion-based baseline model for symbolic music generation, we analyzed the floating-point operations per second (FLOPS) across different layers of the network during training and inference. Table~\ref{tab:flops} provides a breakdown of the total FLOPS and the corresponding CPU utilization metrics for key operations. 

\vspace{-0.5cm}

\begin{table}[h!]
\centering
\caption{FLOPs and CPU Utilization of the Diffusion Baseline}
\label{tab:flops}
\medskip
\resizebox{\columnwidth}{!}{%
\begin{tabular}{|l|c|c|c|c|}
\hline
\textbf{Operation} & \textbf{Calls} & \textbf{Total FLOPs} (MFLOPs) & \textbf{Self CPU \%} & \textbf{CPU Total Time (ms)} \\ \hline
\texttt{aten::addmm} & 145 & 620,622.774 & 37.36\% & 736.399 \\ \hline
\texttt{aten::bmm} & 48 & 12,884.902 & 6.91\% & 126.123 \\ \hline
\texttt{aten::mul} & 1 & 377.487 & 1.08\% & 21.089 \\ \hline
\texttt{aten::add} & 97 & 203.424 & 1.25\% & 14.255 \\ \hline
\texttt{DataParallel::forward} & 50 & 104.858 & 2.14\% & 18.332 \\ \hline
\texttt{aten::expand} & 243 & 0.00 & 0.56\% & 55.675 \\ \hline
\texttt{aten::reshape} & 245 & 0.00 & 0.88\% & 56.436 \\ \hline
\end{tabular}%
}
\end{table}

\textbf{FLOP estimation for LZMidi:} Assuming a binary alphabet, the LZ78 parser creates at most $N_{\text{LZ}}\!\le\! k n/\log n$ phrases for a sequence $x^{n}$; for a general alphabet of size $\mathcal{X}$ this scales to $n|\mathcal{X}|/\log n$. In the worst case the dictionary tree collapses into a line, giving $\sum_{i=1}^{N_{\text{LZ}}-1} i = \mathcal{O}\!\big(N_{\text{LZ}}^{2}\big)=\mathcal{O}\!\big((n|\mathcal{X}|/\log n)^{2}\big)$ node updates and an equivalent training-time FLOP bound.  
During inference, computing the next-symbol probabilities merely scans the $|\mathcal{X}|$ children of the current node, costing $\mathcal{O}(|\mathcal{X}|)$ FLOPs, multiple orders lower than the $6.2\times10^{8}$ FLOPs logged for the diffusion baseline. These bounds underscore the computational advantage of LZMidi over diffusion models.

\vspace{-3pt}

\subsection{Discussion:}
Our results demonstrate the versatility and efficiency of the LZ78 transform in various data-driven tasks, underscoring its theoretical robustness and computational advantages. The LZ78 transform, achieves universality and showcases remarkable performance in classification, discrete filtering, and generative modeling tasks, particularly in resource-constrained environments. Empirical results from genomics classification and symbolic music generation highlight its practical efficacy and competitive performance against computationally intensive transformer-based models with a much lower inference and training overhead. Notably, the genomic classification results approach or surpass the accuracy of deep learning baselines with a significantly lower compute cost, positioning the LZ78 transform as a compelling choice for sustainable AI applications.

These findings prompt several avenues for future research. A detailed complexity analysis of the LZ78 transform remains essential, particularly to better understand performance across diverse tree structures. Exploring more advanced internal SPAs, such as n-gram models or Context Tree Weighting (CTW)~\cite{willems1995CTW}, could further enhance accuracy, especially if these SPAs converge to estimates below zero-order empirical entropy. Achieving such convergence could potentially strengthen existing theoretical guarantees or establish novel theoretical results. 

More specifically, in genomic classification, theoretical explorations similar to Ziv-Merhav cross-parsing~\cite{zivMerhav1993CrossParsing} could establish formal guarantees about the LZ78 scheme’s universality as a relative entropy estimator. Empirical comparisons with established n-gram SPAs would help benchmark the practical gains offered by the LZ78 transform. In the case of symbolic music, work will also explore leveraging universal‐type enumeration~\cite{Seroussi2004Universal} under LZ78 parsing, using efficient enumeration of universal‐type classes to preserve higher‐order statistical structure during generation.
Finally, expanding evaluations to additional genomic (and symbolic music) datasets and tasks would solidify the broader applicability and robustness of this approach across the genomics domain.

%
\section{Conclusion}
In this work, we investigated the trade-offs between information and computation within nonlinear transform-based compression frameworks, for various information-processing tasks across image, textual, spatial and symbolic domains. 
Initially, we examined two emerging nonlinear representation techniques: Implicit Neural Representations (INRs) and 2D Gaussian Splatting (GS). Our analysis highlighted essential distinctions, with INRs offering compact and resolution-adaptive representations capable of capturing intricate high-frequency details at significant computational costs, whereas GS provided rapid, low-latency encoding and decoding with inherently intuitive spatial interpretations. We then introduced a novel textual transform domain for ultra-low bitrate compression, which simultaneously enhances perceptual quality and facilitates effective semantic denoising. By leveraging the human-readability of the textual representation, we showed how noise can be effectively filtered without degrading any essential information.

Next, we presented the novel Lempel–Ziv (LZ78) ``transform", which we formalized as a universal transformation over sequential probability models. The LZ78 transform provably upgrades simple zero-order SPAs into universal estimators competitive with any finite-memory model. We then show that the this transform is highly effective in practical scenarios, including in sequence classification, discrete filtering, and generative modeling tasks, highlighting its broad applicability across various domains. In classification, the LZ78 transform yields strong results across both general and domain-specific datasets, including genomics, where it rivals transformer-based baselines with significant gains in computational efficiency. In universal filtering, we showed that LZ-transformed SPAs enable robust, model-agnostic denoising under unknown channels, with performance guarantees across causal, delayed, and lookahead regimes. In sequence generation, particularly in case of symbolic music, the LZ78-SPA achieves remarkable perceptual fidelity while drastically reducing training and inference overhead compared to state-of-the-art diffusion models. 

Collectively, our findings articulate fundamental insights into the trade-offs intrinsic to nonlinear transformations. By integrating these transforms within contemporary computational paradigms, we demonstrate significant potential for advancing performance in compression, classification, denoising, and generative AI applications. Together, these results lay a possible foundation for ``bits for sustainable AI" -- paving the way for sustainable information processing across various data-driven tasks.



\newpage
\bibliographystyle{RS}
\bibliography{references}
\newpage
\appendix
\section{Proofs: LZ Transform}\label{app:lz-transform-proofs}
\textbf{Lemma \ref{lem:lz-transform-part-1}.}

    If SPA $q$ satisfies
    \[\limsup_{n\to\infty} \max_{x^n \in \Acal^n} \left(\frac{1}{n}\log \frac{1}{q(x^n)} - \mu_0(x^n)\right) \leq 0,\]
    then, for $\hat{q}= \lztrans\{q\}$,
    \[\limsup_{n\to\infty} \max_{x^n \in \Acal^n} \left(\frac{1}{n}\log \frac{1}{\hat{q}(x^n)} - \frac{1}{n} \sum_{z \in \Zcal(x^n)} \left|\mathcal{Y}\{x^n, z)\} \right| \mu_0\left( \mathcal{Y}\{x^n, z)\} \right)\right) \leq 0\]

    \begin{proof}
        Dividing the log loss of $\hat{q}$ according to the LZ78 context of each symbol and then applying the definition of the LZ transform from \prettyref{def:lz-transform},
        \begin{equation}
            \begin{split}
                \frac{1}{n}\log \frac{1}{\hat{q}(x^n)} = \frac{1}{n}\sum_{t=1}^n \log \frac{1}{\hat{q}(x_t|x^{t-1})} &= \frac{1}{n}\sum_{z\in\Zcal(x^n)} \sum_{\{t \,:\, z_c(x^{t-1}) = z\}}  \log \frac{1}{\hat{q}(x_t|x^{t-1})} \\
                &\stackrel{(a)}{=} \frac{1}{n}\sum_{z\in\Zcal(x^n)} \log \frac{1}{q\left(\mathcal{Y}\{x^n, z\}\right)},
            \end{split}
            \label{eqn:split-q-hat-into-phrases}
        \end{equation}
        where (a) directly applies \prettyref{def:lz-transform}.

        The assumption on $q$ can be rewritten as: $\forall y^m \in \Acal^*$,
        \begin{equation}
            \frac{1}{m}\log \frac{1}{q(y^m)} - \mu_0(y^m) \leq \xi(m),\label{eqn:rewritten-q-assumption}
        \end{equation}
        where $\xi(m)$ is a function purely of $m$ and not $y^m$ and $\limsup_{m\to\infty} \xi(m) \leq 0$.
        $|\xi(m)|$ is also bounded above, i.e., $\xi(m) \leq B, \forall m \geq 1$, as the assumption on $q$ cannot hold if the loss ever grows unbounded.
        
        Applying \eqref{eqn:split-q-hat-into-phrases} and \eqref{eqn:rewritten-q-assumption},
        \begin{align*}
            &\max_{x^n \in \Acal^n} \left(\frac{1}{n}\log \frac{1}{\hat{q}(x^n)} - \frac{1}{n} \sum_{z \in \Zcal(x^n)} \left|\mathcal{Y}\{x^n, z)\} \right| \mu_0\left( \mathcal{Y}\{x^n, z)\} \right)\right) \\
            &\hspace{40px} = \max_{x^n \in \Acal^n} \frac{1}{n}\sum_{z \in \Zcal(x^n)} \left|\mathcal{Y}\{x^n, z)\}\right| \left(\frac{1}{\left|\mathcal{Y}\{x^n, z)\}\right|} \log \frac{1}{q(\mathcal{Y}\{x^n, z)\})} -  \mu_0\left( \mathcal{Y}\{x^n, z)\}\right) \right) \\
            &\hspace{40px}\leq \max_{x^n \in \Acal^n} \frac{1}{n}\sum_{z\in\Zcal(x^n)} \left|\mathcal{Y}\{x^n, z)\}\right| \xi\left(\left|\mathcal{Y}\{x^n, z)\}\right|\right) \triangleq \frac{1}{n} \max_{x^n \in \Acal^n} \sum_{z\in\Zcal(x^n)} m_z \xi(m_z),
        \end{align*}
        where we define $m_z \triangleq \left|\mathcal{Y}\{x^n, z)\}\right|$ to make notation less cumbersome.
        Note that, although not explicitly stated in the notation, $m_z$ is a function of $x^n$.
        We also define $z_t$ as shorthand for $z_c(x^{t-1})$.

        It now remains to show that:
        \[\limsup_{n\to\infty} \max_{x^n} \frac{1}{n}\sum_{t=1}^n \xi(m_{z_t}) \leq 0 \,\,\iff\,\, \forall \epsilon > 0,\,\exists N > 0 \text{ s.t. } \forall n > N,\, \max_{x^n} \frac{1}{n}\sum_{t=1}^n \xi(m_{z_t}) \leq \epsilon.\]

        By the definition of $\xi$, $\forall \delta > 0$, $\exists M > 0$ s.t., $\forall m > M$, $\xi(m) < \delta$.
        Using this fact, we now show:
        \begin{enumerate}[A.]
            \item For LZ phrases that are long enough, most of the phrase will have $\xi(m) < \delta$. 
            \item  For the remainder of symbols in those large phrases, the total sum of $\xi(m_{z_t})$ is not too large.
            \item For large enough $N$, there are very few phrases that are not ``long enough'' (as per B), and the sum of $\xi(m_{z_t})$ over those is small.
        \end{enumerate}
        We will set constants make the contribution of each of these three components bounded by $\frac{\epsilon}{3}$, leading to a total contribution of $\epsilon$.

        \textbf{A}:
        Set $\delta = \frac{\epsilon}{3}$. 
        Recall that for $m_{z_t} > M$, $\xi(m_{z_t}) > \frac{\epsilon}{3}$.
        Let $T_A$ be the set of timesteps with $m_{z_t} \geq M$.
        Then,
        \[\max_{x^n} \frac{1}{n}\sum_{t \in T_A} \xi(m_{z_t}) \leq \frac{|T_A|}{n}\frac{\epsilon}{3} \leq \frac{\epsilon}{3}.\]

        \textbf{B}:
        For a given node, $z$, of the LZ78 prefix tree, $m_z$ is the number of nodes in the sub-tree for which $z$ is the root.
        So, for every phrase, at most $M$ symbols have a corresponding $m_{z_t} < M$.
        For phrases longer than $L \triangleq \frac{3MB}{\epsilon}$, at most an $\frac{\epsilon}{3B}$ fraction of symbols have $m_{z_t} \leq M$.
        For those symbols, we can use the bound $\xi(m) \leq B$.
        Define $T_B$ to be the set of timesteps in phrases longer than $L$ with $m_{z_t} \leq M$.
        Then,
        \[\max_{x^n} \frac{1}{n}\sum_{t \in T_B} \xi(m_{z_t}) \leq \frac{|T_B|}{n}B \leq \frac{\epsilon}{3B}B = \frac{\epsilon}{3}.\]
        
        \textbf{C}:
        As each phrase is unique, there are at most $\sum_{\ell=1}^L A^\ell + L < LA^L + L$ phrases with length $< L$ (including possibly the last phrase).
        Those phrases comprise at most $L^2(A^L + 1)$ symbols.
        
        Choose $N = \frac{3L^2(A^L+1)B}{\epsilon}$.
        Then, for $n > N$, less than an $\frac{\epsilon}{3B}$ fraction of symbols are in phrases with length $\leq L$.
        Defining the corresponding timesteps as $T_C$, $\forall n > N$,
        \[\max_{x^n} \frac{1}{n}\sum_{t \in T_C} \xi(m_{z_t}) \leq \frac{|T_C|}{n}B \leq \frac{\epsilon}{3B} B = \frac{\epsilon}{3}.\]

        \textbf{A+B+C}:
        As $T_A \cup T_B \cup T_C = \{t\,:\, 1 \leq t \leq n\}$ (i.e., $T_A$, $T_B$, and $T_C$ together span all timesteps), $\forall n > N$,
        \[\max_{x^n} \frac{1}{n}\sum_{t=1}^n \xi(m_{z_t}) \leq \max_{x^n} \frac{1}{n}\sum_{t \in T_A} \xi(m_{z_t}) + \frac{1}{n}\sum_{t \in T_B} \xi(m_{z_t}) + \frac{1}{n}\sum_{t \in T_C} \xi(m_{z_t}) \leq 3\frac{\epsilon}{3} = \epsilon,\]
        so, as $\epsilon$ is arbitrary, $\limsup_{n\to\infty} \max_{x^n} \frac{1}{n}\sum_{t=1}^n \xi(m_{z_t}) \leq 0$.
\end{proof}

\textbf{Lemma \ref{lem:lz-transform-part-2}.}
    For any individual sequence $\xv$,
    \[\limsup_{n\to\infty} \frac{1}{n} \sum_{z \in \Zcal(x^n)} \left|\mathcal{Y}\{x^n, z)\} \right| \mu_0\left( \mathcal{Y}\{x^n, z)\} \right) \leq \mu(\xv).\]

    \begin{proof}
        For notational simplicity, define
        \[(\ast) \triangleq \frac{1}{n} \sum_{z \in \Zcal(x^n)} \left|\mathcal{Y}\{x^n, z)\} \right| \mu_0\left( \mathcal{Y}\{x^n, z)\} \right).\]
        Fix some $k > 0$.
                
        Let $v_\ell^{m_\ell}$ be the ordered subsequence of $x^n$ where the corresponding LZ78 context has length $\ell$.
        Using this notation, $(\ast)$ can be rearranged to get
        \begin{align*}
            (\ast) &= \frac{1}{n} \sum_{\ell = 0}^n\sum_{z \in \Acal^\ell} \left|\mathcal{Y}\{x^n, z)\} \right| \mu_0\left( \mathcal{Y}\{x^n, z)\} \right) \stackrel{(a)}{=}  \sum_{\ell = 0}^n \frac{m_\ell}{n} \mu_\ell(v_\ell^{m_\ell})
            \\
            &= \sum_{\ell=0}^k  \frac{m_\ell}{n} \mu_\ell(v_\ell^{m_\ell}) + \sum_{\ell=k+1}^n  \frac{m_\ell}{n} \mu_\ell(v_\ell^{m_\ell})
            \stackrel{(b)}{\leq} \sum_{\ell=0}^k \frac{m_\ell}{n} \mu_0(v_\ell^{m_\ell}) + \sum_{\ell=k+1}^n  \frac{m_\ell}{n} \mu_k(v_\ell^{m_\ell}).
        \end{align*}
        Here, $(a)$ follows from \prettyref{rem:k-order-markov-loss-sum-zero-order-loss} and $(b)$ is a consequence of the fact that, for any fixed input sequence, $\mu_k(x^n)$ is non-increasing in $k$.
        
        As the SPA that assigns a uniform probability to any symbol has a scaled log loss of $(\log A)$ on any individual sequence, the first term is upper-bounded by $\left(\log A\sum_{\ell=0}^k \frac{m_\ell}{n}\right)$.
        Also, within a phrase in an LZ78 parsing of a sequence, the length of the context increases by $1$ for each symbol in the phrase.
        So, each phrase has at most one symbol with a context length of $\ell$, $\forall \ell \in \Nbb$ and $m_\ell \leq C(x^n)$.
        Therefore,
        \[(\ast) \leq k\frac{C(x^n)}{n} \log A + \sum_{\ell=k+1}^n \frac{m_\ell}{n} \mu_k(v_\ell^{m_\ell}).\]
        \cite{originalLZ78paper} states that, for any individual sequence, $\frac{C(x^n)}{n} = o(1)$, so 
        \[(\ast) \leq \sum_{\ell=k+1}^n \frac{m_\ell}{n} \mu_k(v_\ell^{m_\ell}) + o(1).\]

        As pointwise minimization is concave, Jensen's inequality provides
        \begin{align*}
            \mu_k(x^n) &= \min_{q \in \Mcal_k} \frac{1}{n}\sum_{t=k+1}^n \log \frac{1}{q(x_t|x^{t-1})} \geq \min_{q \in \Mcal_k} \frac{1}{n}\sum_{\ell=k+1}^n \frac{m_\ell}{n} \left(\frac{1}{m_\ell}\log \frac{1}{q(v_\ell^{m_\ell})}\right)\\
            &\geq \sum_{\ell=k+1}^n \frac{m_\ell}{n} \min_{q \in \Mcal_k} \left(\frac{1}{m_\ell}\log \frac{1}{q(v_\ell^{m_\ell})}\right) = \sum_{\ell=k+1}^n \frac{m_\ell}{n} \mu_k(v_\ell^{m_\ell}).
        \end{align*}
        Thus, we can bound $(\ast)$ by
        \[(\ast) \leq \mu_k(x^n) + o(1) \implies \limsup_{n\to\infty} (\ast) \leq \mu_k(\xv),\]
        $\forall k > 0$.
        Then taking $k\to\infty$,
        \[\limsup_{n\to\infty} \,(\ast) \leq \mu(\xv).\]
    \end{proof}

\section{Proofs: Sequence Generation}\label{sec:generation-proofs}

\textbf{Theorem~\ref{thm:LZ78-main}}
Let \(P\) be the law of a process with components taking values in a finite alphabet \(\mathcal{X}\), and let \(Q^m\) be the LZ78-based sequential probability assignment (SPA) constructed using \(m\) i.i.d training sequences from \(P_{X^n}\). Then, for any fixed \(n\),
\[
D\bigl(P_{X^n} \,\big\|\, Q^m_{X^n}\bigr) \;\xrightarrow[m\to\infty]{\text{a.s.}}\; 0,
\]
where \(D(\cdot\|\cdot)\) denotes the Kullback--Leibler divergence.

\begin{definition}
    Let $Q^m$ be the probability model induced by an LZ78 tree built with $m$ equal-length realizations $X^n \simiid P_{X^n}$ sampled from source $P$ over alphabet $\Acal$.
    Let $X^{(i),n}$ denote the $i$\textsuperscript{th} such sequence generated.
    No assumptions are placed on $P$.
\end{definition}
\begin{remark}
    In this setting, the depth of the LZ78 tree is upper-bounded by $n$.
\end{remark}
\begin{definition}[Symbol counts]
    $\Ccal(a|x^n)$ is the number of times that symbol $a$ appears in $x^n$.
\end{definition}
\begin{theorem}
    Assume that the SPA at each node of the LZ78 tree, $q$, satisfies $q(a|y^m) - \frac{\Ccal(a|y^m)}{m} \to 0$, for all individual sequences $\yv$.
    Then as $m\to\infty$,
    \[D(P_{X^n}\lVert Q^m_{X^n}) \convas 0.\]

    \begin{proof}
        First, we define some additional notation:
        \begin{itemize}
            \item When parsing the $m$\textsuperscript{th} sample at index $t$, denote the current node of the LZ78 tree by $z_t^m$.
            The subsequence of symbols seen at $z_t^m$ until time $t$ is denoted $\Scal(z_t^m, m, t)$.
            Denote the length of this subsequence by $\ell(z_t^m, m, t)$.
            \item The LZ78 SPA at sample $m$, step $t$ is denoted $q(\cdot|\Scal(z_t^m, m, t))$.
            The SPA for a node that has not yet see data is denoted $q(\cdot)$.
        \end{itemize} ·
        Consider any $Y^n \in \Acal^n$ such that $P_{X^n}(Y^n) > 0$.
        \[\log \frac{1}{Q_{X^n}^m(Y^n)} = \sum_{t=1}^n \log \frac{1}{q(Y_t|\Scal(z_t^m, m, t))}.\]

        \begin{fact}\label{fact:count-ratio-conv}
            Fix $t \geq 1$, $Y^{t-1} \in \Acal^n$ such that $P_{X^{t-1}}(Y^{t-1}) > 0$.
            Then, $\forall a \in \Acal$,
            \[\frac{\Ccal(a|\Scal(z_t^m, m, t)))}{\ell(z_t^m, m, t)} \convas P_{X_{t}|X^{t-1}}(a|Y^{t})\quad \text{as } m\to\infty.\]

            \begin{proof}
                Fix some $m > 0$, and define
                \[\Ccal_m(Y^{t-1}) = \sum_{i=1}^m \indic{X^{(i),t-1} = Y^{t-1}}.\]
                We can bound $\ell(z_t^m, m, t)$ by a constant plus $\Ccal_m(Y^{t-1})$ on both sides.
                Note that, for $\mathcal{C}_m(Y^{t-1}) \geq t$, no returns to the root occur before reaching $z_t^m$, so $z_t^m$ is a depth-$(t-1)$ corresponding directly to the prefix $Y^{t-1}$.
                Otherwise, $z_t^m$ is some node encountered after a return to the root, in which case we cannot say much about $\ell(z_t^m, m, t)$ relative to $\mathcal{C}_m(Y^{t-1})$.

                By the law of large numbers, $\frac{1}{m} \Ccal_m(Y^{t-1}) \convas P_{X^{t-1}}(Y^{t-1}) > 0$ as $m \to \infty$.
                Therefore, there almost surely exists some $M$ such that $\Ccal_M(Y^{t-1}) \geq t$.
                From this point, consider $m > M$.
                
                For a lower bound on $\ell(z_t^m, m, t)$, the node $z_t^m$ was visited for all but maybe the first $t$ times $Y^t$ was seen.
                For an upper bound, we consider all the possible times $z_t^m$ was visited that do not correspond to $Y^{t-1}$, \ie, times  $z_t^m$ was visited after a return to the root.
                There is exactly one return to the root for every leaf of the tree, so the number of extra visits is bounded.
                So, $\forall m > M$, almost surely $\ell(z_t^m, m, t) = \Ccal_m(Y^{t-1}) + O(1)$.

                By the same logic, $\Ccal(a|\Scal(z_t^m), m, t)) = \Ccal_m(Y^{t-1} \,^\frown a) + O(1)$, where $\,^\frown$ represents sequence concatenation.

                By the law of large numbers, $\frac{1}{m} \Ccal_m(Y^{t-1}) = P_{X^{t-1}}(Y^{t-1}) + o_p(1)$, and analogously for $\frac{1}{m} \Ccal_m(Y^{t-1} \,^\frown a)$.
                As a result,
                \begin{align*}
                    \frac{\Ccal(a|\Scal(z_t^m, m, t)))}{\ell(z_t^m, m, t)} &= \frac{\frac{1}{m}\left(\Ccal_m(Y^{t-1} \,^\frown a) + O(1) \right)}{\frac{1}{m}\left(\Ccal_m(Y^{t-1}) + O(1)\right)} \\
                    &= \frac{P_{X^{t}}(Y^{t-1} \,^\frown a) + o_p(1)}{P_{X^{t-1}}(Y^{t-1}) + o_p(1)} \convas P_{X_{t}|X^{t-1}}(a|Y^{t-1}),
                \end{align*}
                by Slutsky's theorem.
            \end{proof}
        \end{fact}
        \begin{corollary}
            We know that, almost surely, for sufficiently large $m$ $\ell(z_t^m, m, t) = \Ccal_m(Y^{t-1}) + O(1)$.
            So, by the law of large numbers, $\ell(z_t^m, m, t)$ almost surely grows unbounded, for any $Y^{t-1}$ with non-zero measure under $P$.
        \end{corollary}
        \begin{corollary}\label{cor:node-spa-convergence}
            By assumption, $\forall\, Y^{t-1} \in \Acal^{t-1}$ with nonzero measure under $P$, 
            \[q(a|\Scal(z_t^m, m, t)) \to \frac{\Ccal(a|\Scal(z_t^m, m, t)))}{\ell(z_t^m, m, t)}, \quad\text{as}\quad \ell(z_t^m, m, t) \to \infty.\]
            Applying the fact $\ell(z_t^m, m, t)\convas \infty$, along with \prettyref{fact:count-ratio-conv}, as $m\to\infty$,
            \[q(a|\Scal(z_t^m, m, t)) \convas P_{X_{t}|X^{t-1}}(a|Y^{t-1}).\]
        \end{corollary}
        By \prettyref{cor:node-spa-convergence}, the continuous mapping theorem, and Slutsky's theorem (as $n < \infty$ is a fixed quantity), for any fixed $Y^n$ with nonzero measure under $P$,
        \[\log \frac{1}{Q_{X^n}^m(Y^n)} = \sum_{t=1}^n \log \frac{1}{q(Y_t|\Scal(z_t^m, m, t))} \convas \sum_{t=1}^n \log \frac{1}{P_{X_{t}|X^{t-1}}(Y_{t}|Y^{t-1})} = \log \frac{1}{P_{X^n}(Y^n)}.\]
        Applying this to the relative entropy,
        \begin{align*}
            D(P_{X^n}\lVert Q^m_{X^n}) &= \Ebb \left[ \log P_{X^n}(X^n)\right] + \Ebb \left[ \log Q^m_{X^n}(X^n)\right] \\
            &= \Ebb \left[ \log P_{X^n}(X^n)\right] + \sum_{Y^n : P_{X^n}(Y^n) > 0} P_{X^n}(Y^n) \log \frac{1}{Q^m_{X^n}(Y^n)}.
        \end{align*}
        Applying Slutsky's theorem (using the fact that the summation has a fixed, finite number of terms),
        \[\sum_{Y^n : P_{X^n}(Y^n) > 0} P_{X^n}(Y^n) \log \frac{1}{Q^m_{X^n}(Y^n)} \convas \Ebb \left[ \log P_{X^n}(X^n)\right],\]
        and, therefore,
        \[D(P_{X^n}\lVert Q^m_{X^n}) \convas \Ebb \left[ \log P_{X^n}(X^n)\right] - \Ebb \left[ \log P_{X^n}(X^n)\right] = 0.\]
    \end{proof}
\end{theorem}

\end{document}